\documentclass[epj,nopacs,final]{svjour}
\usepackage{graphics}
\usepackage{latexsym}
\usepackage{subfigure,rotating,pdflscape}
\usepackage{epsfig,color,rotating,amsmath,delarray,array}
\usepackage{pifont,float,amssymb}

\newcommand{\be}{\begin{eqnarray}}
\newcommand{\ee}{\end{eqnarray}}
\newcommand{\bc}{\begin{center}}

\newcommand{\beq}{\begin{equation}}
\newcommand{\eeq}{\end{equation}}

\newcommand{\nn}{\nonumber \\ }

\title{The Impact of New Polarization Data 
from Bonn, Mainz and Jefferson Laboratory on \boldmath$\gamma p\to \pi N$ Multipoles
}
\titlerunning{The impact of the new polarization observables
}
\authorrunning{A.V.~Anisovich {\it et al.}}

 \author{
 A.V.~Anisovich\mbox{\inst{1}$^,$\inst{2}},
 R.~Beck\inst{1}\thanks{\it Correspondence to: beck@hiskp.uni-bonn.de},
 M.~D\"oring\inst{3}\mbox{$^,$\inst{4}},
 M.~Gottschall\inst{1},
 J.~Hartmann\inst{1},
 V.~Kashevarov\inst{5},
 E.~Klempt\inst{1},
 Ulf-G.~Mei\ss ner\mbox{\inst{1}$^,$\inst{6}$,$\inst{7}}
 V.~Nikonov\mbox{\inst{1}$^,$\inst{2}},
 M.~Ostrick\inst{5},
 D.~R\"onchen\inst{1}$^,$\inst{6}\thanks{\it Correspondence to: roenchen@hiskp.uni-bonn.de},
 A.~Sarantsev\mbox{\inst{1}$^,$\inst{2}},
 I.~Strakovsky\inst{3},
 A.~Thiel\inst{1},
 L.~Tiator\inst{5},
 U.~Thoma\inst{1},
 R.~Workman\inst{3},
 Y.~Wunderlich\inst{1}
 \\
}                     
\institute{\inst{1}Helmholtz-Institut f\"ur Strahlen- und Kernphysik der Universit\"at Bonn,
Nu\ss allee 14-16, 53115 Bonn, Germany\\
\inst{2} NRS "Kurchatov Institute", PNPI, 188300, Gatchina, Russia\\
\inst{3}Department of Physics, George Washington University, 
725 21st Street, NW, Washington, DC 20052, USA\\
\inst{4}Thomas Jefferson National Accelerator Facility, 12000 Jefferson Avenue, Newport News,VA, USA\\
\inst{5}Institut f\"ur Kernphysik der Universit\"at Mainz, Johann-Joachim-Becher-Weg 45,
55099 Mainz, Germany\\
\inst{6}Bethe Center for Theoretical Physics, Universit\"at Bonn,
53115 Bonn, Germany\\
\inst{7}Institut f\"ur Kernphysik, Institute for Advanced Simulation,  
J\"ulich Center for Hadron Physics, JARA FAME and JARA HPC, 
Forschungszentrum J\"ulich,  52425 J\"ulich, Germany\\
}
\date{Received: date / Revised version: date}

\abstract{
New data on pion-photoproduction off the proton have been included in the partial wave analyses
Bonn-Gatchina and SAID and in the dynamical coupled-channel approach J\"ulich-Bonn. All reproduce the recent new data well: 
the double polarization data
for $E$, $G$, $H$, $P$ and $T$ in $\gamma p\to \pi^0p$ from ELSA, 
the beam asymmetry $\Sigma$ for $\gamma p\to \pi^0p$ and $\pi^+n$ from
Jefferson Laboratory, and the precise new differential cross section and beam asymmetry data  
$\Sigma$ for $\gamma p\to \pi^0p$ from MAMI.  The new fit results for the multipoles are 
compared with  predictions not taking into account the new data. The mutual agreement is 
improved considerably  but still far from being perfect.
}
\begin{document}
\maketitle
%
\section{Introduction}
QCD, Quantum Chromodynamics, is the accepted theory of the strong interactions.
However, the spectrum of the strongly interacting particles, the hadrons, 
remains to be derived from the basic principles of QCD. A good guide to the hadron
spectrum has been the quark model, which include some of the basic features of
QCD like the modelling of the confining force. These can thus be used to estimate
the expected number of states in a given energy range. While in the meson spectrum
one observes more states than given by simple quark-antiquark models, see e.g.
Refs.~\cite{Crede:2008vw,Klempt:2007cp,Agashe:2014kda}, the
situation is very different in the baryon sector: there are many more three-quark states
predicted (see e.g. \cite{Capstick:1986bm,Loring:2001kx,Giannini:2015zia}) than observed.
This is often called the {\it missing resonance} problem. Lattice QCD seems to confirm 
the large number of predicted states, even though
these calculations use a quark mass corresponding to $m_\pi=396$\,MeV 
and decay properties of these states are not investigated \cite{Edwards:2011jj}.
A more recent study of baryons with $J^P=1/2^\pm,3/2^\pm$ varied the pion mass between 255 and
596\,MeV and found good agreement with low-lying existing states \cite{Engel:2013ig}. Furthemore,
the first lattice QCD calculation of pion-nucleon scattering in the $J^P=1/2^-$ channel can be
found in Ref.~\cite{Lang:2012db}.
The prediction of hybrid baryons  increases the number of expected resonances even further
\cite{Capstick:2002wm,Dudek:2012ag}.
Several alternatives have been suggested to understand the problem of 
the {\it missing resonances}. Here, we just mention  the quark-diquark model,
\cite{Anselmino:1992vg}, the dynamically generation of baryon 
interaction of mesons and octet or decuplet (ground-state) baryons, see e.g. 
\cite{Kaiser:1995cy,Meissner:1999vr,Kolomeitsev:2003kt,Sarkar:2009kx,Oset:2009vf,Bruns:2010sv}, 
the AdS/QCD model~\cite{Forkel:2008un,deTeramond:2014asa,Klempt:2010du} or Schwinger-Dyson
models~\cite{Segovia:2014aza,Roberts:2015tha}.
Recent surveys of the
field can be found in \cite{Klempt:2009pi,Klempt:2012fy,Crede:2013kia,Plessas:2015mpa}.

A realistic alternative solution of the problem of {\it missing resonances} is the possibility
that these resonances have escaped observation due to a small $\pi N$ coupling. In $\pi N$ elastic
scattering, this coupling constant enters in the entrance and the exit channel, and resonances with
weak $\pi N$ coupling remain unobserved. To remedy this situation, experiments have been and 
are being performed at Bonn \cite{CBELSA}, Mainz \cite{CBMAMI}, and Newport News \cite{CLAS} 
in which high-energy photons
induce reactions off nucleons. In the case of photoproduction of pseudoscalar mesons, 
several (at least eight) independent observables with different settings of
beam and target polarization and with detection of the recoil polarization of the outgoing
nucleon need to be measured to define the four complex amplitudes governing the reaction
\cite{Barker:1975bp,Chiang:1996em,Fasano:1992es}. For an analysis aiming to determine the partial 
waves of lowest angular momenta, 
fewer observables are sufficient provided the statistical power of the experiments is 
sufficiently high \cite{Wunderlich:2014xya}. For example, in the region below the 2$\pi$ 
threshold, the Watson theorem provides an additional constraint which reduces the number of 
necessary observables. It has been shown that a measurement of the differential cross
section d$\sigma$/d$\Omega$ and of the photon beam asymmetry $\Sigma$ in $\pi^0$ photoproduction 
are sufficient to determine the electric contribution (or the E2/M1 ratio) to the 
(dominantly magnetic) $\Delta(1232)3/2^+$  excitation \cite{Beck:1997ew,Blanpied:1997zz}.
Additional data on the polarization observable $E$~\cite{Gottschall:2014uha}, 
$G$~\cite{AThiel:2012uha} and on $T, P$, and  $H$~\cite{Hartmann:2014mya}, allowed for a determination of a D-wave amplitude exciting the $N(1520)3/{2^-}$.

The four complex amplitudes for pion photoproduction are determined for each bin in energy 
and angle.  For each of these bins, one overall phase remains undetermined. 
This phase is therefore generally allowed to vary with energy and angle.
Hence, multipoles cannot be extracted from the four complex amplitudes
(known up to the overall phase), since the angular variation of the phase
prohibits the calculation of the partial wave projection integrals needed
for this purpose. Instead, (model-dependent) multipoles can be determined directly in a
truncated partial wave analysis, utilizing the full angular distributions of
the polarization observables 
\cite{Omelaenko:1981cr}. Then, there remains one
unknown overall phase, which now
depends only on the energy~\cite{Grushin:1989inbk}. The multipoles obtained
from such an analysis can be fitted within models which
return the wanted quantities: masses, widths, the helicity
amplitudes of the contributing resonances, and the background
contribution. At the resonance poles, these quantities can be uniquely defined and compared to conventional Breit-Wigner parameters \cite{Workman:2013rca}. 
The $\pi N$ coupling constants need to be 
determined from fits to $\pi N$ elastic scattering data.

At present, it is not yet possible to determine the four complex amplitudes uniquely 
for any reaction, at least not over a range from threshold to above 2\,GeV in mass. 
Instead, (model-dependent) multipoles can be determined
from a fit to the data. One of the best studied reactions is pion photoproduction off protons,
i.e. $\gamma p\to \pi^0p$ and $\gamma p\to \pi^+n$. Several experiments with different 
polarization settings have been performed recently which should lead to a better determination
of the multipoles. In this paper we compare the multipoles from leading partial wave analysis 
groups before and after inclusion of the new data.
The hope (and expectation)  is that the new data enforce a convergence of the different 
partial wave analyses, 
and that a more unified picture of the spectrum of $N^*$ and $\Delta^*$ resonances will emerge.

The paper is organized as follows. In Section~\ref{PWA} we outline the different partial 
wave analysis (PWA)  approaches -- Bonn-Gatchina (BnGa), J\"ulich-Bonn (J\"uBo),
MAID, and SAID -- used to
derive the multipoles. The four PWA groups exploit different data sets: MAID used the data
available in 2007 only, BnGa, J\"uBo and SAID make extensive use of most available data 
on pion and photo-induced reactions. The data are listed in Section~\ref{Data}. 
The three groups BnGa, J\"uBo, and SAID made new fits based on
additional new polarization data; these are also listed in Section~\ref{Data}.
In Section~\ref{Gamma} we present the results on the photoproduction multipoles 
from the energy dependent fits
to the data on $\gamma p\to \pi N$. The paper concludes with a discussion and a 
summary in Section~\ref{Sum}.

\section{\label{PWA}The partial wave analysis}
Ideally, a partial wave analysis should respect all constraints
imposed by theoretical considerations like gauge invariance, analyticity,
unitarity, crossing symmetry, and chiral symmetry. All reactions
should be fitted simultaneously, and the program should be fast enough
to allow for systematic studies. However, at the moment this is technically not
achievable. A self-consistent method satisfying
all (or at least most) constraints derived from theory leads
to highly complex equations and requires, consequently, high computational
efforts and long fitting times, like e.g. in the J\"ulich-Bonn approach.  Alternatively, an approach that allows for fast
computing enables one to change models and to test the presence of new
resonances easily. In this case, approximations have to be made. In the following,
the PWA models used to fit the data and the reactions they address
are outlined briefly. There are many further approaches which fit photoproduction
data. We mention here recent results of EBAC (Argonne-Osaka) \cite{Kamano:2013iva}, of the Gie\ss en group 
\cite{Shklyar:2012js}, and of the Kent State group \cite{Shrestha:2012ep}. 

\subsection{BnGa}
The frame of the BnGa  partial wave analysis code has been documented in
a number of publications. The approach relies on a  fully relativistically invariant
operator expansion method. It  combines the analysis of different reactions imposing 
analyticity and unitarity constraints directly. The code calculates amplitudes for the 
production and decay of baryon resonances and includes some $u$ and $t$-channel exchange diagrams
\cite{Anisovich:2004zz,Anisovich:2006bc,Anisovich:2009zy}. Here, we recall its main features.

Originally the partial wave amplitudes were described in the
K-matrix/P-vector approach \cite{Chung:1995dx}. For photoproduction
reactions, amplitudes representing the $t$ and $u$-channel exchanges
are added to the resonant part. The multipoles representing the $\pi
N$ elastic amplitudes are calculated exploiting a technique described
in \cite{Anisovich:2009zy}. In the analysis of new data sets, new
resonances with masses above 2.2~GeV were included as relativistic
multi-channel Breit-Wigner amplitudes.

Recently, the K-matrix/P-vector approach was changed to a
dispersion-relation approach based on the N/D technique. 
The real part of the two body loop diagrams was then
calculated using the regularization subtraction procedure with a
subtraction point taken at the channel threshold. In this simplified
version, the N/D-method corresponds to the K-matrix approach with
real parts of the loop diagrams taken into account. For three-body
final states, only the imaginary part of the loop diagrams is taken into
account, which is calculated as the spectral three-body integral.

A multi-channel amplitude {\boldmath$\hat{{\rm A}}(s)$} with the
matrix ele\-ments ${A}_{ab}(s)$ defines the transition amplitude from a
state 'a' to state 'b' where the initial and the final channels are,
e.g., $\gamma N$, $\pi N$, $\eta N$, $K\Lambda$, $\pi \Delta$. The
partial wave amplitude depends on the isospin $I$, the total angular
momentum $J$, and the parity $P$; the quantities $I (J^P)$ are
suppressed here. Transitions between different channels are taken
into account explicitly in the K-matrix; the amplitude is given by
 \be
 \mathbf{\hat A}(s) \;=\; \mathbf{\hat K}\;(\mathbf{\hat I}\;-
 \mathbf{\hat B \hat K})^{-1}\,. \label{k_matrix}
\ee
where $\mathbf{\hat K}$ is the K-matrix, $\mathbf{\hat I}$ is the
unity matrix and $\mathbf{\hat B}$ is a diagonal matrix of the
respective loop diagrams. The imaginary part of the elements $B_j$ is
equal to the corresponding phase spaces
\be
\hat B_j= Re B_j+\;i\rho_j\,.
\ee
If the real part of the loop diagram is neglected, this method
corresponds to the classical K-matrix approach. For two-particle
states (for example $\pi N$), the phase space for $J=L+1/2$
states is equal to (see \cite{Anisovich:2006bc}):
\be
\label{psr_plus}
\rho_+(s)=\frac{\alpha_L}{2L+1} \frac{2|\vec
k|}{\sqrt{s}}\frac{k_{10}+m_N}{2m_N} \frac {|\vec
k|^{2L}}{F(L,r,k^2)}
\ee
and for states with $J=L-1/2$, it is given by
\be
\label{psr_minus}
\rho_{-}(s)=\frac{\alpha_L}{L} \frac{2|\vec
k|}{\sqrt{s}}\frac{k_{10}+m_N}{2m_N} \frac {|\vec
k|^{2L}}{F(L,r,k^2)}\,.
\ee
Here, $s$ is the total energy squared, $k$ is the relative momentum
between baryon and meson, $\vec k$ its three-vector component,
$k_{10}$ is the energy of the baryon (with mass $m_N$) calculated in
the c.m.s. of the reaction, and $L$ is the orbital
angular momentum of the baryon-meson system. The
coefficient $\alpha_L$ is equal to:
\be
\alpha_L=\prod\limits_{n=1}^L\frac{2n-1}{n}\,.
\ee
The phase volume is regularized at large energies by a standard
Blatt-Weisskopf form factor normalized as\\
$F(L,r,k^2)\to k^{2L}\,(s\to\infty)$.  We use $r=0.8$\,fm as range
of the interaction. The functions $F(L,r,k^2)$ are given explicitely in
Ref.~\cite{Anisovich:2004zz}.

Using these phase-space volumes, the elements of the diagonal matrix $\hat
B$ are calculated as:
\be
&&B_{j}(s)=b_j+(s-(m_{1j}+m_{2j})^2)\times \nn
&&\int\limits_{(m_{1j}+m_{2j})^2}^\infty\!\!\!\! \frac{ds'}{\pi}
\frac{\rho_j(s')}{(s'-s-i\varepsilon)(s'-(m_{1j}+m_{2j})^2)},
\ee
where $b_j$ are subtraction constants and $m_{1j}, m_{2j}$ are
masses of the particles in the channel $j$. The exact
formulas for the three-body phase-space volume are given in
\cite{Anisovich:2006bc}.

The K-matrix $\mathbf{\hat K}$ is cast into the form
 \be
 K_{ab}\;=\;\sum_\alpha \frac{g_a^{(\alpha)} g_b^{(\alpha)}} {M^2_\alpha - s} \;+\; f_{ab}.
 \label{Kmat}
 \ee
$M_\alpha$ and $g_a^{(\alpha)}$ are the mass and the coupling
constant of the resonance $\alpha$; $f_{ab}$ describes a
direct (non-resonant) transition from the initial state $a$ to the
final state $b$, e.g. from $\pi N\to\Lambda K$.
For most partial waves it is sufficient to assume that $f_{ab}$ are
constants. The $S_{11}$ and $S_{31}$ waves require a slightly
more complicated structure, we use
\be
 f_{ab} =\frac{f_{ab}^{(1)}+f_{ab}^{(2)}\sqrt s}{s-s_0^{ab}}\,.
\ee
In this case, the $f_{ab}^{(i)}$ and $s_0^{ab}$ are constants
which are determined in the fits. In the case of the $S_{11}$ wave,
this more flexible parameterization is required to describe $S$-wave transitions
$\pi N\to \pi N$, $\pi N\to \eta N$, and $\eta N\to \eta N$. This form was also
tried for $P$-wave amplitudes but it did not improve the quality of the fit.
Let us note that this form is similar to the one used by SAID
\cite{Arndt:2006bf}.

The helicity-dependent photoproduction amplitude to produce the
final state '$b$' from the initial state '$a$' is then given by
 \be
  A_a \;=\; \hat P_b\;(\hat I\;-\;\hat B \hat K)^{-1}_{ba}\,,
\ee
where the production  vector $\mathbf{\hat P}$ is written in the form
 \be
 P_{b}\;=\;\sum_\alpha \frac{ g_{\gamma \rm N}^{(\alpha)} g_b^{(\alpha)}}{M^2_\alpha - s} \;+\;
 \tilde f_{(\gamma N)b}\,.
 \label{Pvect}
 \ee
The coefficients $g_{\gamma\rm N}^{(\alpha)}$ are the photo-couplings of the
resonance $\alpha$ and the non-resonant production of a final
state $b$ is re\-presented by $\tilde f_{(\gamma N)b}$. These are functions of $s$ but in practice,
constant values $\tilde f_{(\gamma N)b}$ are sufficient to obtain a good fit.

Due to its weak coupling, the $\gamma N$ interaction can be taken
into account in the form of a P-vector. No loops due to virtual
decays of a resonance into $\gamma N$ and back into the resonance are
required. A similar approach can be used to describe decay modes
with a weak coupling in the form of a D-vector amplitude. Then the
transition from K-matrix channel $a$ to the final channel $f$ can be
described as:
\be
\label{amplitude}  A_{af} \;=\; \hat D_{af} + [\hat K (\hat I\;-\;\hat
B \hat K)^{-1}\,\hat B ]_{ab} \hat D_{bf}\;.
\ee
The parametrization of the
D-vector follows the one for the P-vector:
\be
 D_{bf}\;=\;\sum_\alpha \frac{g_b^{(\alpha)}g_{f}^{(\alpha)} }{M^2_\alpha - s} \;+\;
 \tilde d_{bf}\,.
 \label{Dvect}
 \ee
Here, $g_{f}^{(\alpha)}$ is the coupling of a resonance to the final
state and $\tilde d_{bf}$ represents non-resonant transitions from
the K-matrix channel $b$ to the final state $f$. The  D-vector approach 
is used for the
channels with three-body final states (e.g. $f_0(980) N$,
$N(1535)\pi$). In the present analysis the non-resonant transitions for these final states are not needed to get a good description of the data.
In cases where both initial and final coupling
constants are weak, we use an approximation which we call PD-vector.
In this case the amplitude is given by
\be
A_{f} \;=\; \hat G_{f} + \hat P_{a}[(\hat I\;-\;\hat B \hat
K)^{-1}\,\hat B ]_{ab} \hat D_{bf}\;\;.
\ee
$\hat G_{f}$ corresponds to a tree diagram for the transition from
initial channel ($\gamma N$ in the case photoproduction)  to the
state '$f$':
\be
 G_{f}\;=\;\sum_\alpha \frac{g_{\gamma N}^{(\alpha)}g_{f}^{(\alpha)} }{M^2_\alpha - s} \;+\;
 \tilde h_{(\gamma N)f}\,.
 \label{PDvect}
 \ee
Here, $g_{\gamma\rm N}^{(\alpha)}$ is the resonance photoproduction
coupling and $\tilde h_{(\gamma N)f}$ represents direct non-resonant
transitions from the initial photon-nucleon system to the different
final states.

At high energies, angular distributions of photo-pro\-duced mesons
exhibit clear peaks in the forward direction. These peaks originate
from meson exchanges in the $t$-channel. Their contributions are
parameterized as $\pi$, $\rho(\omega)$, $\rm K$ or $\rm K^*$
exchanges.
The corresponding exchange
amplitudes are written in the form of the exchange of Regge trajectories.
The invariant part
of the $t$-channel exchange amplitude can then be written as
\cite{Anisovich:2008zz}:
\be
A=g(t)\frac{1+\xi\exp(-i\pi\alpha(t))}{\sin(\pi\alpha(t))} \left
(\frac{\nu}{\nu_0} \right )^{\alpha(t)} \;.
\ee
We use $g(t)=c\exp(-bt)$  as vertex function and form factor.
Further, $\alpha(t)$  describes the trajectory,
$\nu=\frac 12 (s-u)$, $\nu_0$ is a normalization factor, and $\xi$
the signature of the trajectory. Note that the Pomeron, $f_0$ and $\pi$ 
have a positive, whereas the $\rho$, $\omega$ and $a_1$ exchanges have 
a negative signature. The Reggeon propagators are written as
\be
R(+,\nu,t)=\frac{e^{-i\frac{\pi}{2}\alpha(t)}} {\sin
(\frac{\pi}{2}\alpha(t))} \left (\frac{\nu}{\nu_0}\right
)^{\alpha(t)}\;, \nonumber\\ R(-,\nu,t)=\frac{ie^{-i\frac{\pi}{2}\alpha(t)}}
{\cos (\frac{\pi}{2}\alpha(t))} \left (\frac{\nu}{\nu_0}\right
)^{\alpha(t)}  \;.
\label{7}
\ee
where '+' and '-' indicate the signature of the Regge-trajectories,
'+' for natural parity exchange ($J^{P}=0^+$, $1^-$, $2^+, \ldots$) and '-' for
unnatural parity exchange ($J^{P}=1^+, {2^-}, 3^+, \ldots$).
To eliminate the poles at $t<0$ additional $\Gamma$-functions are
introduced in Eq.~(\ref{7}). In the case of the Pomeron trajectory
\be
\sin \left (\frac{\pi}{2}\alpha(t)\right ) \to \sin \left
(\frac{\pi}{2}\alpha(t)\right ) \; \Gamma \left (\frac{
\alpha(t)}{2}\right ) \;.
\label{pomeron}
\ee
For $\rho$ and $\omega$ exchanges the poles at $t<0$ start from
$\alpha=-1$ and therefore
\be
\cos \left (\frac{\pi}{2}\alpha(t)\right ) \to \cos \left
(\frac{\pi}{2}\alpha(t)\right )  \; \Gamma \left (\frac
{\alpha(t)}{2} +\frac 12\right )\, .
\label{rho}
\ee

For pion production, e.g., we use $\rho$ and $\rho'$ exchanges with
the following trajectories:
\be
\rho:~\qquad\qquad\alpha(t)&=&0.5+(0.85/GeV^2) t\nonumber \\
\rho':\qquad\qquad\alpha(t)&=-&0.75+(0.85/GeV^2)t\,,
\ee
for $t$ given in units of GeV$^2$.
With these amplitudes, large data sets have been fitted. We quote here a few recent
papers
\cite{Anisovich:2011fc,Gutz:2014wit,Sokhoyan:2015fra,Sokhoyan:2015eja,TheCBELSA/TAPS:2015ula}.

\subsection{J\"uBo}
The J\"ulich-Bonn model has been developed over 
the years, see Refs.~\cite{Doring:2010ap,Ronchen:2012eg,Ronchen:2014cna,Ronchen:2015vfa} and 
references therein. It should be
noted that, strictly speaking, J\"uBo is not a PWA but rather a dynamical 
coupled-channel (DCC) approach, it also aims at a microscopic description of reaction dynamics.  
The J\"uBo aproach allows to extract the baryon spectrum based on a 
simultaneous analysis of pion- and photon-induced reactions. 
Theoretical constraints of the $S$-matrix like unitarity and analyticity are manifestly 
implemented or at least approximated in case of three-body unitarity. 
Left-hand cuts as well as the correct structure of 
complex branch points are taken into account. The formalism allows for the determination of 
resonance states in terms of poles in the complex energy plane of the scattering matrix together 
with the corresponding residues and helicity couplings. Note that such a determinations is
independent of any assumption on the resonance line shape, such as the commonly used Breit-Wigner parameterization \cite{Workman:2013rca}.

The scattering process of pion-induced reactions is described by a Lippmann-Schwinger 
equation that reads, after projection to the partial-wave basis,
\begin{multline}
T_{\mu\nu}(q,p',E)=V_{\mu\nu}(q,p',W) \\
+\sum_{\kappa}\int\limits_0^\infty dp\,
 p^2\,V_{\mu\kappa}(q,p,W)G^{}_\kappa(p,W)\,T_{\kappa\nu}(p,p',W) \ ,
\label{eq:scattering_juelich}
\end{multline}
where $\mu$ ($\nu$, $\kappa$) denote the outgoing (incoming, intermediate) meson-baryon channels 
$\pi N,\, \eta N, \,K\Lambda,\, K\Sigma$ and the effective three-body channels $\pi\Delta$, 
$\sigma N$, and $\rho N$. The re\-so\-nant sub-amplitudes in the latter channels describe the 
corresponding phase shifts of $\pi\pi$ and $\pi N$ scattering~\cite{Schutz:1998jx,Doring:2009yv}. 
Explicit expressions for the two- and three-body propagators encoded in $G_\kappa$ can be found in 
Ref.~\cite{Doring:2009yv}. Non-analyticities from complex branch points~\cite{Ceci:2011ae} are also
included. In Eq.~(\ref{eq:scattering_juelich}), $W=\sqrt s$ denotes the scattering energy  and
$q\equiv|\vec q\,|$ ($p'\equiv |\vec p\,'|$) is the
 modulus of the outgoing (incoming)  three-momentum that can be on- or off-shell. The scattering 
potential $V_{\mu\nu}$ is constructed from an effective Lagrangian using time-ordered perturbation theory. 
Resonance states are included in $V_{\mu\nu}$ in the form of  $s$-channel processes, while $t$- and 
$u$-channel exchanges of light mesons and baryons constitute the non-resonant part of the potential, 
in addition to contact interaction terms~\cite{Ronchen:2015vfa} to simulate heavy exchanges that do 
not behave as dynamical degrees of freedom.

The explicit inclusion of $t$- and $u$-channel diagrams is, on the one hand, necessary to fulfill 
constraints from three-body unitarity~\cite{Aaron:1969my}. On the other hand, the scattering amplitude 
has left-hand cuts with associated branch points. To respect these non-analyticities at the correct 
position, $u$-channel diagrams are included. The strengths of these cuts, however, cannot be derived 
in the J\"uBo framework. Therefore, the associated vertices can be varied using form factors, in order 
to let the data determine their strengths. For some $t$-channel quantum numbers $(\rho,\,\sigma)$, 
however, pseudo-data  for $N\bar N\to\pi\pi$ exist, which allows to determine their strength 
using crossing~\cite{Schutz:1998jx,Schutz:1994ue}. The amplitude contains also the nucleon pole 
at the physical position and with the physical residue, obtained through the renormalization 
process described, e.g., in Ref.~\cite{Ronchen:2015vfa}.

The potential $V_{\mu\nu}$ of Eq.~(\ref{eq:scattering_juelich}) is written as
\begin{equation}
V_{\mu\nu} = \sum_{i=0}^{n}
\frac{\gamma^a_{\mu;i}\,\gamma^c_{\nu;i}}{W-m_i^b} + V^{\rm NP}_{\mu\nu} 
+\frac{1}{m_N}\gamma^{{\rm CT};a}_{\mu}\gamma^{{\rm CT};c}_\nu\;.
\label{eq:vhad_juelich}
\end{equation}
The first term denotes the sum of all $s$-channel resonance graphs in a given partial wave. 
The bare \underline{c}reation (\underline{a}nni\-hi\-la\-tion) vertices $\gamma^c_{\mu;i}$ ($\gamma^a_{\nu;i}$) 
of a resonance $i$ with bare mass $m_i^b$ are constructed from an effective Lagrangian that can be 
found in Table~8 of Ref.~\cite{Doring:2010ap}. Explicit expressions of the $\gamma^c_{\mu;i}$ and 
$\gamma^a_{\nu;i}$ are given in Refs.~\cite{Doring:2010ap,Ronchen:2012eg}. Note that no SU(3) 
assumptions are made for the $s$-channel resonances. The $t$- and $u$-channel exchange diagrams 
comprised in $V^{\rm NP}$ are also derived from an effective Lagrangian and here, SU(3) flavor symmetry 
is used to relate the coupling constants of the different meson-baryon channels~\cite{Ronchen:2012eg}. 
Form factors allow for (moderate) breaking of SU(3) symmetry. An overview of the various hadron 
exchanges included in the approach is given in Ref.~\cite{Ronchen:2012eg}.
The contact vertex functions $\gamma^{{\rm CT};c}_\mu$ ($\gamma^{{\rm CT};a}_\mu$) are described in 
detail in Ref.~\cite{Ronchen:2015vfa}. In the J\"ulich-Bonn approach, the dynamical generation of 
resonances via the interplay of the $t$- and $u$-channel terms alone, without the need for a bare resonance, 
is possible and realized for some resonances~\cite{Ronchen:2012eg}. Note also, that the parameters
of the bare resonances have no physical meaning but only the ones of the fully dressed states.

The photoproduction of pseudoscalar mesons is studied using the approach developed in 
Ref.~\cite{Ronchen:2014cna} (for a manifestly gauge-invariant variant of the approach, 
see Ref.~\cite{Huang:2011as}). While in Ref.~\cite{Ronchen:2014cna} the hadronic final-state 
interaction is provided by the J\"ulich-Bonn DCC model, the photoproduction inter\-action kernel is 
parameterized by energy-de\-pen\-dent polynomials, avoiding any further input from Born graphs.

The multipole amplitude $M_{\mu\gamma}$ of the photoproduction process is given by~\cite{Ronchen:2014cna}
\begin{multline}
M_{\mu\gamma}(q,W)=V_{\mu\gamma}(q,W) \\+\sum_{\kappa}\int\limits_0^\infty dp\,p^2\,
T_{\mu\kappa}(q,p,W)G^{}_\kappa(p,W)V_{\kappa\gamma}(p,W)\,.
\label{eq:multi_juelich}
\end{multline}
Here, the index $\gamma$ denotes the initial $\gamma N$ state and $T_{\mu\kappa}$ is the hadronic 
half-off-shell $T$-matrix introduced in Eq.~(\ref{eq:scattering_juelich}).
The photoproduction kernel $V_{\mu\gamma}$ of Eq.~(\ref{eq:multi_juelich}) is written as
\be
V_{\mu\gamma}(p,W)=\alpha^{\rm NP}_{\mu\gamma}(p,W)+\sum_{i} \frac{\gamma^a_{\mu;i}(p)\,
\gamma^c_{\gamma;i}(W)}{W-m_i^b} \ .
\label{eq:vgam_juelich}
\ee
The vertex function $\gamma^c_{\gamma;i}$ describes the tree-level coupling of the $\gamma N$ channel to 
the nucleon and the excited baryonic states with resonance number $i$. The photon coupling to the 
non-pole part of the photoproduction kernel is represented by $\alpha^{\rm NP}$. The hadronic resonance 
annihilation vertex $\gamma^a_{\mu;i}$ is the same that is used in the construction of the hadronic 
scattering potential in Eq.~(\ref{eq:vhad_juelich}). Note that, so far, the combined fits of hadronic 
scattering and photoproduction data have not led to any resonance pole that was not already present
in the analysis of hadronic scattering alone, but kaon photoproduction data are not yet included.

\subsection{MAID}
In the spirit of a dynamical approach to pion photo- and electroproduction, the $t$-matrix of the
unitary isobar model MAID is set up by the ansatz~\cite{Tiator:2011pw}
\begin{equation}
t_{\gamma\pi}(W,Q^2)=t_{\gamma\pi}^B(W,Q^2) + t_{\gamma\pi}^{R}(W,Q^2)\,,\label{eq:DM}
\end{equation}
with a background and a resonance $t$-matrix, each of them constructed in a
unitary way. Here, $Q^2$ defines the virtuality of the photon.


For a specific partial wave $\alpha = \{j,l,\ldots\}$, the background
$t$-matrix is set up by a potential multiplied by the pion-nucleon scattering
amplitude in accordance with the K-matrix approximation,
\begin{equation}
 t^{B,\alpha}_{\gamma\pi}(W,Q^2)=v^{B,\alpha}_{\gamma\pi}(W,Q^2)\,[1+it_{\pi N}^{\alpha}(W) ]\, ,
\label{eq:Kmatrix}
\end{equation}
where only the on-shell part of pion-nucleon rescattering is maintained and the
off-shell part from pion-loop contributions is neglected. Whereas this
approximation would fail near the threshold for
$\gamma,\pi^0$~\cite{Bernard:1991rt,Kamalov:2001qg}, it is well justified in
the resonance region because the main contribution from pion-loop effects is
absorbed by the nucleon resonance dressing.

The background  potential $v_{\gamma\pi}^{B,\alpha}(W,Q^2)$ is described by
Born terms obtained with an energy-dependent mixing of
pseudovector-pseudoscalar $\pi NN$ coupling and $t$-chan\-nel vector meson
exchanges. The mixing para\-meters and coupling constants are determined by an
analysis of nonresonant multipoles in the appropriate energy
regions~\cite{Maid98}. In the latest version MAID2007~\cite{MAID07}, the $S$,
$P$, $D$, and $F$ waves of the background contributions are unitarized as
explained above, with the pion-nucleon elastic scattering amplitudes,
$t^{\alpha}_{\pi N}=[\eta_{\alpha} \exp(2i\delta_{\alpha})-1]/2i$, described by
phase shifts $\delta_{\alpha}$ and the inelasticity parameters $\eta_{\alpha}$, are
taken from the GWU/SAID analysis~\cite{Arndt:1995ak}.

For the resonance contributions we follow Ref.~\cite{Maid98} and assume
Breit-Wigner forms for the resonance shape,
\begin{eqnarray}
t_{\gamma\pi}^{R,\alpha}(W,Q^2) &=& {\bar{\cal A}}_{\alpha}^R(W,Q^2)\, \label{eq:BW}\\
&\times&  \frac{f_{\gamma N}(W)\Gamma_{tot}(W)\,M_R\,f_{\pi N}(W)}{M_R^2-W^2-iM_R\,
\Gamma_{tot}(W)}\, e^{i\phi_R(W,Q^2)}\,,\nonumber
\end{eqnarray}
where $f_{\pi N}(W)$ is the usual Breit-Wigner factor describing the decay of a resonance with
total width $\Gamma_{tot}(W)$ and partial $\pi N$ width $\Gamma_{\pi N}(W)$. The vertex functions
for $\gamma N$ and $\pi N$ and the partial widths for the $\pi N$, $\pi\pi N$ and $\eta N$ channels
are parametrized in a common way, which may differ in some model-dependent parts. All details of
the parametrization and fitted values are found in Ref.~\cite{MAID07}. 
%
The phase $\phi_R(W,Q^2)$ in Eq.~(\ref{eq:BW}) is introduced to
adjust the total phase such that the Fermi-Watson theorem is fulfilled below
two-pion threshold.


The\,current\,version,\,MAID2007,\,describes\,all\,13\,four-star resonances below
$W=2$~GeV: $\Delta(1232)3/2^+$, $N(1440){1/2^+}$, $N(1520){3/{2^-}}$, $N(1535){1/2^-}$, $\Delta(1620){1/2^-}$,
$N(1650){1/2^-}$, $N(1675){5/2^-}$, $N(1680){5/2^+}$, $\Delta(1700){3/2^-}$, $N(1720){3/2^+}$,
$\Delta(1905){5/2^+}$,
$\Delta(1910){1/2^+}$, and $\Delta(1950){7/2^+}$. In a coming update of MAID, this list of 
dominant resonances
is no longer sufficient, due to the high accuracy of the data and the availability of many
polarization observables with single and double polarization. With these data, also the weaker
three-star and two-star resonances can be analyzed and will be included in the model.

In most cases, the resonance couplings $\bar{\mathcal A}_{\alpha}^R(W,Q^2)$ are assumed to be
independent of the total energy. However, for the $\Delta(1232)3/2^+$ it was critical to also introduce
an energy dependence in terms of the virtual photon three-momentum $k(W,Q^2)$. For all other
resonances discussed here, we assume a simple $Q^2$ dependence, $\bar{\mathcal A}_{\alpha}(Q^2)$,
parameterized as functions of $Q^2$ by an ansatz
\begin{equation}
\bar{\mathcal A}_{\alpha}(Q^2) =\bar{\mathcal A}_{\alpha}(0) (1+a_1 Q^2+a_2 Q^4 + \cdots)\, e^{-b_1
Q^2}\,. \label{eq:ffpar}
\end{equation}

The electromagnetic couplings of the resonance excitations, $\bar{\mathcal A}_{\alpha}^R(W,Q^2)$ are
for most cases energy-independent. For real photons, $Q^2=0$, they are fitted to the pion
photoproduction data. For virtual photons they are parametrized as functions of $Q^2$ and
are fitted to the world data of pion electroproduction. For further details see
Ref.~\cite{Tiator:2011pw,MAID07}.

A special advantage in the MAID ansatz is that it is applicable for all isospin channels and also
for electroproduction. It is fitted up to photon virtualities of $Q^2\approx 5$~GeV$^2$, in the
isoquartet $J^P=3/2^+$ partial wave of the $\Delta(1232)3/2^+$ region even up to $Q^2=8$~GeV$^2$.

\subsection{SAID}
The SAID fit parametrization recently transitioned from a form, 
motivated~\cite{Workman:2005eu}  by a 3-channel Heitler K-matrix, similar to the MAID approach, 
to a form~\cite{Workman:2012jf} based on the same Chew-Mandelstam
K-matrix used in the SAID $\pi N$ elastic scattering fits~\cite{Workman:2012hx}.
The form used in $\pi N$ elastic scattering is
\begin{align}
\label{eqn:CM_piN}
T_{\alpha \beta} = [1 - \bar{K} C ]_{\alpha \sigma}^{-1} \bar{K}_{\sigma \beta},
\end{align}
where the indices label channels $\pi N$, $\pi \Delta$, $\rho N$, and $\eta N$ and $C$ is the
Chew-Mandelstam function described in Refs.~\cite{Workman:2012jf,Workman:2012hx,Paris:2010tz,Arndt:1985vj}.
The extension used to include an electromagnetic channel is
\begin{align}
\label{eqn:CM_photo}
T_{\alpha \gamma} = [1 - \bar{K} C]_{\alpha \sigma}^{-1} \bar{K}_{\sigma \gamma}.
\end{align}
The added elements $\bar{K}_{\sigma \gamma}$ provide the $\gamma N$ coupling to the hadronic
channels. The energy dependence is parameterized in terms of an energy polynomial, similar
to that used in the J\"ulich-Bonn analysis. For the $\gamma N\to \pi N$ channels, these
elements are further constrained to approach the Born terms at threshold.

The SAID formalism differs from all other approaches described here in the way resonance
properties are deduced. Only the $\Delta(1232)3/2^+$ is explicitly introduced as a Chew-Mandelstam
K-matrix pole. All other poles arise from the factor $[1-\bar{K}C]^{-1}$ which is common to
the $\pi N$ scattering and photoproduction analyses. Two important facts follow. First, the pion
photoproduction analysis will have the same pole/cut structure as in the $\pi N$ scattering analysis.
Second, as the poles are generated - not added by hand - only those necessary to fit this process
are produced. This is then the minimal required set.

The Chew-Mandelstam K-matrix parameters for hadro\-nic channels
are obtained from fits
to the SAID $\pi N$ elastic and $\eta N$ production database~\cite{SAID_Data}.
The K-matrix elements associated with the $\pi \Delta$ and $\rho N$ channels serve to
provide quasi-two-body branch points in a unitary approach but are not fitted to
reaction data. The hadronic parameters are fixed in the analysis of photoproduction data.
Only the polynomial elements of $\bar{K}_{\sigma \gamma}$ in Eq.~\eqref{eqn:CM_photo}
have been varied in the present study.

\subsection{Comparison of the methods}

All partial wave analyses/dynamically coupled-channel approaches discussed here 
aim at representing the physical reality,
however, at different levels of sophistication. All models allow for
coupled channel effects. All models respect -- at least approximately -- two-body
unitarity. And in all four aproaches, the amplitudes are analytic functions of the
invariant mass.

The J\"uBo model is a dynamical coupled-channel (DCC) model based on a hadronic
scattering potential. The potential is iterated in a Lippmann-Schwinger equation formulated in time-ordered perturbation theory;
two-body unitarity is thus fulfilled automatically. The three-body $\pi\pi N$ final state
is represented by the channels $\rho N$, $\sigma N$ and $\pi\Delta(1232)3/2^+$.
$t$- and $u$-channel exchanges of known mesons and baryons are derived from
an effective Lagrangian, these contributions form the
non-resonant part of the amplitude. Data on pion and photo-induced reactions are
fitted simultaneously, eleven $N^*$ and ten $\Delta^*$ resonances are
included in the fit, and their properties are derived from fits to the data.
In addition, three dynamically generated states are found. 
The J\"uBo model is the most ambitious
approach. The price one has to pay is that it requires the largest computing power.

The main focus of the BnGa model is the search for new $N^*$ and $\Delta^*$
resonances, the so-called {\it missing resonances}, which are expected at
masses above 1.7\,GeV. Constraints governing the low-energy regime from chiral
perturbation theory or from the Watson theorem are not imposed; however,
the fit solution satisfies the Watson theorem to a few degrees. In the model, 
resonances in a given partial wave are added in a K-matrix. The $t$- and $u$-channel
exchanges are used to parametrize the main part of the background, smaller
phenomenological background contributions are added within the K-matrix. The
data on pion and photo-induced reactions are fitted in a common fit. The data include
three-body final states which are fitted event by event in a likelihood fit.
The fit determines the properties of nineteen $N^*$ and nine $\Delta^*$ resonances.
The program is optimized to yield fast
response: excluding three-body final states, a few hundreds of fits can be performed
to validate the existence and the quantum numbers of new states. When
three-body final states are included, the number of fits is limited to
a few dozens.

The MAID model uses multi-channel Breit-Wigner amplitudes to describe
the resonant part of the photoproduction amplitude. The contribution of
background channels (Born terms and $t$-channel exchanges) to the photoproduction
cross section is large. The additional resonant contributions interfere destructively
and bring the cross section down to the observed level. The phases of the
total amplitude are modified by additional phases which are adjusted
to guarantee the Watson theorem for the mass range below
the $2\pi$ production. At present, the seven four-star $N^*$ and the
six four-star $\Delta^*$ resonances are included in the fit. Their masses,
widths and $\pi N$ couplings are taken from the RPP.

The SAID model is the only approach in which no resonances (except the $\Delta(1232)3/2^+$)
are put in {\it a priori}. Instead, the K-matrix elements in Eq.~(\ref{eqn:CM_piN})
are expanded as polynomials in $W-W_{\rm threshold}$. The K-matrix elements
describe elastic and charge exchange scattering and inelastic channels like
$\eta N$, $\rho N$, and $\pi\Delta(1232)3/2^+$. The degree of the polynomials varies
with the mass range to be covered from small values up to 5. The approach is very
efficient to fit real and imaginary parts of partial wave amplitudes for $\pi N$ (quasi-)
elastic scattering or the reaction $\pi N\to \eta N$. Photoproduction data
are fitted once masses, widths, and $\pi N$ couplings are known from fits
to the $\pi N$ data.

\section{\label{Data}Data used in the analysis}
\begin{figure*}[pt]
\centering
\includegraphics[width=0.9\textwidth,height=0.74\textheight]{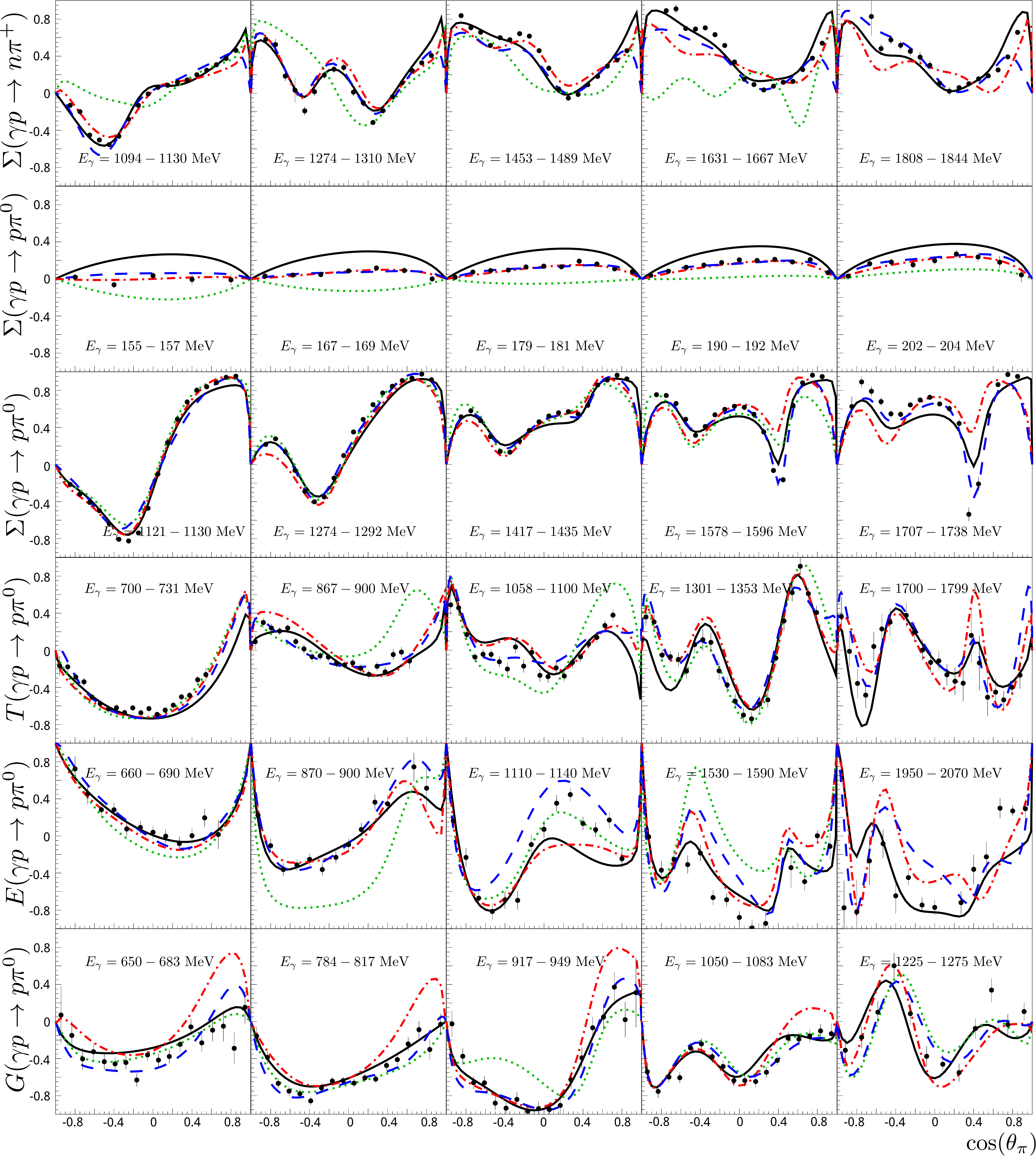}
\caption{\label{prefit}
Selected data and the predictions from the four different PWAs: black solid line: BnGa2011-02, 
blue dashed: J\"uBo2015B, green dotted: MAID2007, red dashed dotted: SAID CM12. 
The predictions are based on fits which did not yet use these new data. The new data are shown 
for the beam asymmetry $\Sigma$ for $\gamma p\to \pi^+\;n$ \cite{Dugger:2013crn} (1$^{\rm st}$ row), for
the beam asymmetry $\Sigma$ in the low-energy region \cite{Hornidge:2012ca}
 and at higher energies (2$^{\rm nd}$ row)  for $\gamma p\to \pi^0\;p$, (2$^{\rm nd}$ and 3$^{\rm rd}$ row).
The next three rows show $T$ \cite{Hartmann:2015kpa},
$G$ \cite{AThiel:2012uha,Thiel:2015tbd}, and $E$ \cite{Gottschall:2014uha,Gottschall:2015tbd} for $\gamma p\to \pi^0\;p$.
Note that the data from Refs.~\cite{Hornidge:2012ca} and  \cite{Dugger:2013crn}  
are included in the fits of J\"uBo2015B and SAID CM12.
 }
\end{figure*}

The four partial wave analyses groups (BnGa, J\"uBo, MAID, SAID) use a large body of pion and photo-induced reactions. Only the SAID group fits the data
on $\pi N$ elastic scattering and determines in energy-independent
fits the real and imaginary part of the $\pi N$ partial wave amplitudes
(given in slices of the invariant mass). The other three PWA groups use 
these amplitudes, jointly with further data, in energy-dependent fits. The 
{\it predictions} of the different PWAs resulting from fits to early data and these
{\it predictions} are shown to the
recent new data in Fig.~\ref{prefit}. The recent new data include measurements of 
the beam asymmetry $\Sigma$ \cite{Hornidge:2012ca,Dugger:2013crn}, on $T, P$ and $H$
\cite{Hartmann:2015kpa}, $G$ \cite{AThiel:2012uha,Thiel:2015tbd}, and $E$ \cite{Gottschall:2014uha,Gottschall:2015tbd}
for the reaction $\gamma p\to \pi^0 p$.   
For the reaction $\gamma p\to \pi^+ n$, new data are available on the beam asymmetry 
$\Sigma$ \cite{Dugger:2013crn}. These data are (mostly) not used in the {\it predictions}.

\begin{figure*}[pt]
\centering
\epsfig{file=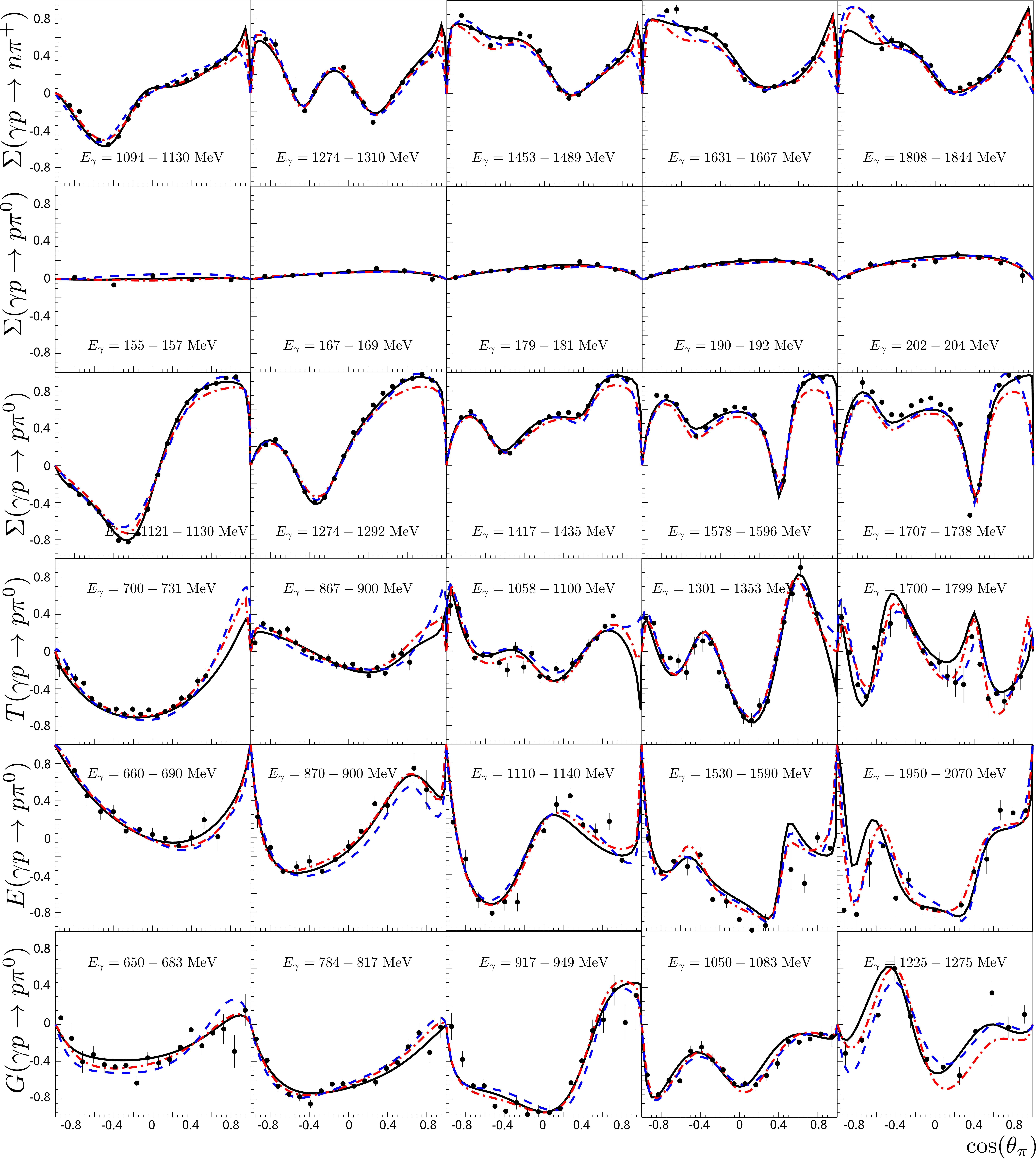,width=0.9\textwidth,height=0.74\textheight}
\caption{\label{postfit}
The new fit results of the different PWAs in comparison with the new data: 
black solid line: BnGa, blue dashed: J\"uBo, red dashed dotted: SAID. New data are shown 
for the beam asymmetry $\Sigma$ for $\gamma p\to \pi^+\;n$ \cite{Dugger:2013crn} (1$^{\rm st}$ row), for
the beam asymmetry $\Sigma$ in the low-energy region \cite{Hornidge:2012ca}
 and at higher energies (2$^{\rm nd}$ row)  for $\gamma p\to \pi^0\;p$, (2$^{\rm nd}$ and 3$^{\rm rd}$ row).
The next three rows show $T$ \cite{Hartmann:2015kpa},
$G$ \cite{AThiel:2012uha,Thiel:2015tbd}, and $E$ \cite{Gottschall:2014uha,Gottschall:2015tbd} for $\gamma p\to \pi^0\;p$.
The BnGa fit did not yet use the data on the beam asymmetry $\Sigma$ for 
$\gamma p\to \pi^0\;p$ in the low-energy region \cite{Hornidge:2012ca}. Nevertheless, the new fit is fully
consistent with the new data.
}
\end{figure*}
\paragraph{SAID:} The SAID group
uses these partial wave amplitudes to determine masses $M_i$, widths
$\Gamma_i$, and $\pi N$ branching ratios BR$_i^{N\pi}$ of the contributing
resonances \cite{Arndt:2006bf}.
In earlier fits, the full GWU/SAID data base for $\gamma p\to\pi^0p$ and $\pi^+n$ was used by SAID.
The fits allowed for a
renormalization of angular distributions, with a $\chi^2$ penalty determined by
the overall systematic error. In the new fits, this freedom was removed for the new
polarized measurements \cite{Hartmann:2015kpa,Thiel:2015tbd,Gottschall:2015tbd}.
A weighting factor of four in the $\chi^2$ fit was applied
to these data, while the set of previous measurements remained unweighted. The included
energy range, in $E_{\gamma}$,
was 155~MeV to 2.7~GeV, the lower limit being chosen to avoid the $\pi^+ n$ threshold.
In order to avoid fitting mutually inconsistent cross section experiments from the
full SAID database, a choice of relatively recent and non-overlapping cross section
experiments was made for both charged and neutral pion production
\cite{Dugger:2007bt,Dugger:2009pn,Ahrens:2004pf,Ahrens:2006gp,Beck:2006ye}.


\paragraph{J\"uBo:} J\"uBo \cite{Ronchen:2012eg} fits the amplitudes of the SAID WI08 energy-dependent
solution jointly with data on other pion-induced reactions,
$\pi^-p\to\eta n$, $K^0\Lambda$, $K^+\Sigma^-$, $K^0\Sigma^0$, and
$\pi^+p\to K^+\Sigma^+$. A detailed account of the data, listing also
the weighting applied to different data sets, is given in Ref.~\cite{Ronchen:2012eg}.
The data base for $\gamma p\to\pi^0p$ and $\pi^+n$ used by J\"uBo is explained in 
Ref.~\cite{Ronchen:2014cna}, while the data taken into account for $\gamma p\to\eta p$ are 
listed in Ref.~\cite{Ronchen:2015vfa}. There, 
also the data on $T$ and $F$ from a recent MAMI measurement on $\eta$
photoproduction~\cite{Akondi:2014ttg} were included.
Altogether almost 30,000 data points from photoproduction reactions were fitted. 
A complete account of the corresponding references can be found online~\cite{data_juelich_photo}.
The solution including all the mentioned data is named Fit~B
in Ref.~\cite{Ronchen:2015vfa}.

\paragraph{BnGa:} The BnGa group uses the amplitudes of the SAID WI08 energy-dependent
solution and a comparable set of data on pion induced reactions as J\"uBo does. Furthermore,
the data on $\pi^-p\to\pi^0\pi^0n$ are included. For photoproduction,
the data base includes the reactions $\gamma p\to \pi N$, $\eta p$, $K^+\Lambda$,
$K^+\Sigma^0$, $K^0\Sigma^+$, and three-body final states $2\pi^0p$
\cite{Sokhoyan:2015fra,Sokhoyan:2015eja,Thoma:2007bm,Sarantsev:2007aa,Assafiri:2003mv,Kashevarov:2012wy}
and $\pi^0\eta p$ \cite{Gutz:2014wit}.
The data sets used and the weights given to them in the fits are listed in
\cite{Anisovich:2011fc,Anisovich:2013vpa}.

\paragraph{MAID:} The MAID group uses data on photo- and electroproduction of 
$\pi^0p$, $\pi^+n$, and $\eta p$ available
at the time when MAID2007 was reported  \cite{Drechsel:2007if}.
The MAID homepage covers also KAON MAID \cite{Mart:1999ed}.
Data on $\pi N$ and $\eta p$ (and $K^+\Lambda$) are fitted independently.

\paragraph{Particle properties:} The SAID and MAID PWA groups use photoproduction
reactions to determine the dynamics of the reaction
and to determine the helicity amplitudes of contributing
resonances while $M_i$, $\Gamma_i$, and BR$^{N}_i$ are imposed from the
fits to $\pi N$ elastic scattering and charge exchange reactions (SAID) or directly
from the Review of Particle Properties, RPP, (MAID).
The BnGa and J\"uBo groups use pion and photo-induced reactions
and determine the properties of the contributing resonances in global fits to
all included data. 

\paragraph{New data:} To study the impact of the new data from Bonn, JLab, and Mainz on 
the photoproduction multipoles,
the PWA groups agreed to perform new fits incorporating the new data. In the fits called 
{\it predictions} below, the new data were (mostly) not yet included. The new fits took 
into account also the data listed in Table~\ref{NEW}. The new fits, shown in Fig.~\ref{postfit}, 
are all capable to reproduce the new data reasonably well.
This does, however, not imply that the  multipoles resulting from these fits become
identical. The partly still different data bases and differences in the formalism used to describe
the new data may be responsible for the remaining differences in the multipoles shown in Figs.~\ref{part1}-\ref{part4}.

\section{\label{Gamma}The photoproduction multipoles}
In Figs.~\ref{part1} - \ref{part4}, the real and imaginary parts of the low-$L^\pi$
multipoles for the photoproduction of pions
from BnGa, J\"uBo, MAID, and SAID are presented. For BnGa, J\"uBo, and SAID
the multipoles are shown from fits to data before and after the inclusion
of the new data on polarization variables (see Table~\ref{NEW}). The main aim of this section is to
demonstrate the impact of the new polarization data on the resulting
multipoles.

\begin{table*}[pt]
\caption{\label{NEW}New data sets introduced recently into the PWA data bases and used
for the new fits are marked with a $\surd$. Data marked with
a $\dagger$ were included already in the predictions (J\"uBo2015B and SAID CM12). }
\renewcommand{\arraystretch}{1.4}
{\scriptsize
\begin{tabular}{|l|ccccccc|ccccccc|}
\hline\hline
Reaction            &      Obs.       &Range (MeV)&$N_{\rm data}$& Ref.
&\hspace{-3mm}BnGa\hspace{-3mm}&\hspace{-3mm}J\"uBo\hspace{-3mm}&\hspace{-3mm}SAID\hspace{-3mm}&
                          Obs.        &Range (MeV)&$N_{\rm data}$& Ref.                 &\hspace{-3mm}BnGa\hspace{-3mm}&\hspace{-3mm}J\"uBo\hspace{-3mm}&\hspace{-3mm}SAID\hspace{-3mm}\\ \hline
$\gamma p\to \pi^0p$&$d\sigma/d\Omega$& 147 - 219 &600&\cite{Hornidge:2012ca}&-&$\dagger$ & $\dagger$
                    &$P,C_x,C_z$&   1845    &1&\cite{Luo:2011uy}     &-&   -     &$\surd$\\
                    & $\Sigma$        & 147 - 206 &220&\cite{Hornidge:2012ca}&-&$\dagger$ &$\dagger$                    
                    & $\Sigma$        & 1102 - 1862&  700       &\cite{Dugger:2013crn}  &$\surd$&$\dagger$&$\dagger$\\
                    &$T$              & 684 - 1891  &494&\cite{Hartmann:2015kpa}&$\surd$&$\surd$&$\surd$
                    & $G$             & 633 - 1300&318&\cite{AThiel:2012uha,Thiel:2015tbd}  &$\surd$&$\surd$        &$\surd$\\
                    &$E$          & 730 - 2100  &455&\cite{Gottschall:2014uha,Gottschall:2015tbd}&$\surd$& $\surd$       & $\surd$
                    & $C_x$        & 463 -- 1338 &  45 &\cite{Sikora:2013vfa}  &$\surd$&    -    &$\surd$\\
                    & $P$         & 684 - 917  &  158       &\cite{Hartmann:2015kpa} &$\surd$&$\surd$&$\surd$
                    &$H$          & 684 - 917  &158&\cite{Hartmann:2015kpa}&$\surd$&$\surd$&$\surd$\\
                                        \hline
$\gamma p\to \pi^+n$&$\Sigma$        & 1112 - 1862&  386 &\cite{Dugger:2013crn}  &$\surd$&$\dagger$&$\dagger$
                    &                &                   &                       &       &         &   \\  
\hline
$\gamma p\to K^+\Lambda$&$d\sigma/d\Omega$&  &&\cite{Jude:2013jzs}&$\surd$    &    -    & -
                                            &                 &            &      &                       &       &        &\\
$\gamma p\to K^+\Sigma^0$&$d\sigma/d\Omega$&  &&\cite{Jude:2013jzs}&$\surd$    &   -     & -
                                            &                 &            &      &                       &       &        &\\
\hline\hline
\end{tabular}
}
\renewcommand{\arraystretch}{1.0}
\end{table*}

For convenience, we recall in Table~\ref{waves}
the quantum numbers of the final $\pi N$ state resulting from
a given multipole. We also remind the reader that in case of a
simple isolated Breit-Wigner resonance, the imaginary part of the multipole
peaks at the nominal mass while the real part goes through zero.

\begin{table}[pb]
\caption{\label{waves}Photoproduction multipoles and partial waves.
The partial waves are characterized by $L^\pi_{2I,2J}$, with $L^\pi$ 
the orbital angular momentum, $I$ the isospin
and $J$ the total spin of the $\pi N$ system.     
In general, two multipoles belong to one spin-parity wave $J^P$.}
\begin{center}
\renewcommand{\arraystretch}{1.2}
\begin{tabular}{cccccccc}
\hline\hline \multicolumn{2}{c}{Multipoles} &$L^\pi$&
\multicolumn{2}{c}{$\pi N$ partial waves}&$J^P$\\
&&& $I=1/2$& $I=3/2$& \\
\hline
$E_{0+}$&   -      &0& $S_{11}$ & $S_{31}$&$1/2^-$\\
 -      &$M_{1-}$ &1& $P_{11}$ & $P_{31}$&$1/2^+$\\
$E_{1+}$& $M_{1+}$ &1& $P_{13}$ & $P_{33}$&$3/2^+$\\
$E_{2-}$& $M_{2-}$ &2& $D_{13}$ & $D_{33}$&$3/2^-$\\
$E_{2+}$& $M_{2+}$ &2& $D_{15}$ & $D_{35}$&$5/2^-$\\
$E_{3-}$& $M_{3-}$ &3& $F_{15}$ & $F_{35}$&$5/2^+$\\
$E_{3+}$& $M_{3+}$ &3& $F_{17}$ & $F_{37}$&$7/2^+$\\
$E_{4-}$& $M_{4-}$ &4& $G_{17}$ & $G_{37}$&$7/2^-$\\
$E_{4+}$& $M_{4+}$ &4& $G_{19}$ & $G_{39}$&$9/2^-$\\
\hline\hline
\end{tabular}
\renewcommand{\arraystretch}{1.0}
\end{center}
\end{table}

\subsection{The multipoles}
\begin{figure*}
\centerline{\epsfig{file=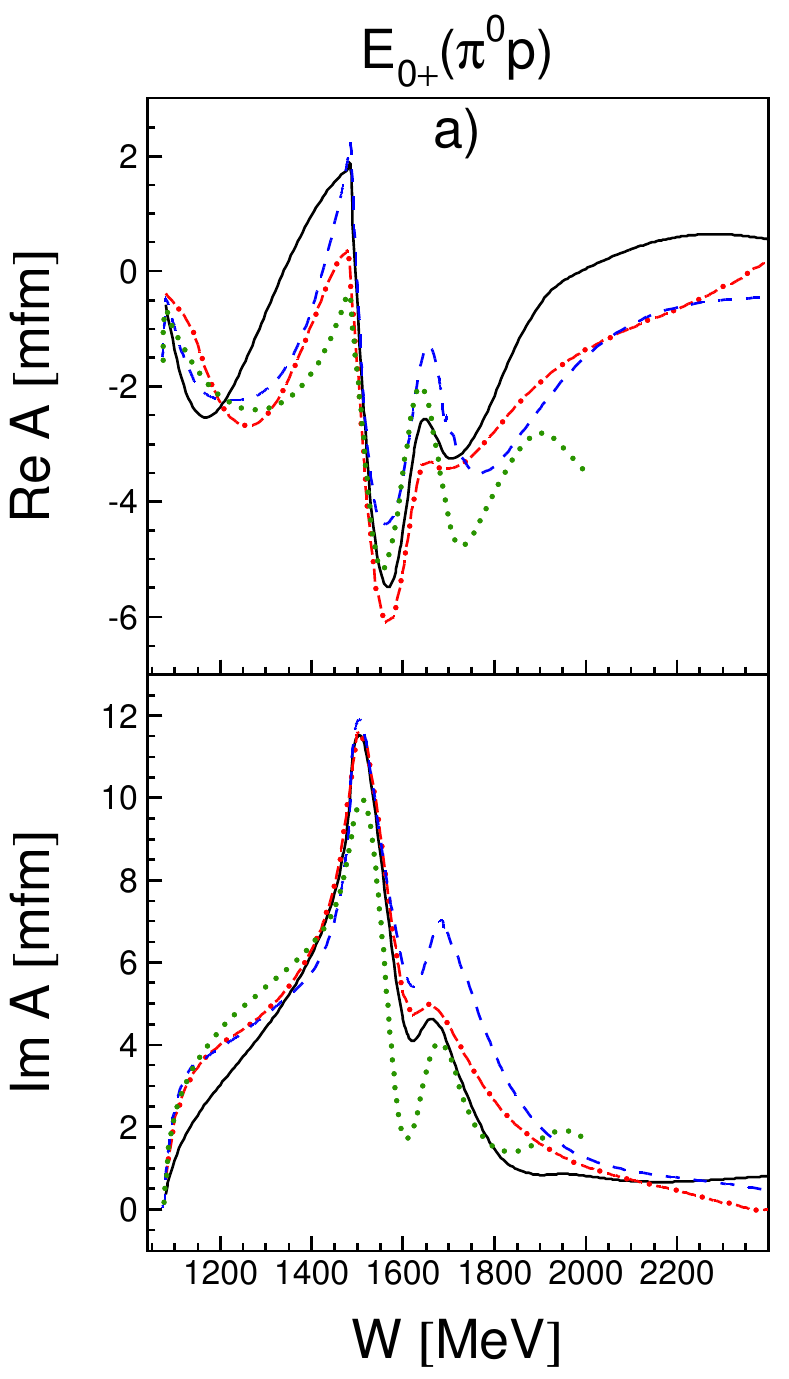,width=0.28\textwidth,height=0.42\textwidth}
\hspace{-12mm}\epsfig{file=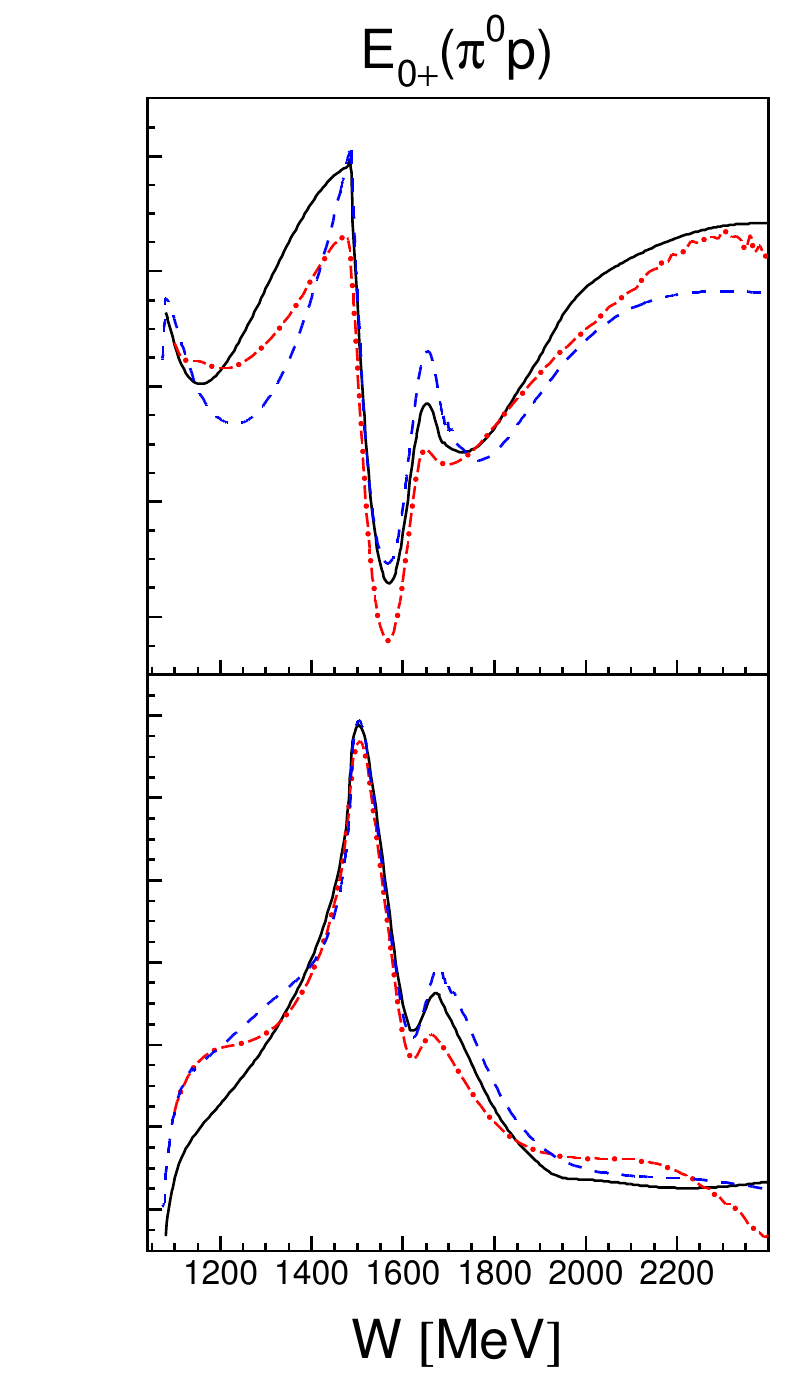,width=0.28\textwidth,height=0.42\textwidth}
              \epsfig{file=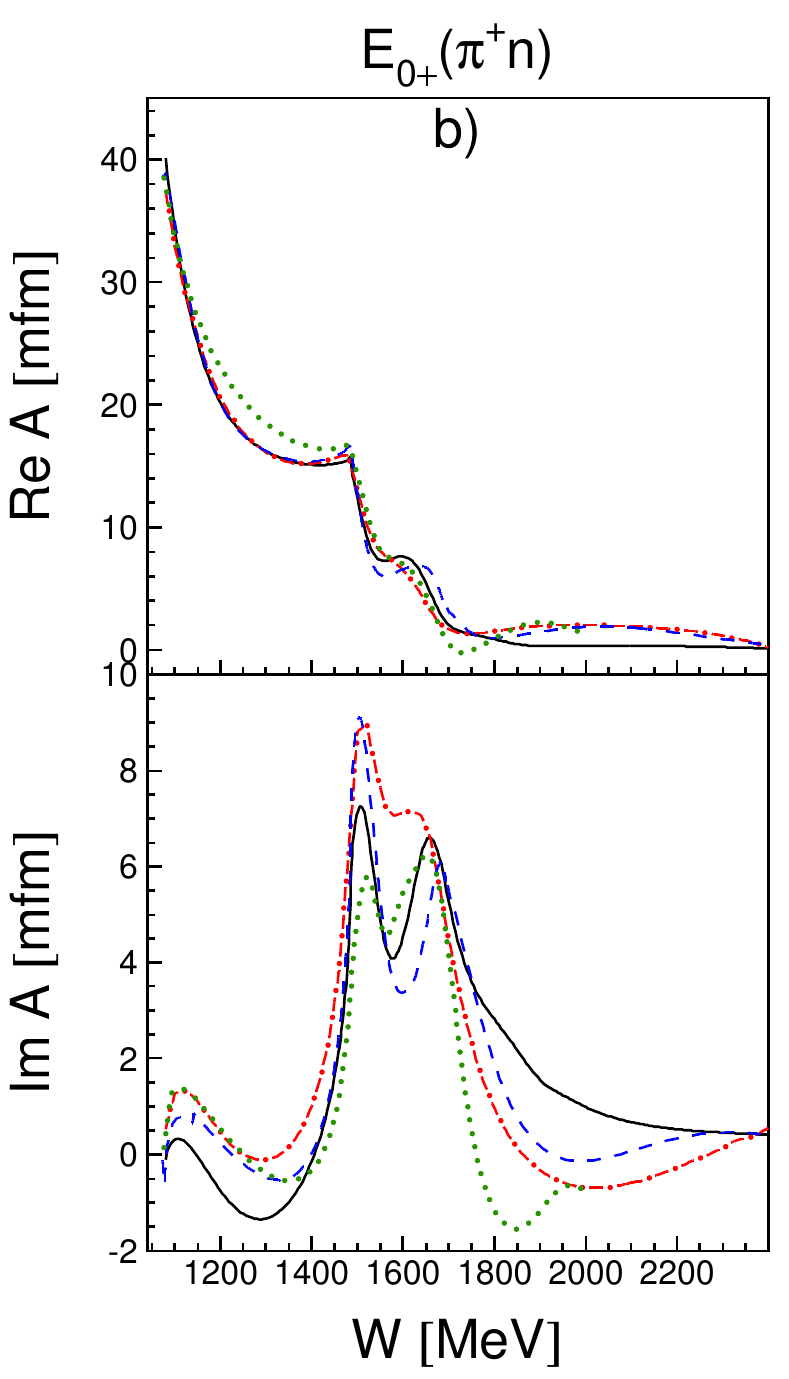,width=0.28\textwidth,height=0.42\textwidth}
\hspace{-12mm}\epsfig{file=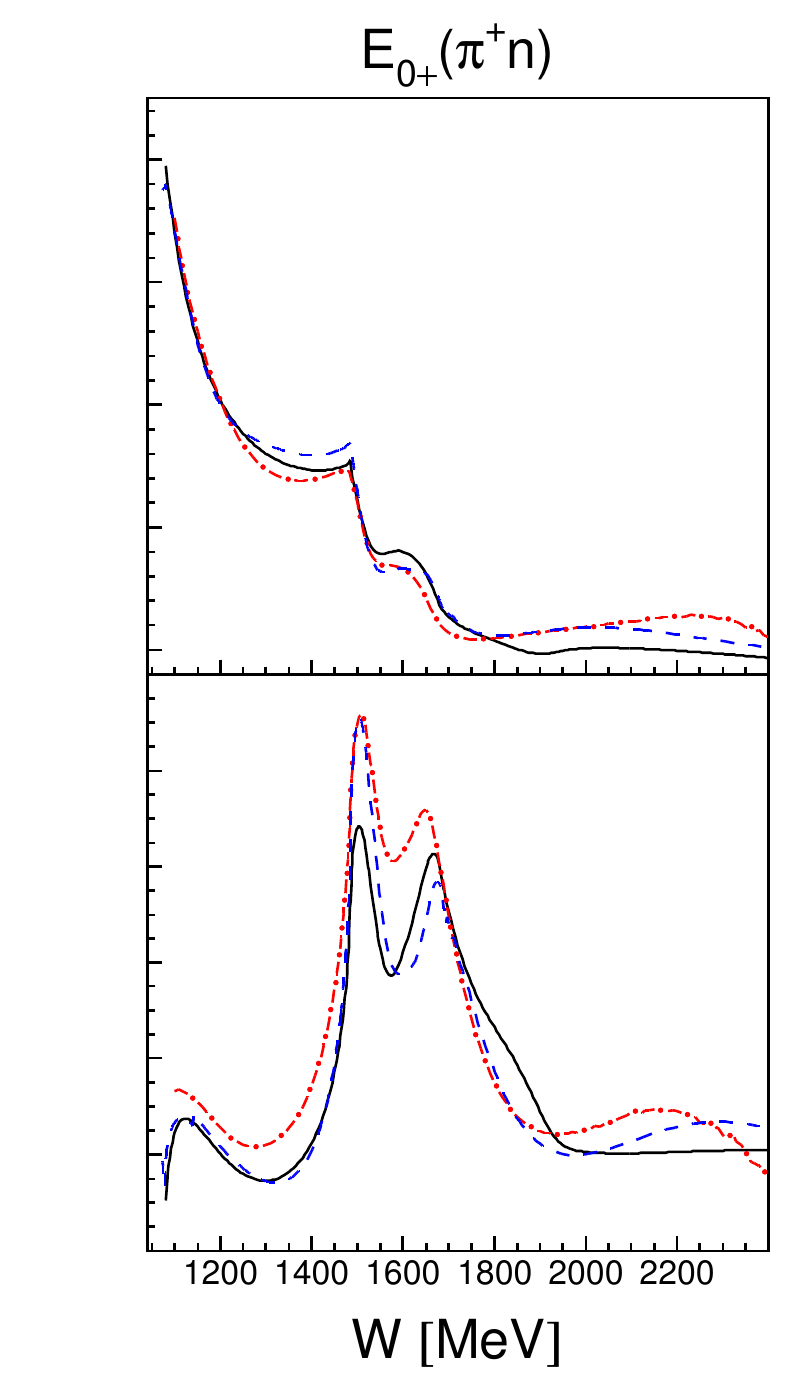,width=0.28\textwidth,height=0.42\textwidth}
}
\centerline{\epsfig{file=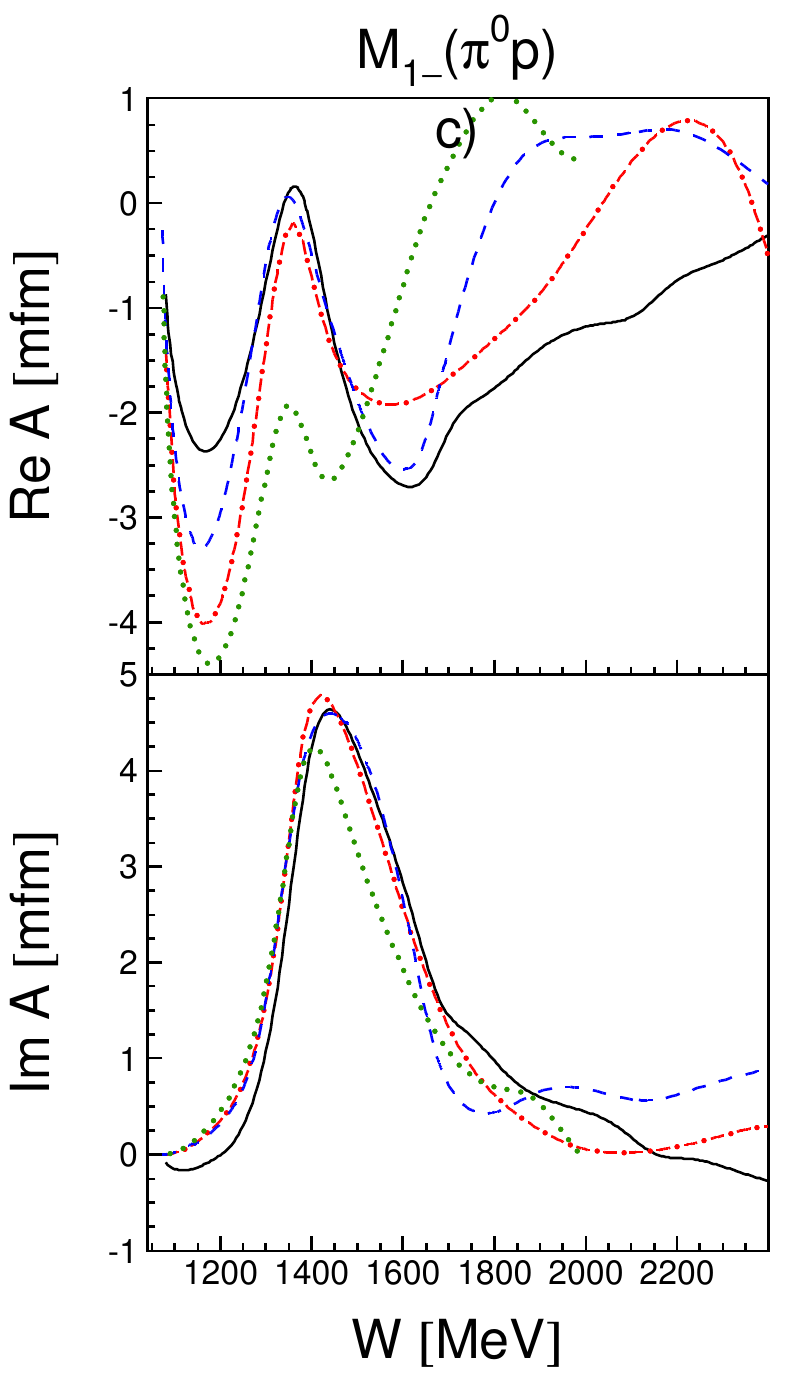,width=0.28\textwidth,height=0.42\textwidth}
\hspace{-12mm}\epsfig{file=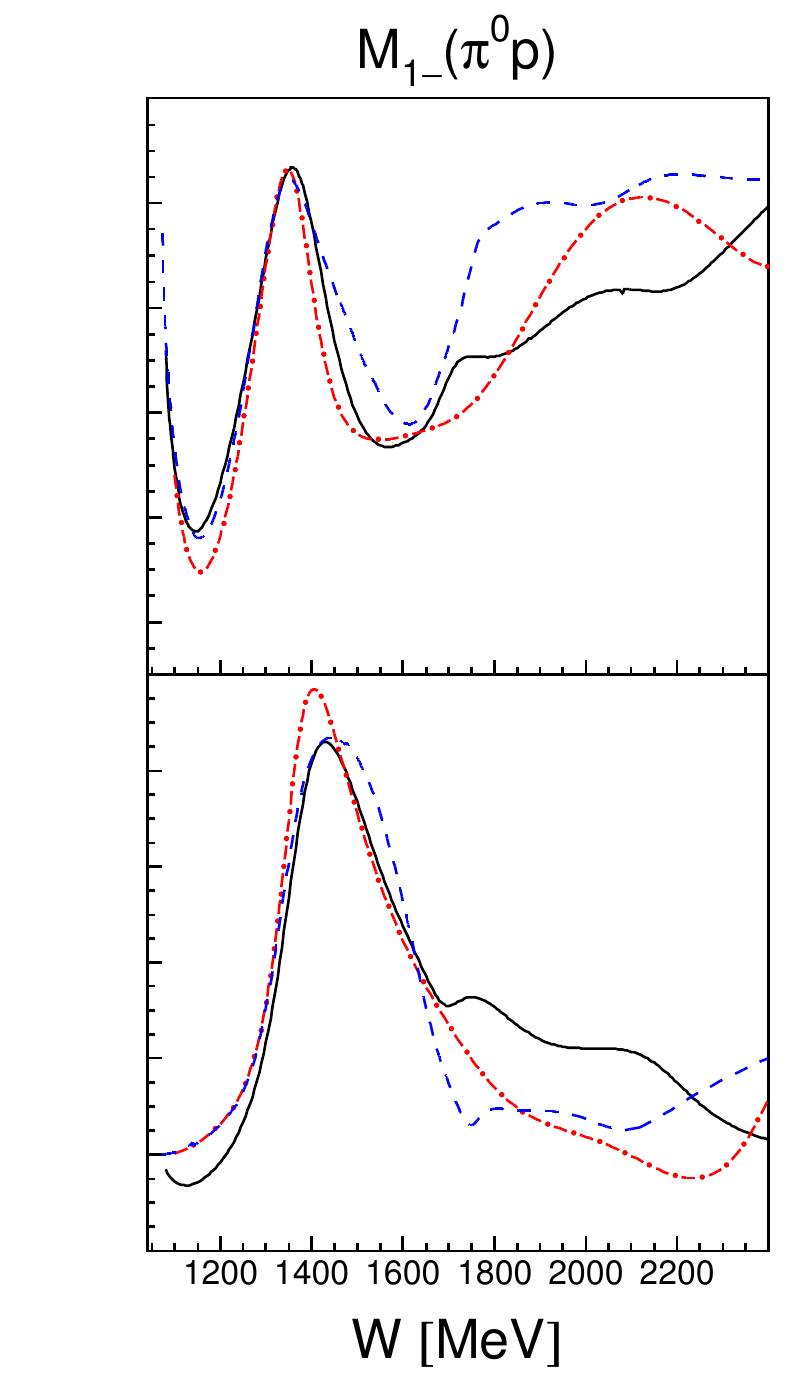,width=0.28\textwidth,height=0.42\textwidth}
             \epsfig{file=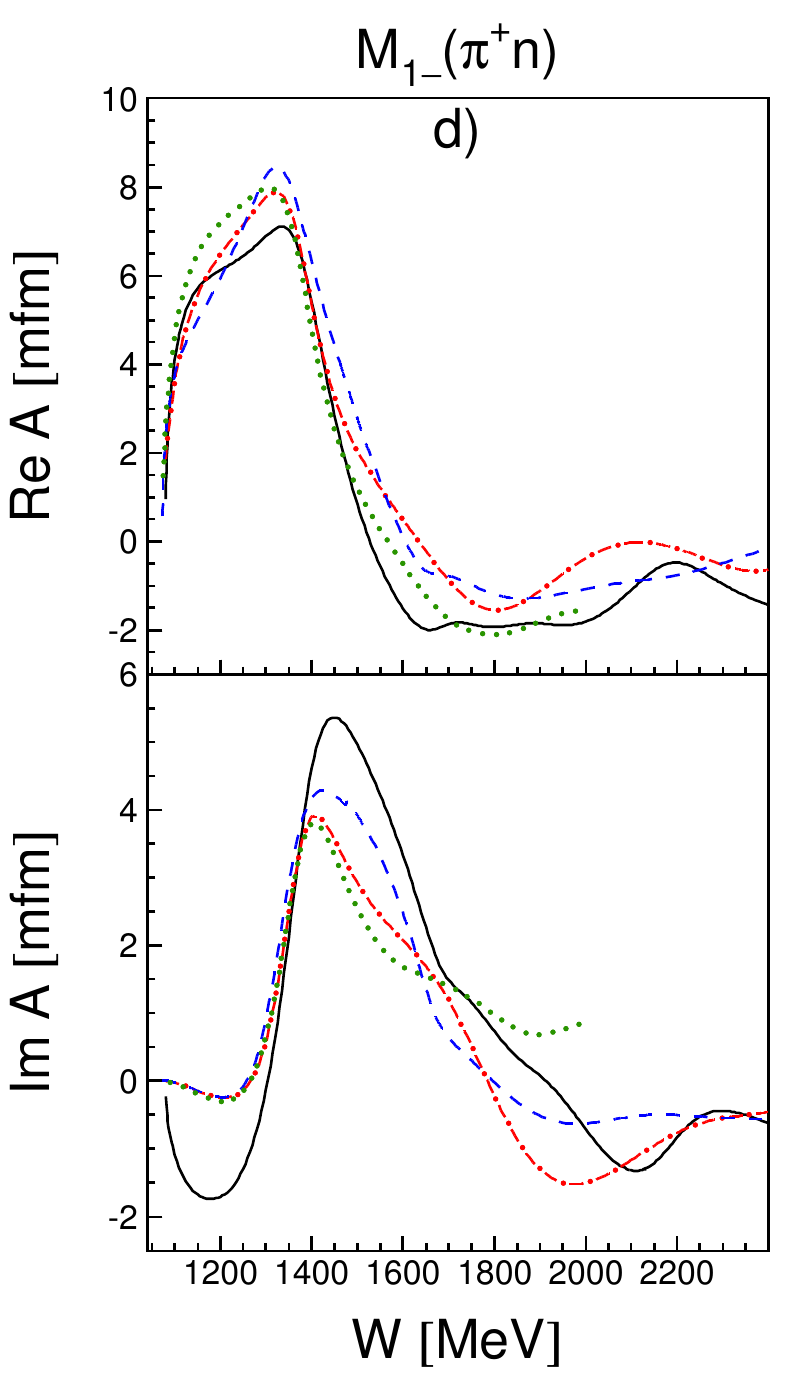,width=0.28\textwidth,height=0.42\textwidth}
\hspace{-12mm}\epsfig{file=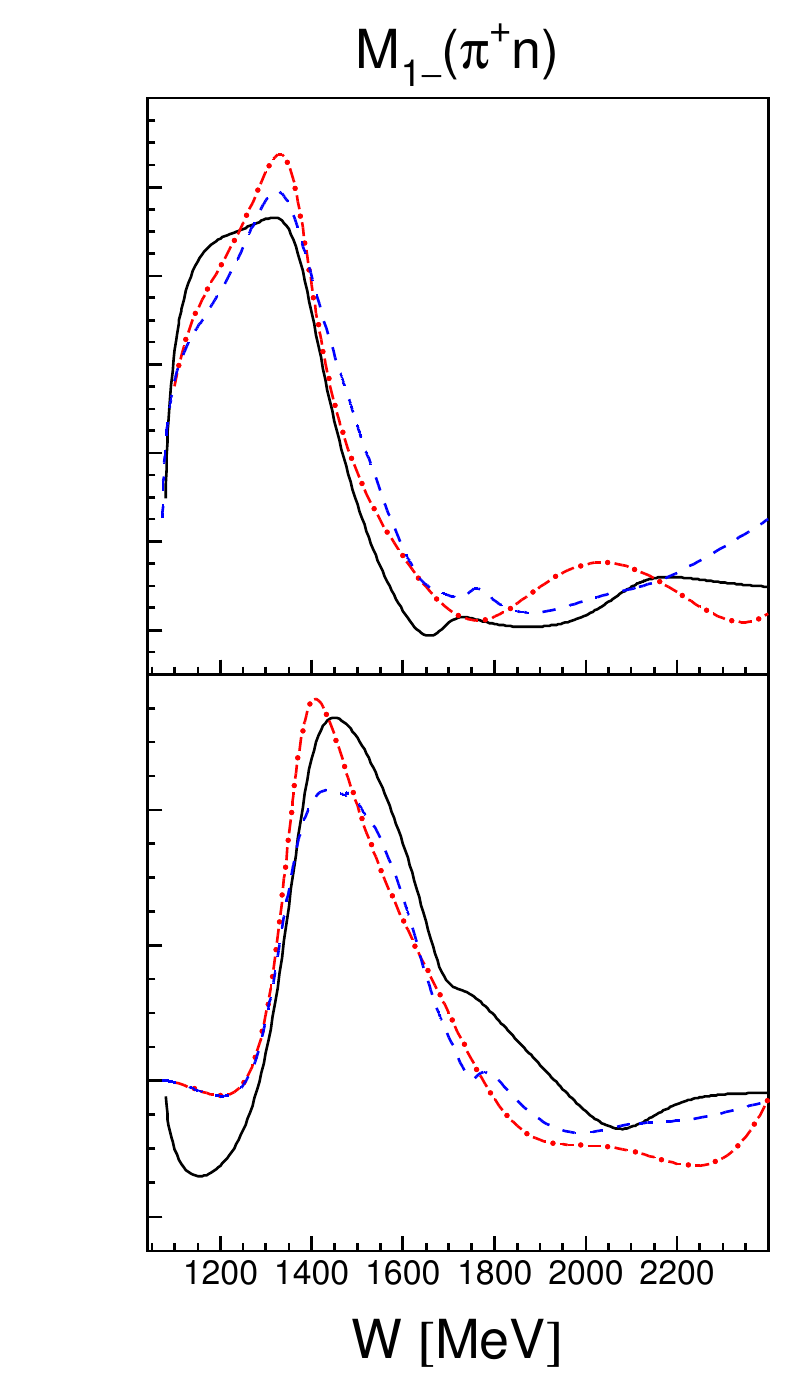,width=0.28\textwidth,height=0.42\textwidth}
}
  \centerline{\epsfig{file=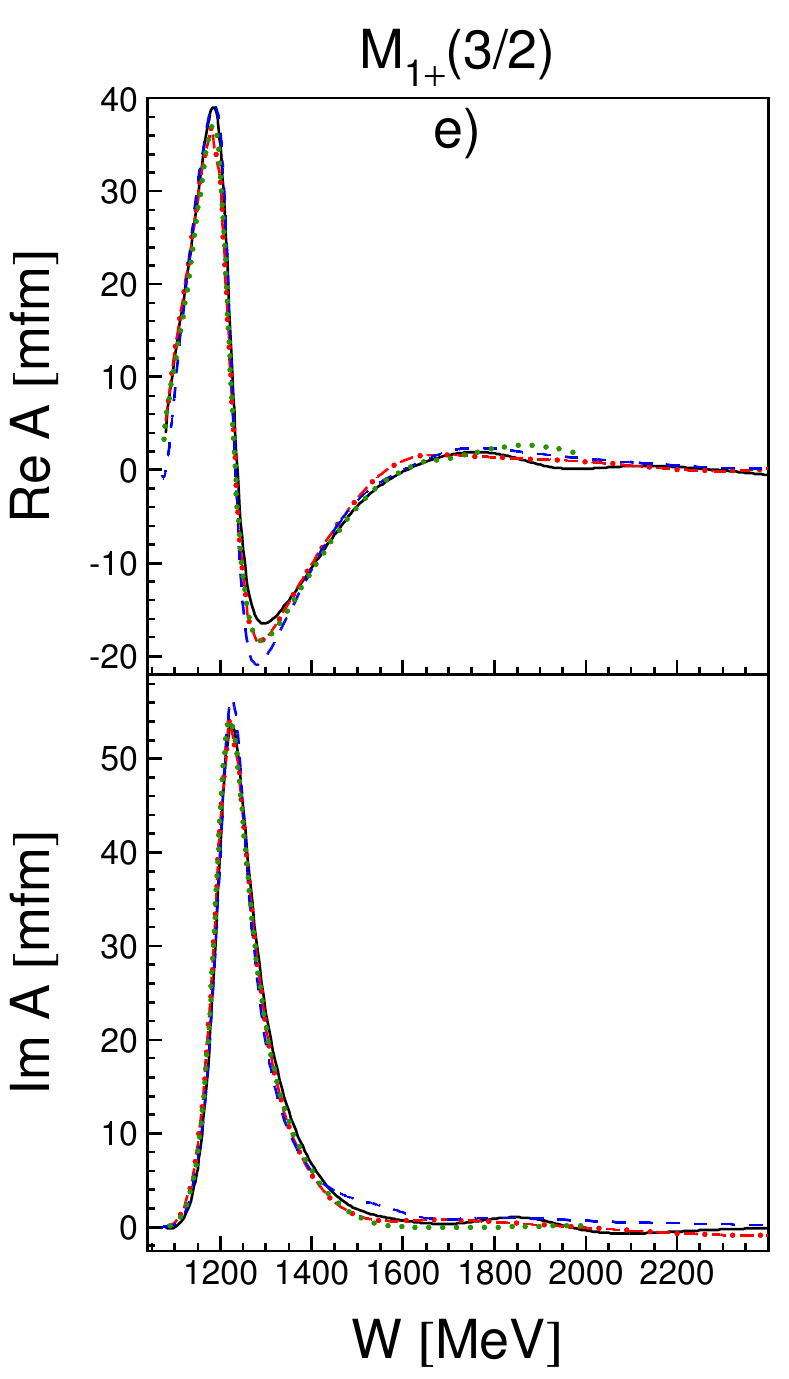,width=0.28\textwidth,height=0.42\textwidth}
\hspace{-12mm}\epsfig{file=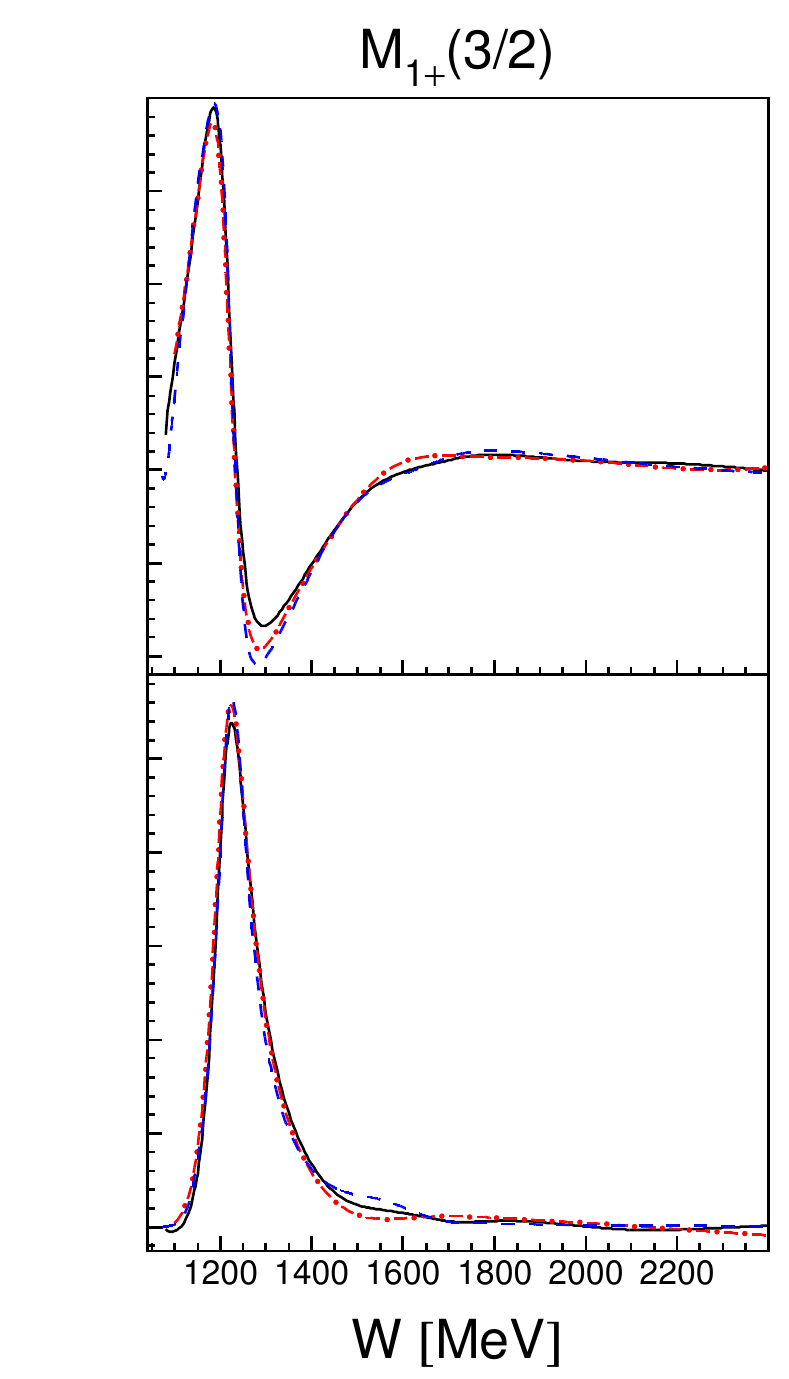,width=0.28\textwidth,height=0.42\textwidth}
              \epsfig{file=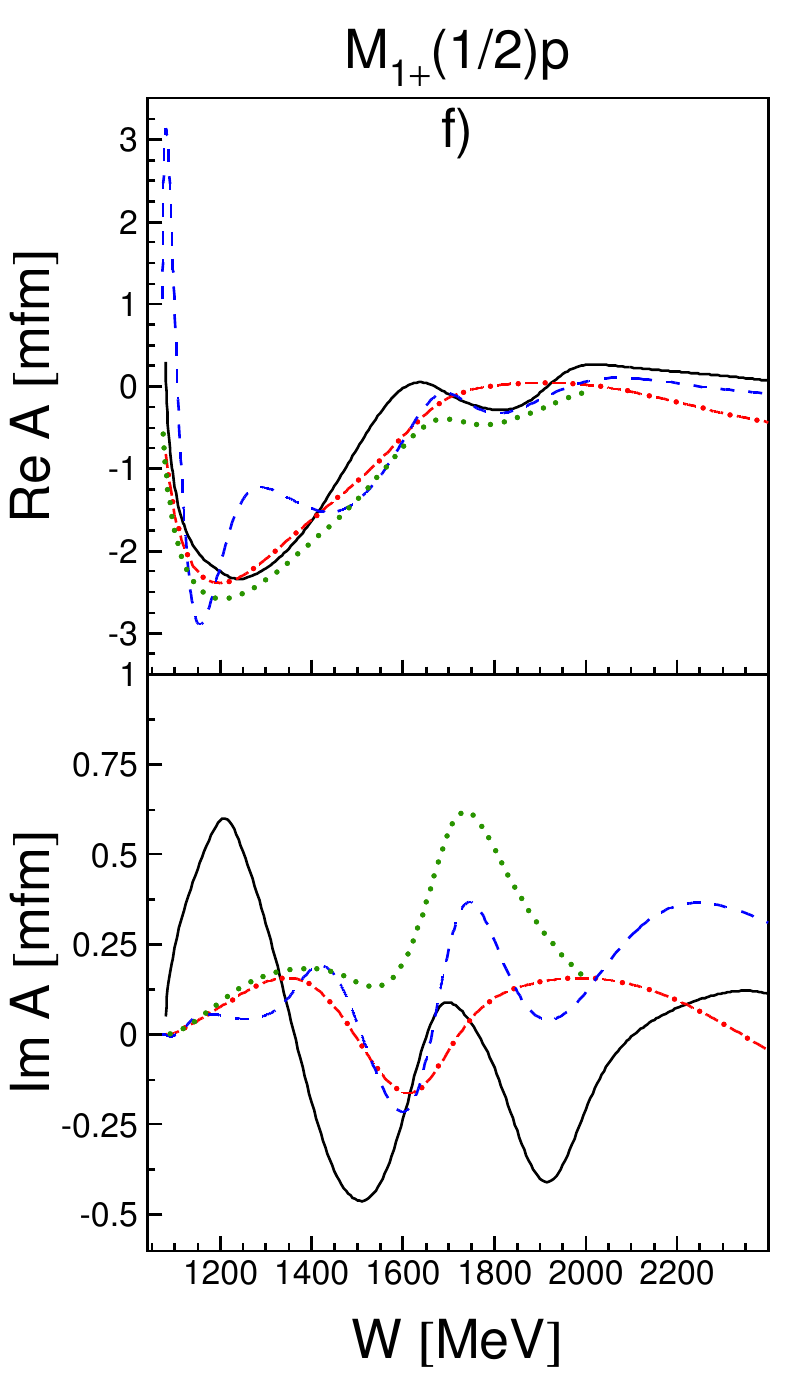,width=0.28\textwidth,height=0.42\textwidth}
\hspace{-12mm}\epsfig{file=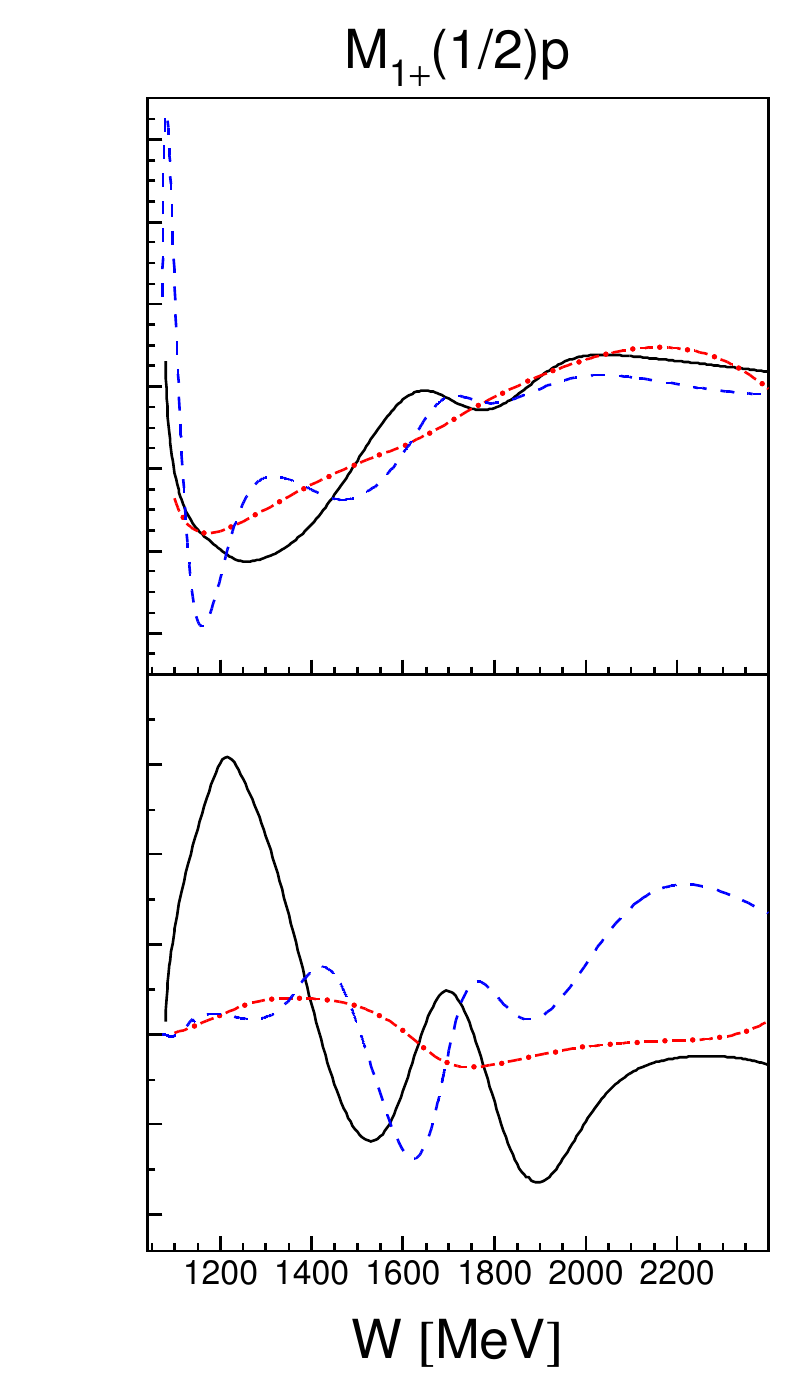,width=0.28\textwidth,height=0.42\textwidth}
}
\caption{\label{part1}Each block presents the real (top) and imaginary (bottom) part of multipoles for
$\gamma p\to \pi N$, before (left) and after (right) including new data. Black solid line: BnGa, blue dashed: J\"uBo, red dashed dotted: SAID, green dotted: MAID. Blocks a and c show the $\gamma p\to \pi^0 p$ multipoles, b and d those for $\gamma p\to \pi^+ n$.
Block e (f) presents the $I=3/2$ ($I=1/2$) multipoles.}
\end{figure*}

\begin{figure*}
  \centerline{\epsfig{file=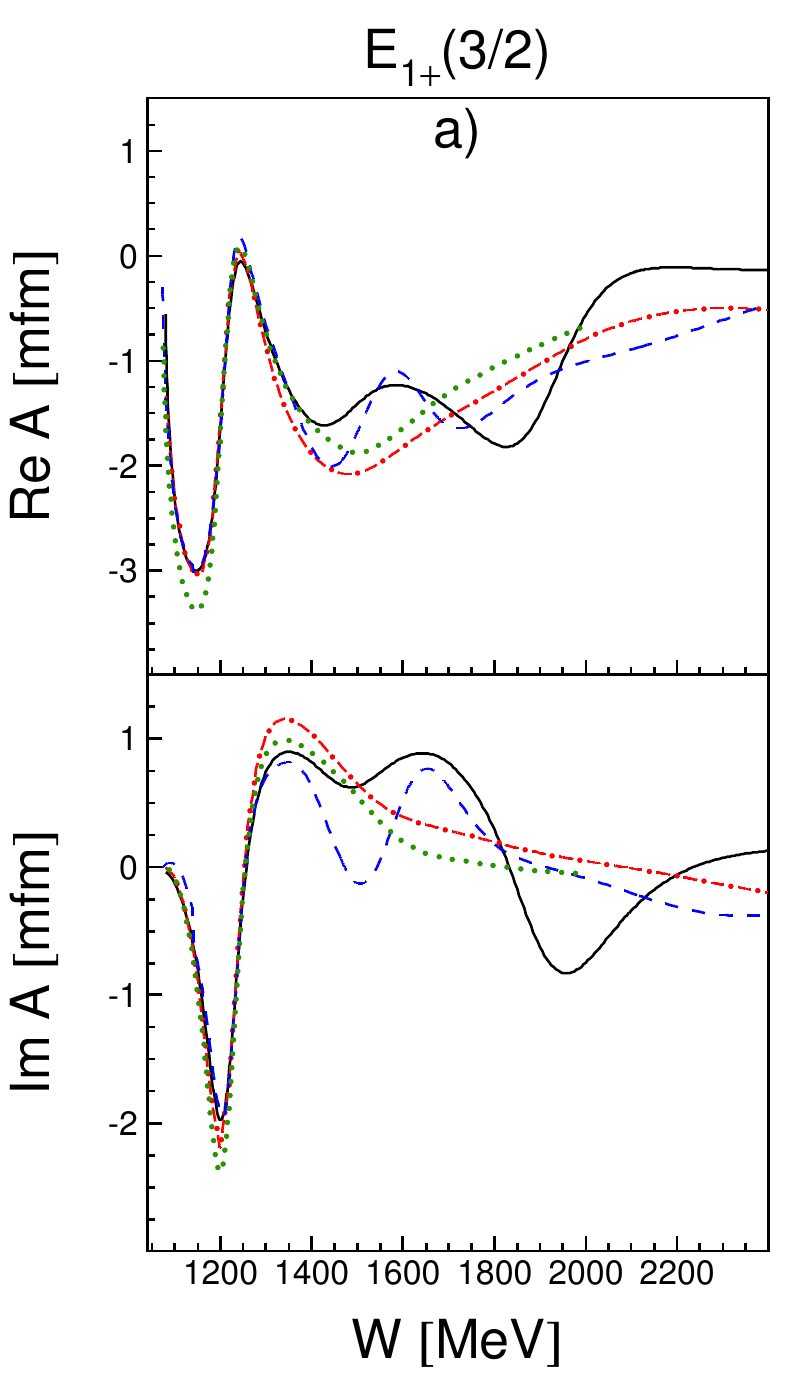,width=0.28\textwidth,height=0.42\textwidth}
\hspace{-12mm}\epsfig{file=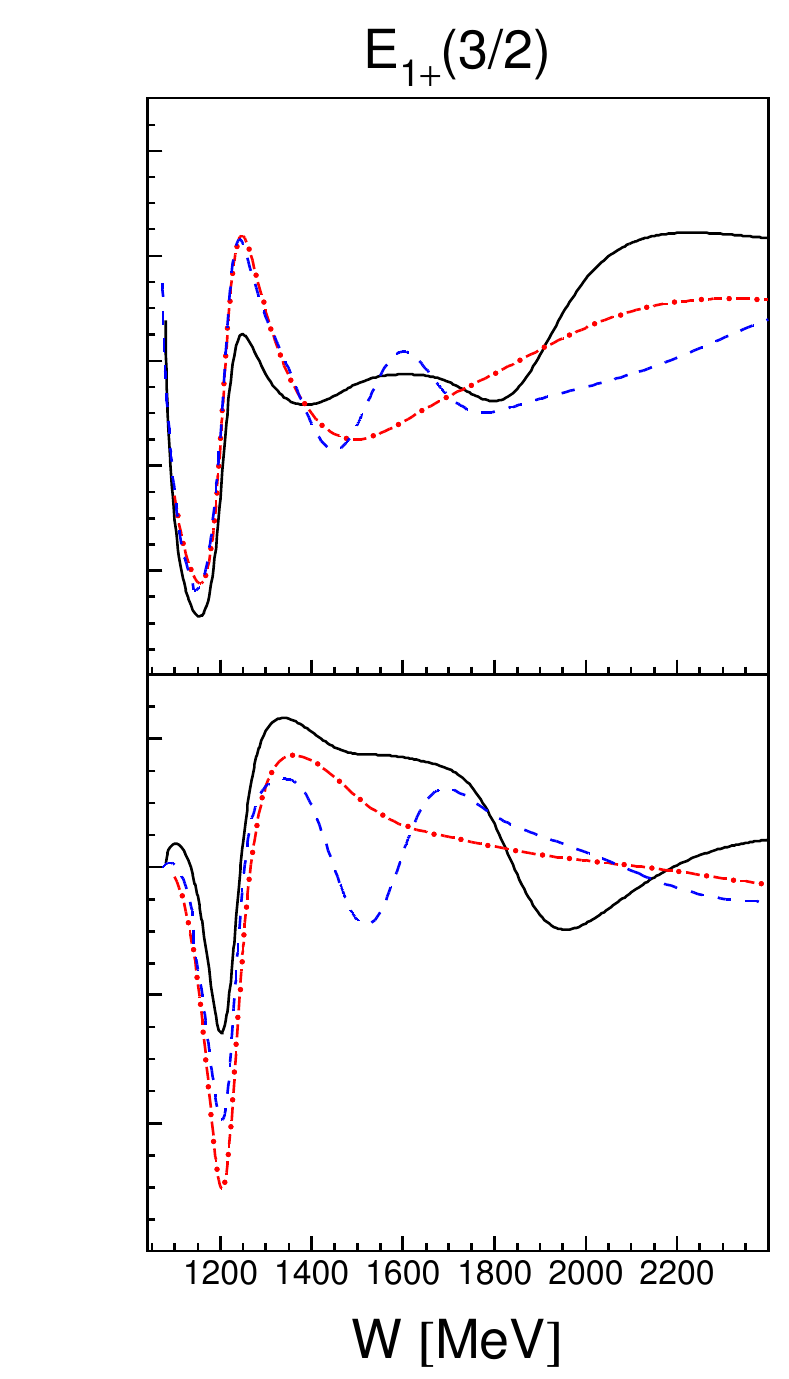,width=0.28\textwidth,height=0.42\textwidth}
              \epsfig{file=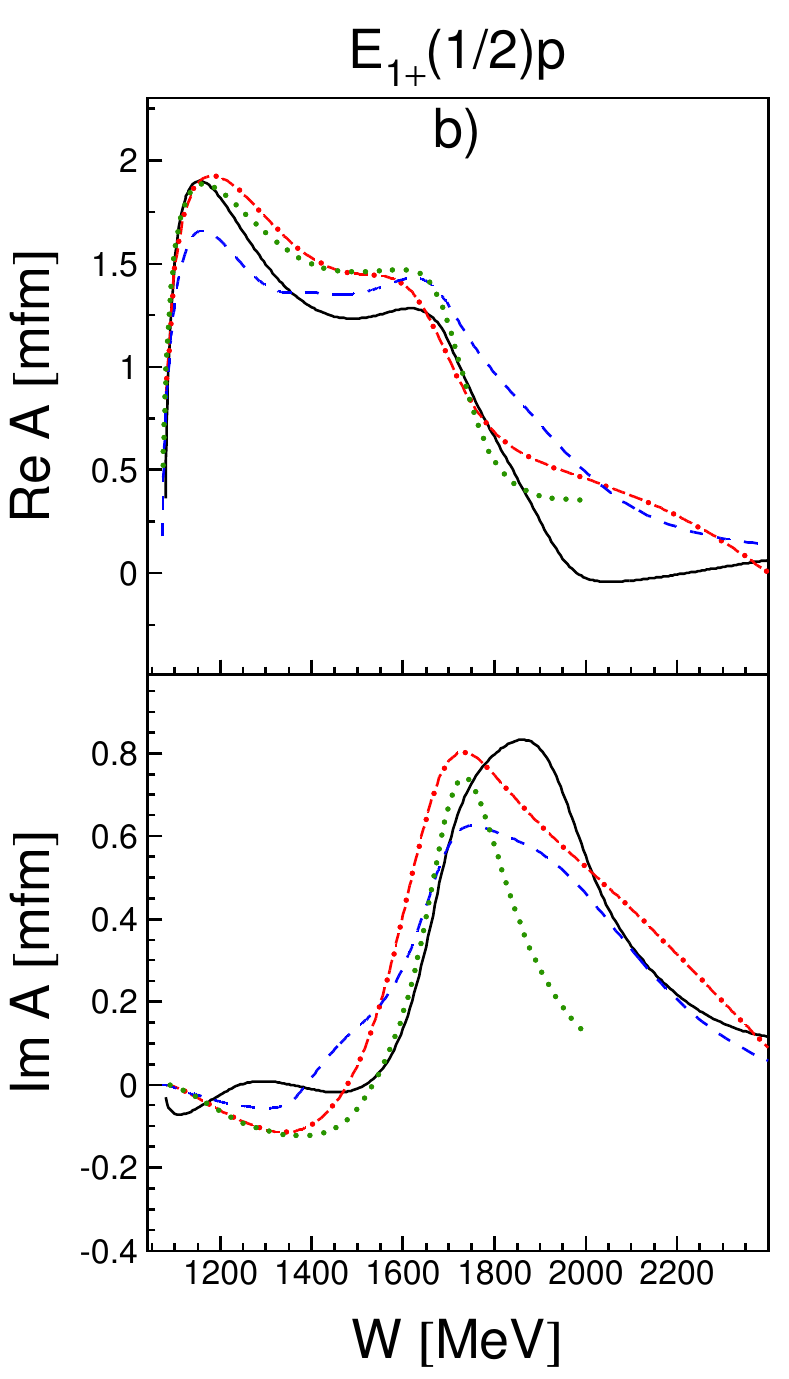,width=0.28\textwidth,height=0.42\textwidth}
\hspace{-12mm}\epsfig{file=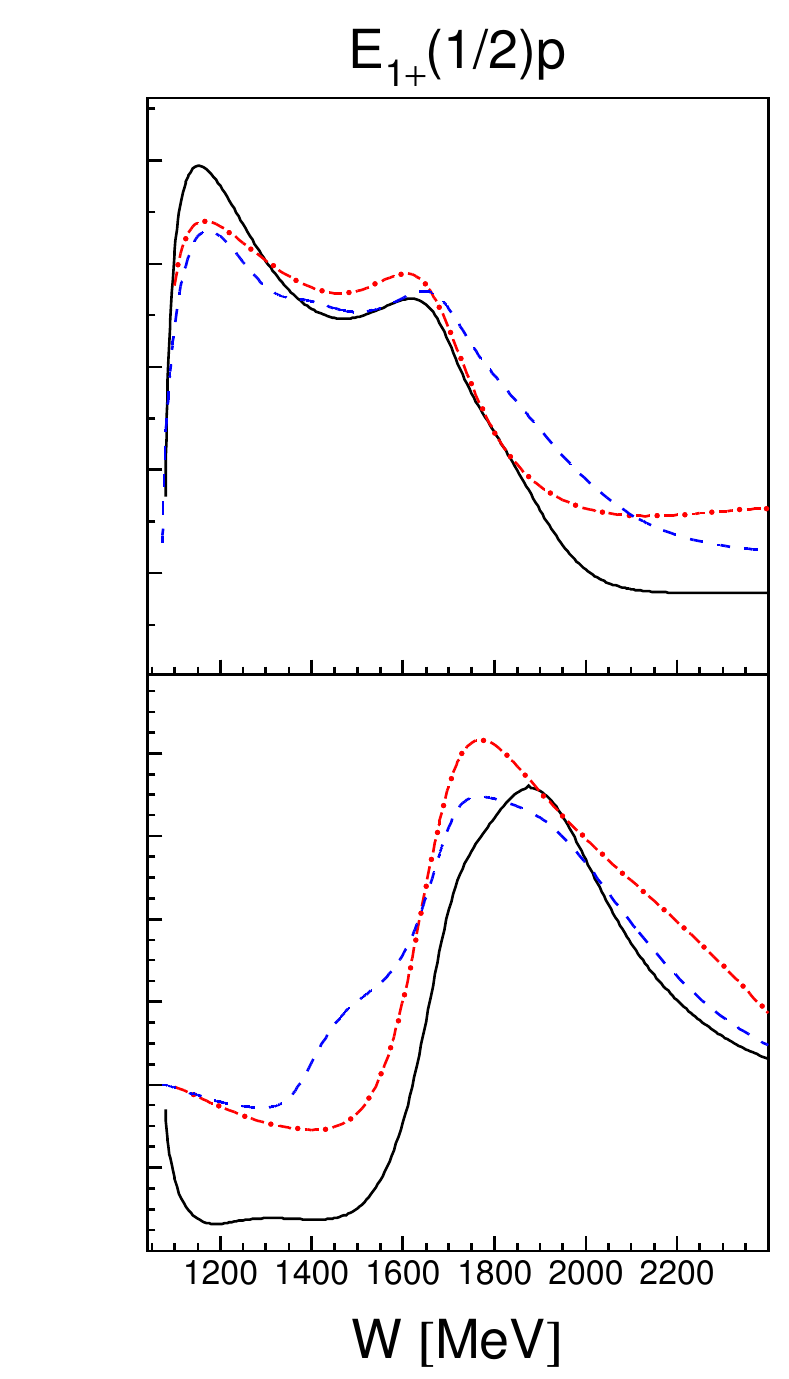,width=0.28\textwidth,height=0.42\textwidth}
}
\centerline{\epsfig{file=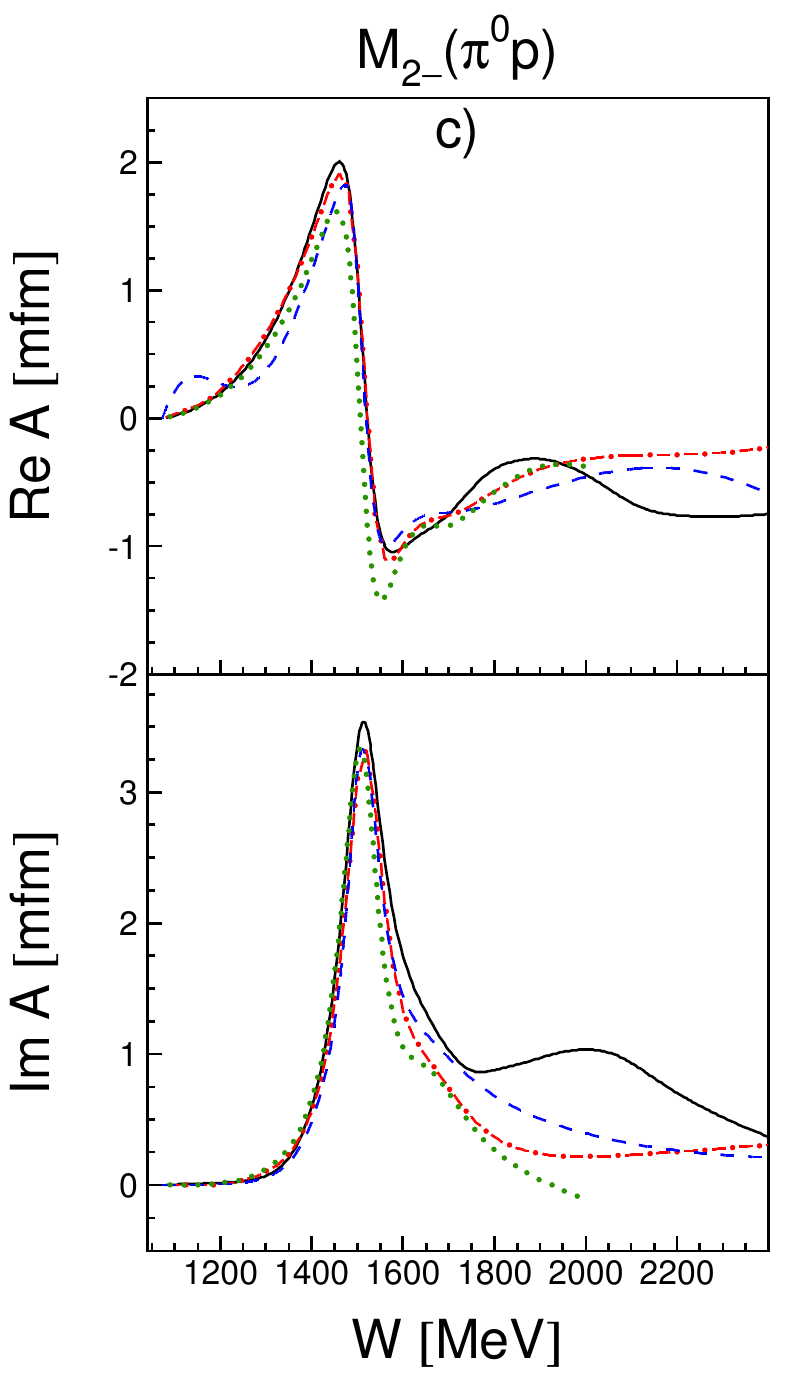,width=0.28\textwidth,height=0.42\textwidth}
\hspace{-12mm}\epsfig{file=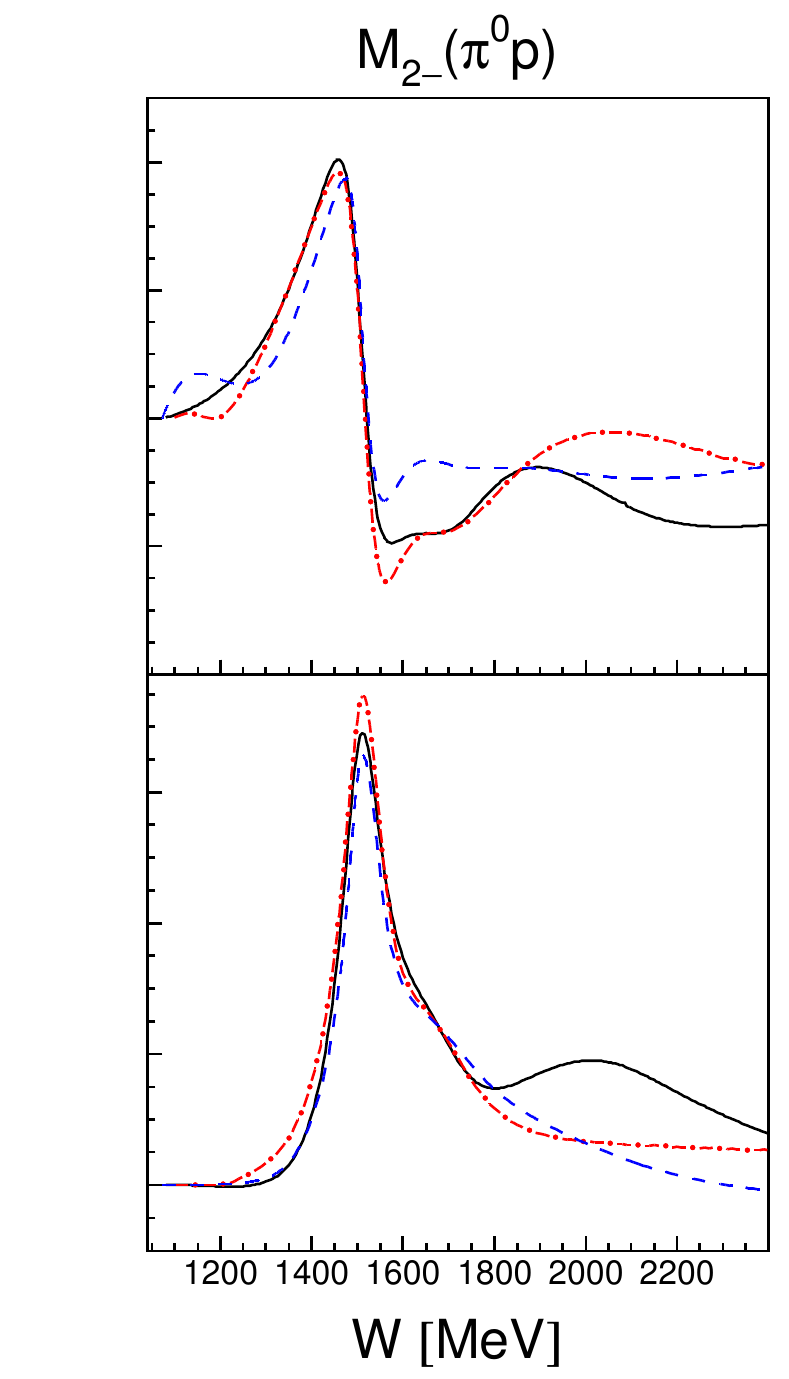,width=0.28\textwidth,height=0.42\textwidth}
             \epsfig{file=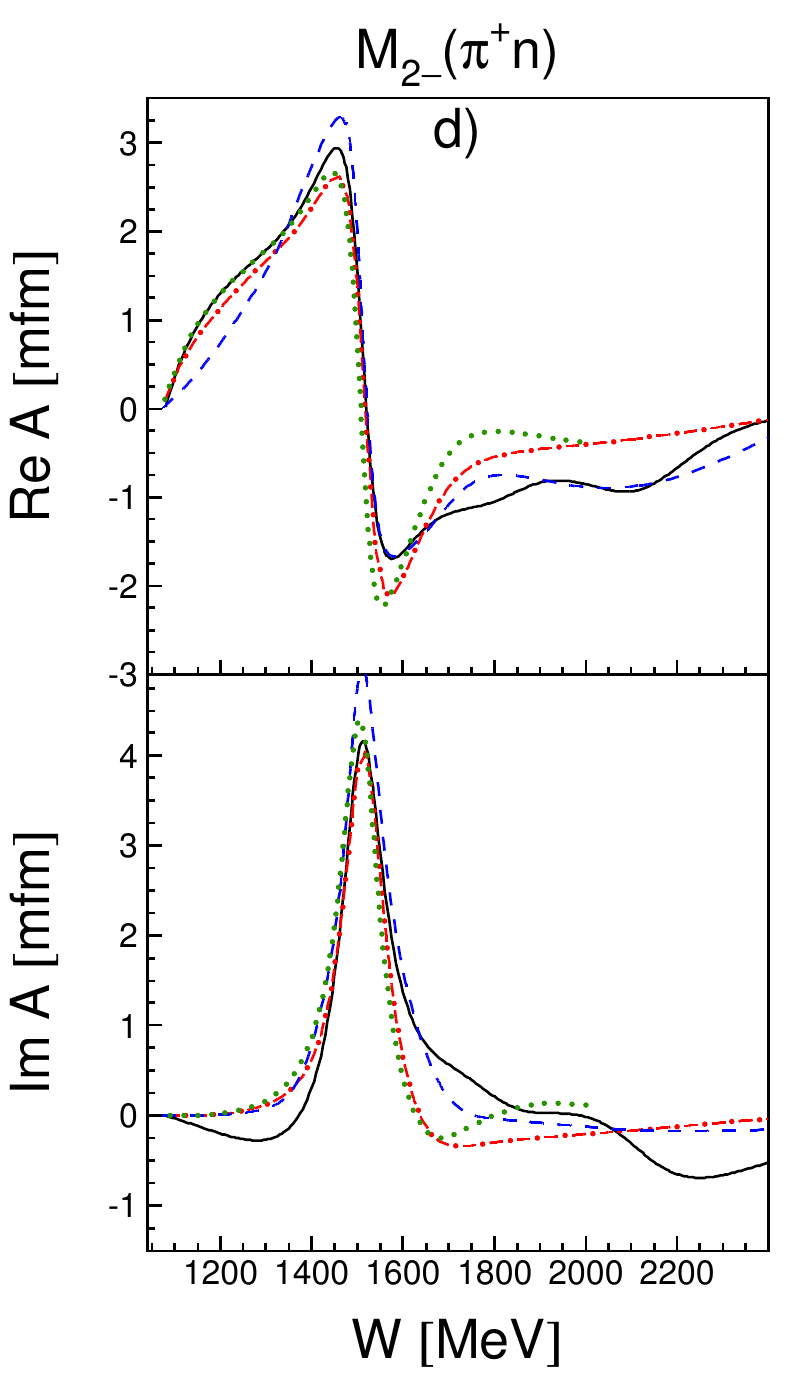,width=0.28\textwidth,height=0.42\textwidth}
\hspace{-12mm}\epsfig{file=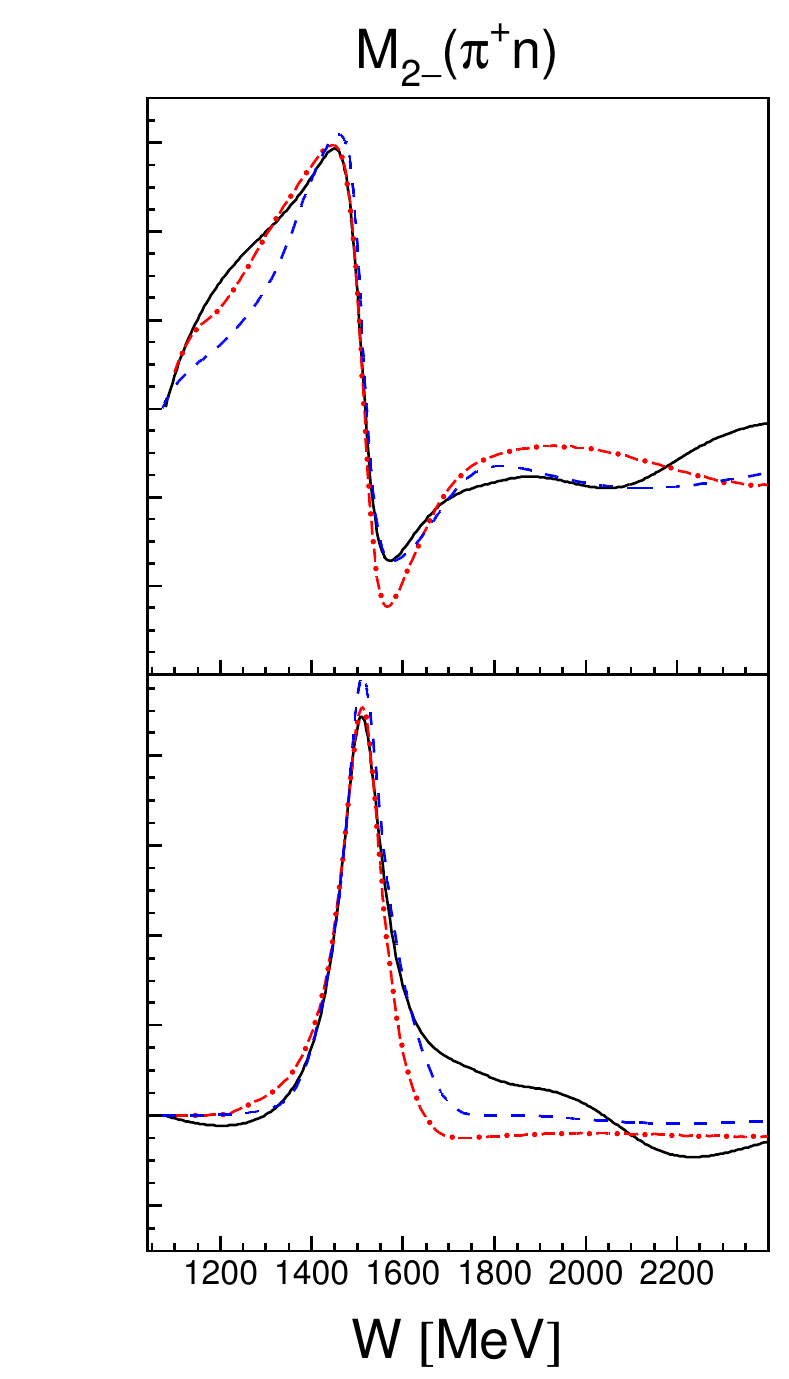,width=0.28\textwidth,height=0.42\textwidth}
}
\centerline{\epsfig{file=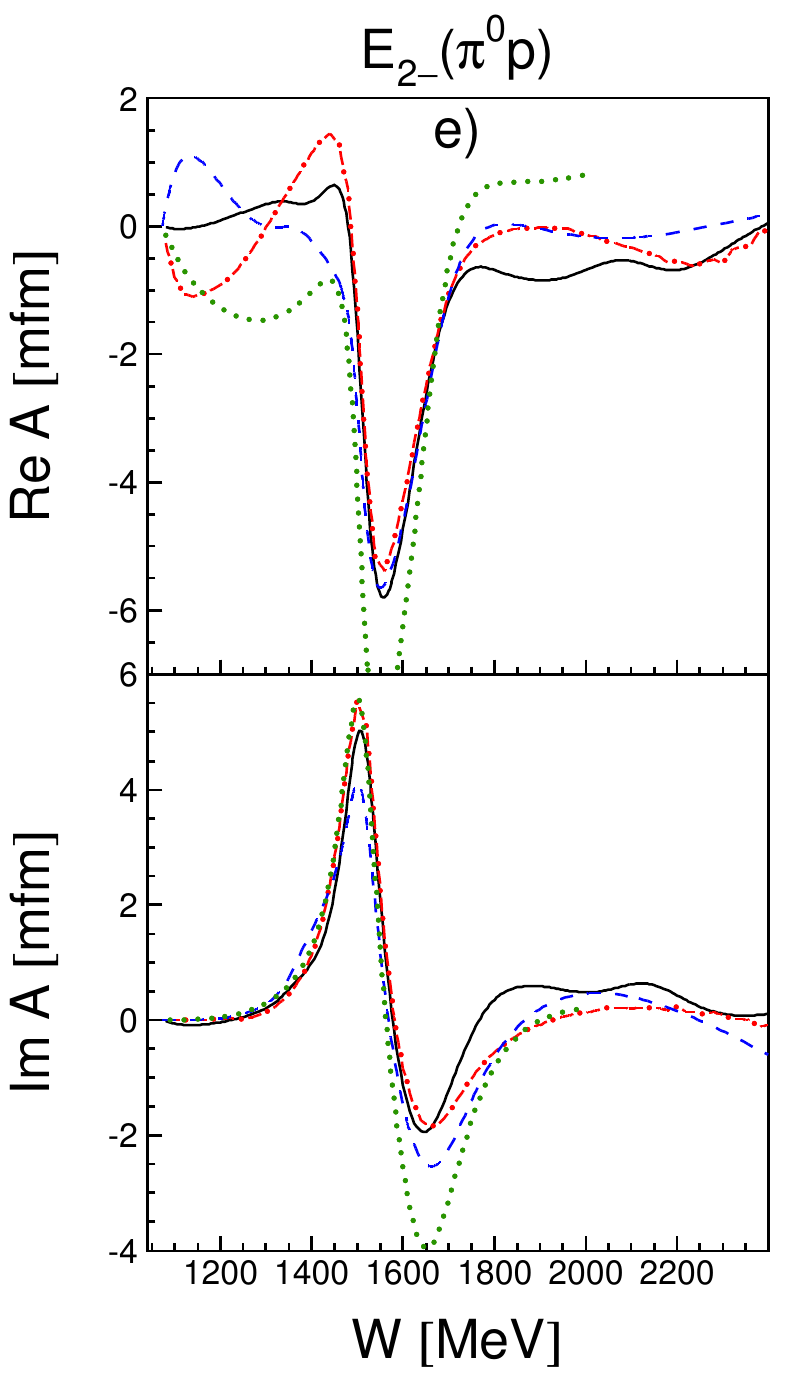,width=0.28\textwidth,height=0.42\textwidth}
\hspace{-12mm}\epsfig{file=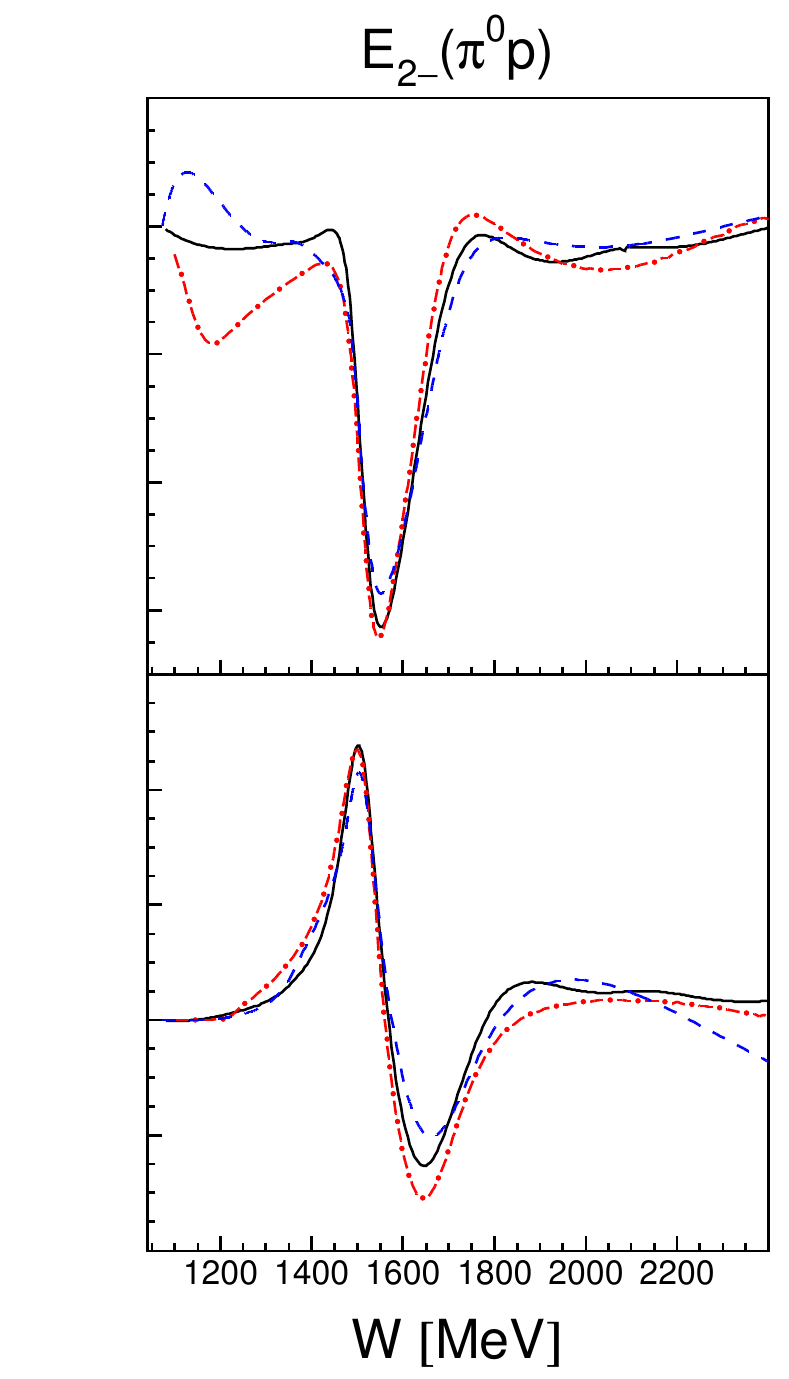,width=0.28\textwidth,height=0.42\textwidth}
              \epsfig{file=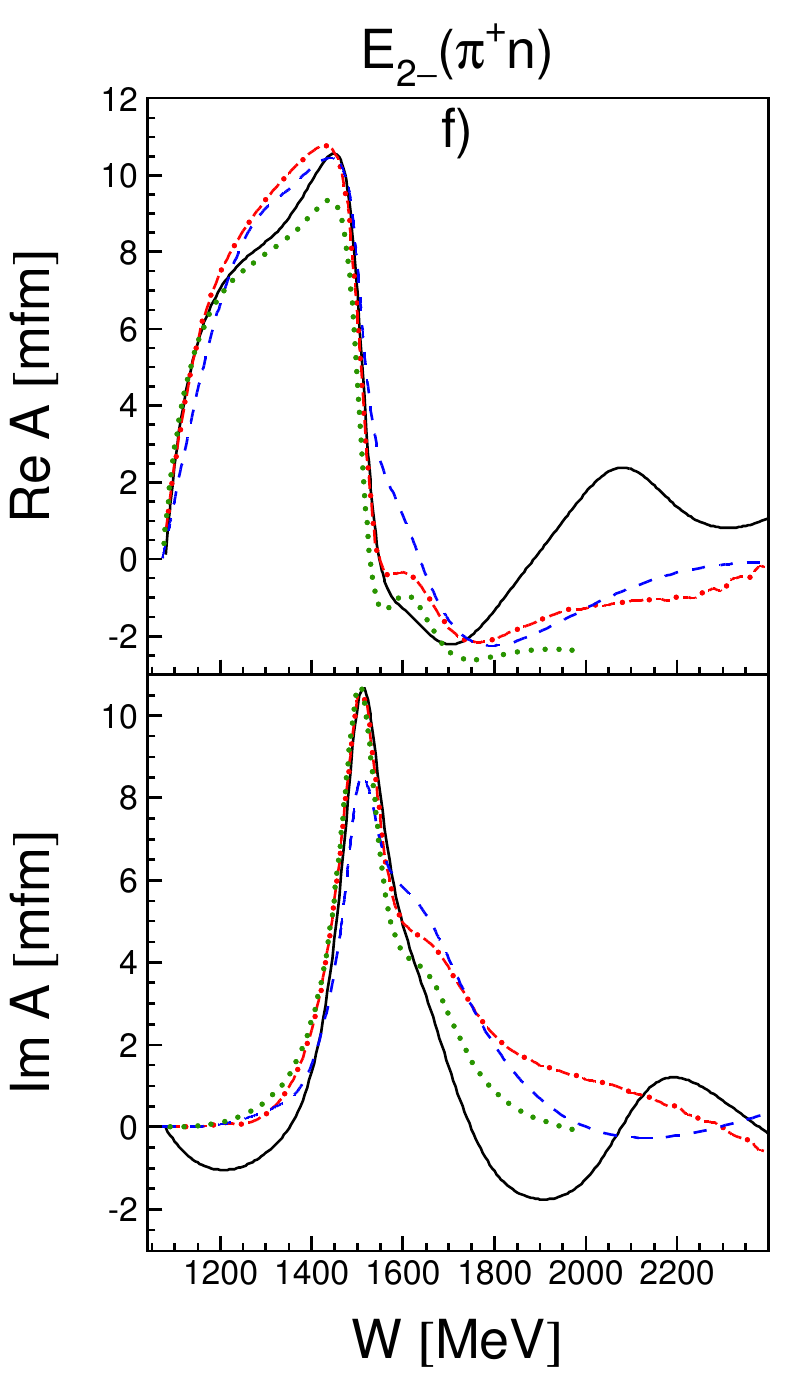,width=0.28\textwidth,height=0.42\textwidth}
\hspace{-12mm}\epsfig{file=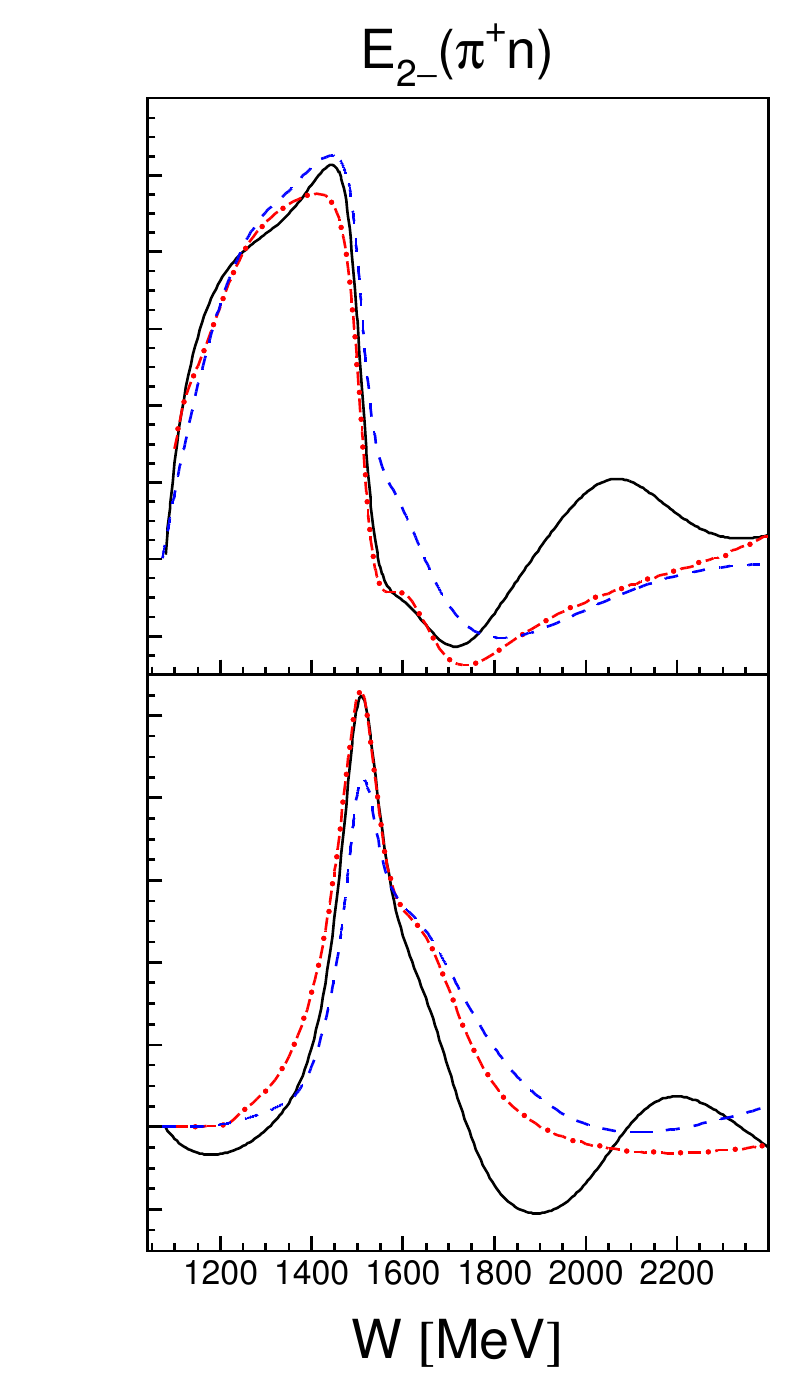,width=0.28\textwidth,height=0.42\textwidth}
}
\caption{\label{part2}Each block presents the real (top) and imaginary (bottom) part of multipoles for
$\gamma p\to \pi N$, before (left) and after (right) including new data. Black solid line: BnGa, blue dashed: J\"uBo, red dashed dotted: SAID, green dotted: MAID. Block a (b)  presents the $I=3/2$ ($I=1/2$) multipole; block c and e
show the $\gamma p\to \pi^0 p$ multipoles, d and f those for $\gamma p\to \pi^+ n$.
}
\end{figure*}

\begin{figure*}
 \centerline{\epsfig{file=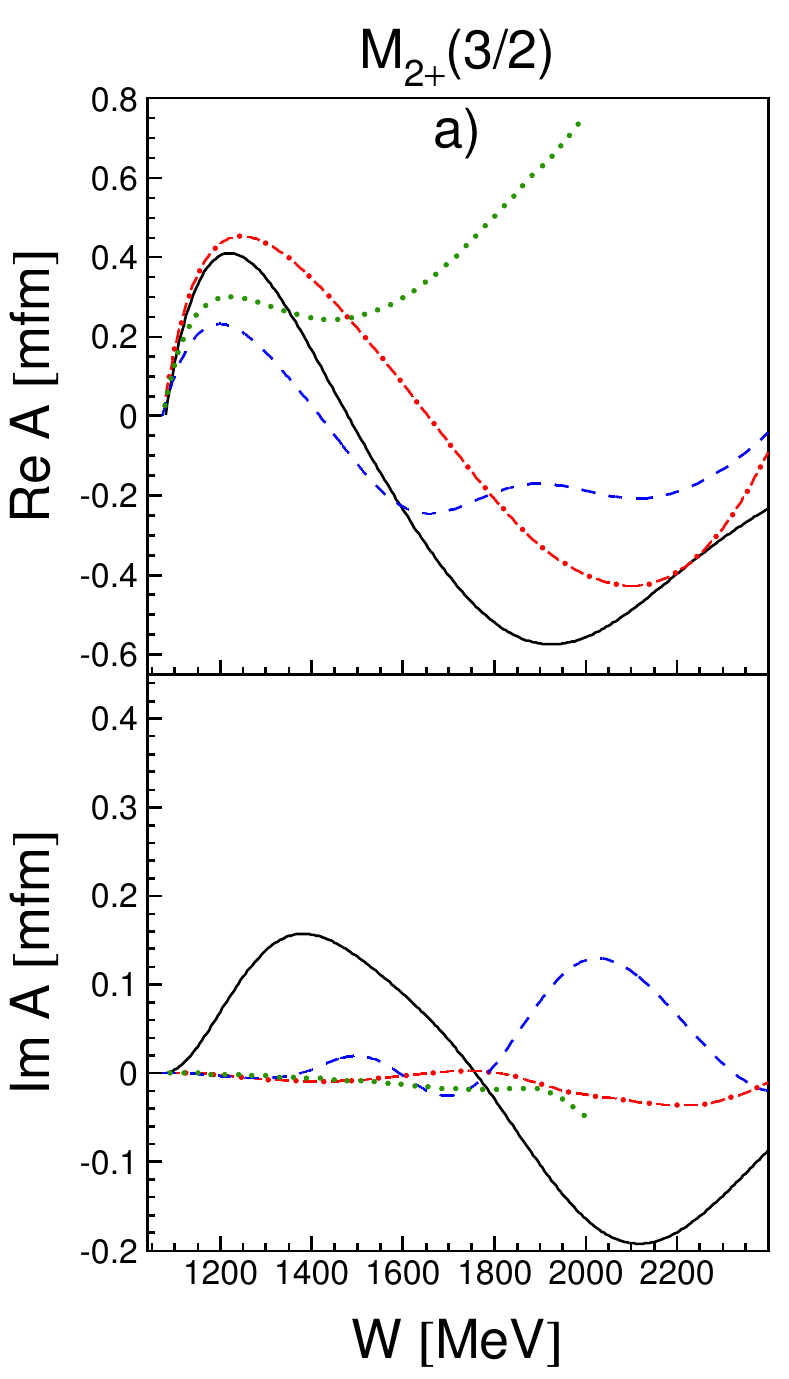,width=0.28\textwidth,height=0.42\textwidth}
\hspace{-12mm}\epsfig{file=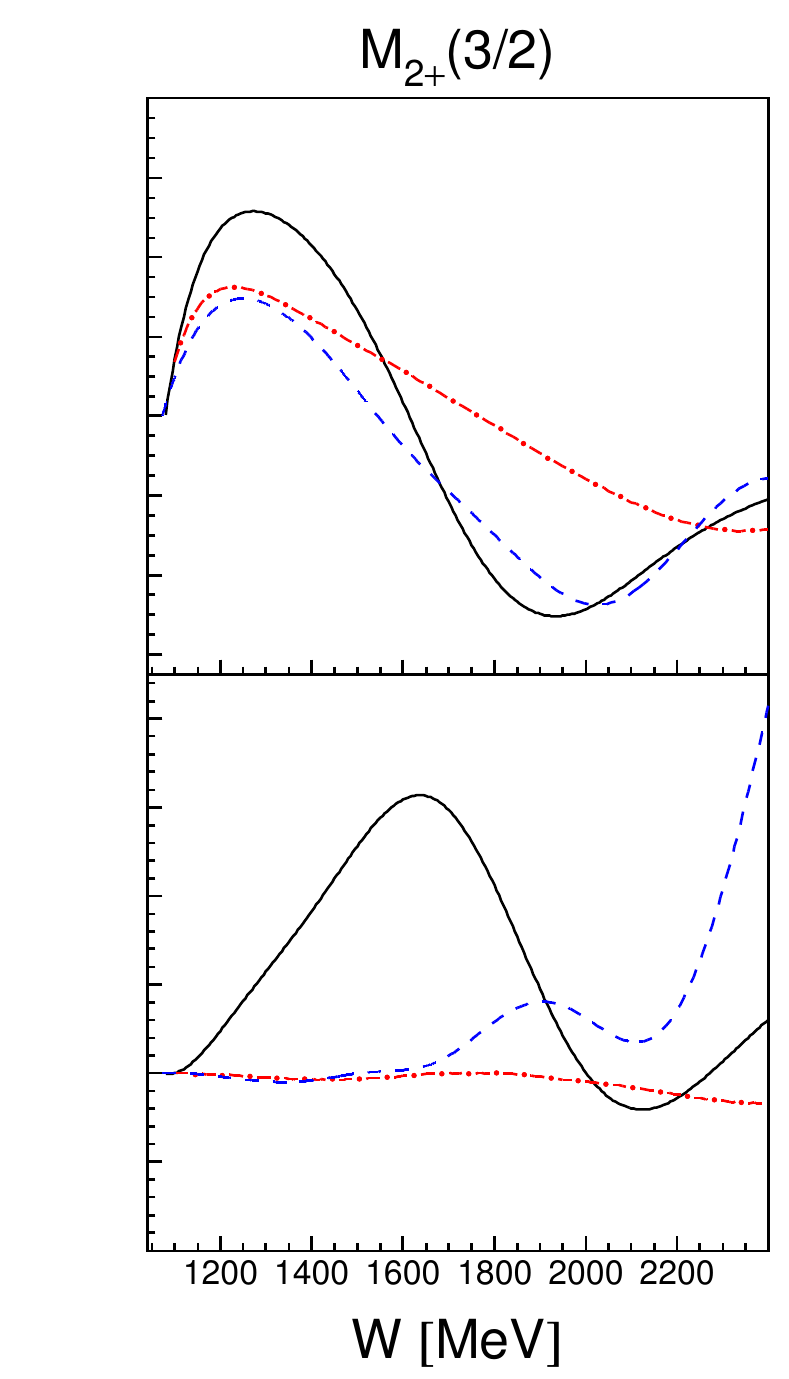,width=0.28\textwidth,height=0.42\textwidth}
             \epsfig{file=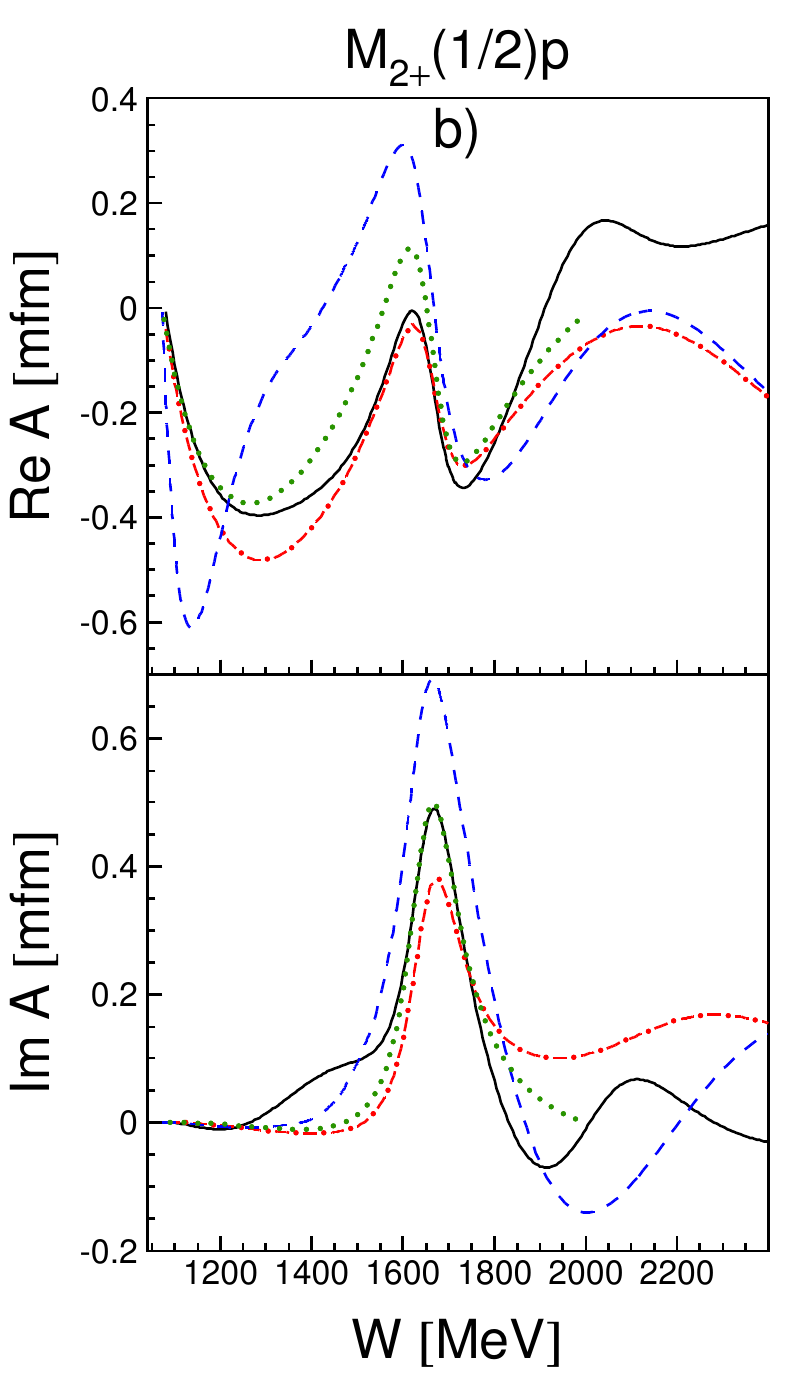,width=0.28\textwidth,height=0.42\textwidth}
\hspace{-12mm}\epsfig{file=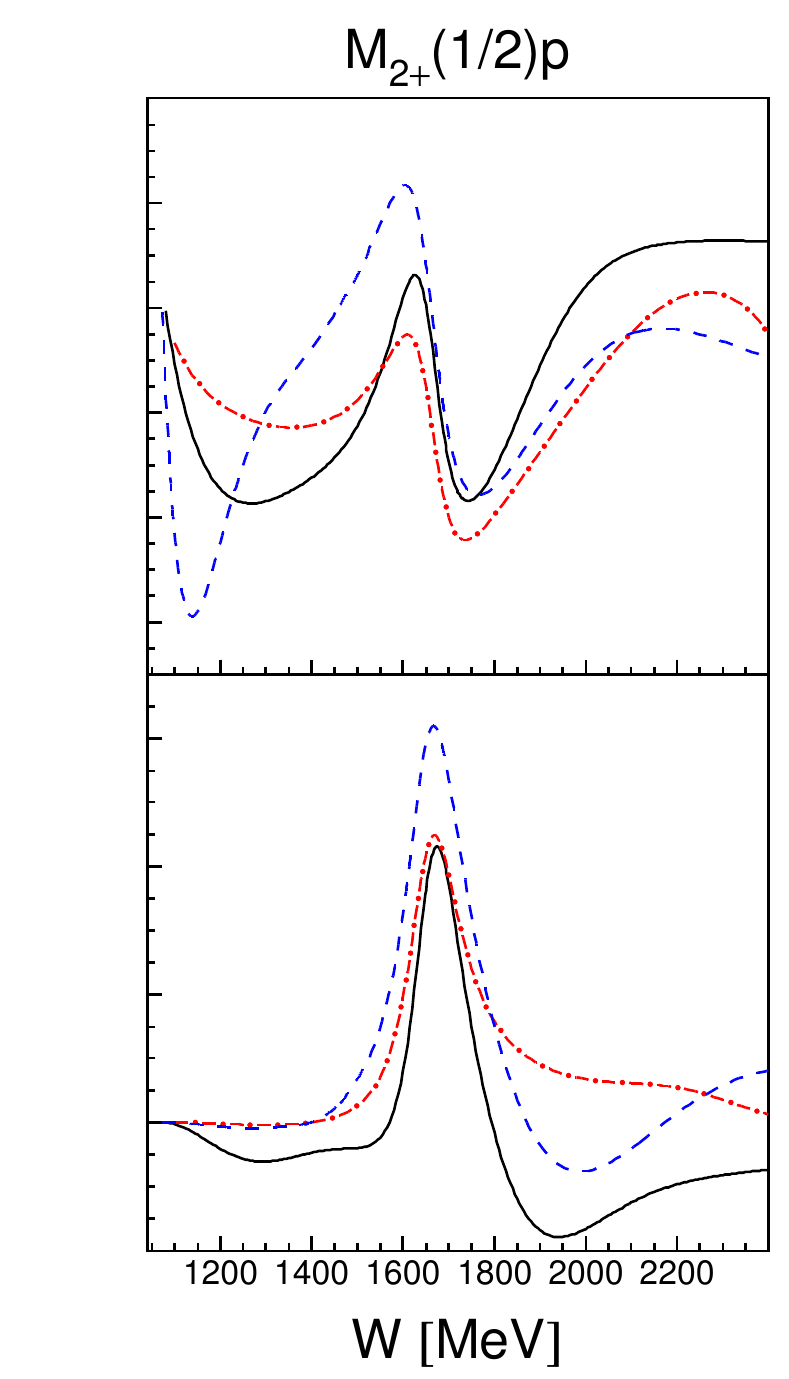,width=0.28\textwidth,height=0.42\textwidth}
}
 \centerline{\epsfig{file=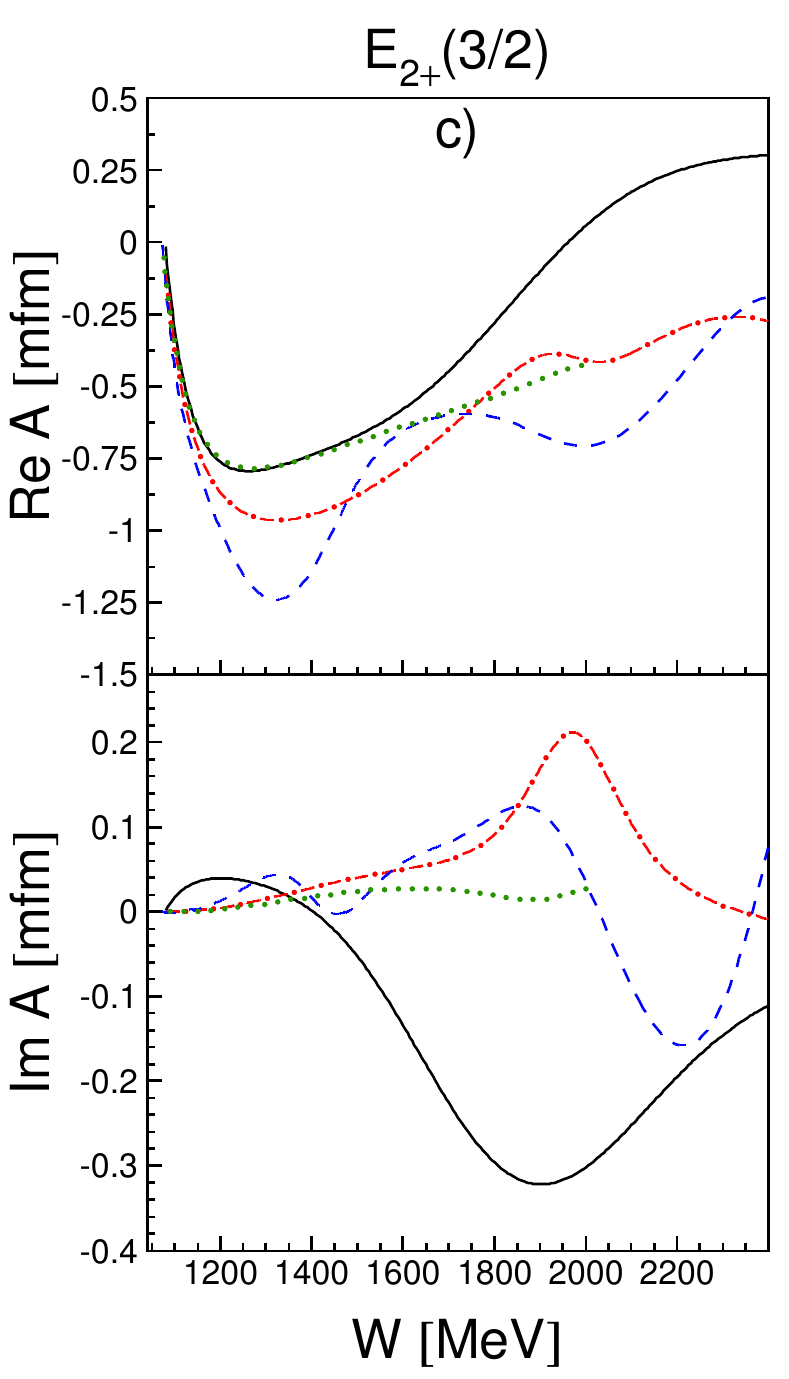,width=0.28\textwidth,height=0.42\textwidth}
\hspace{-12mm}\epsfig{file=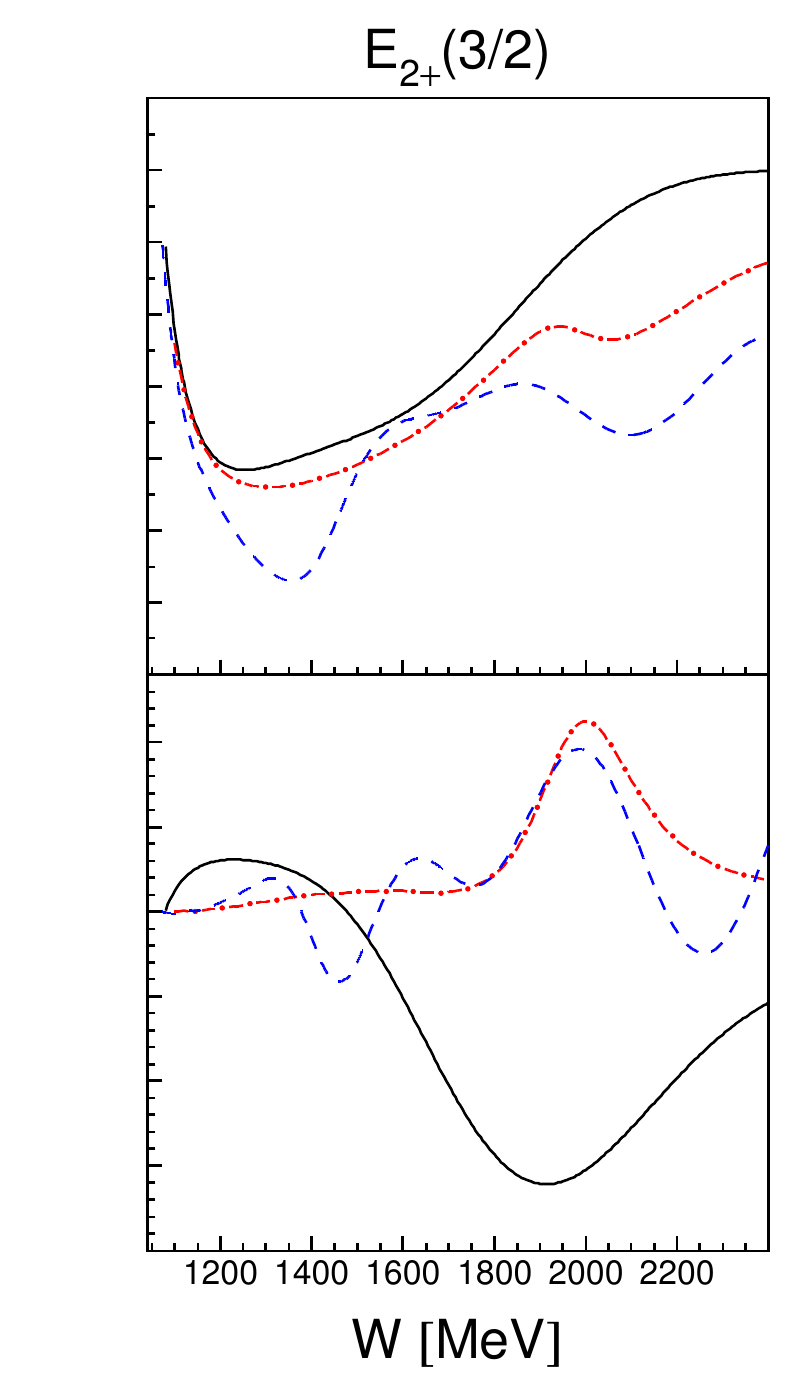,width=0.28\textwidth,height=0.42\textwidth}
              \epsfig{file=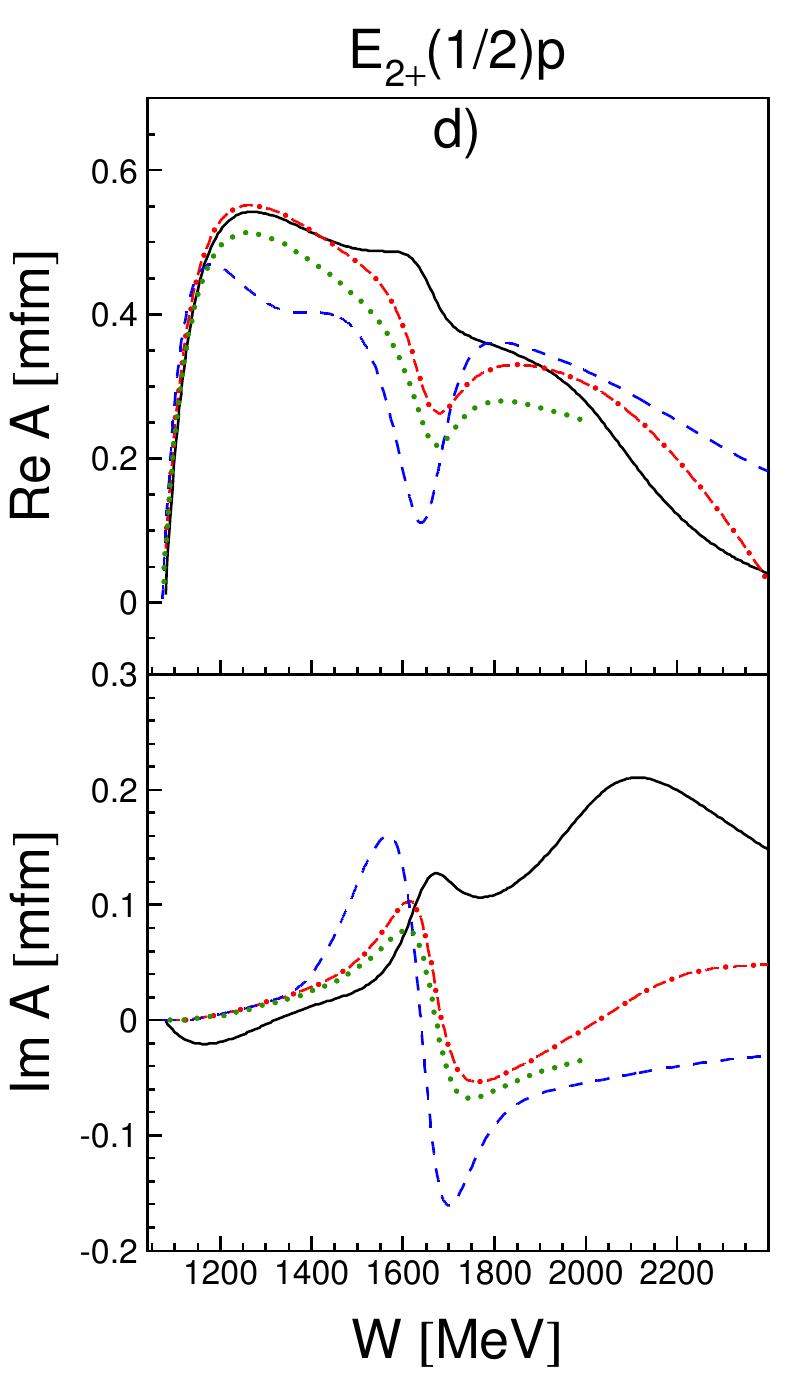,width=0.28\textwidth,height=0.42\textwidth}
\hspace{-12mm}\epsfig{file=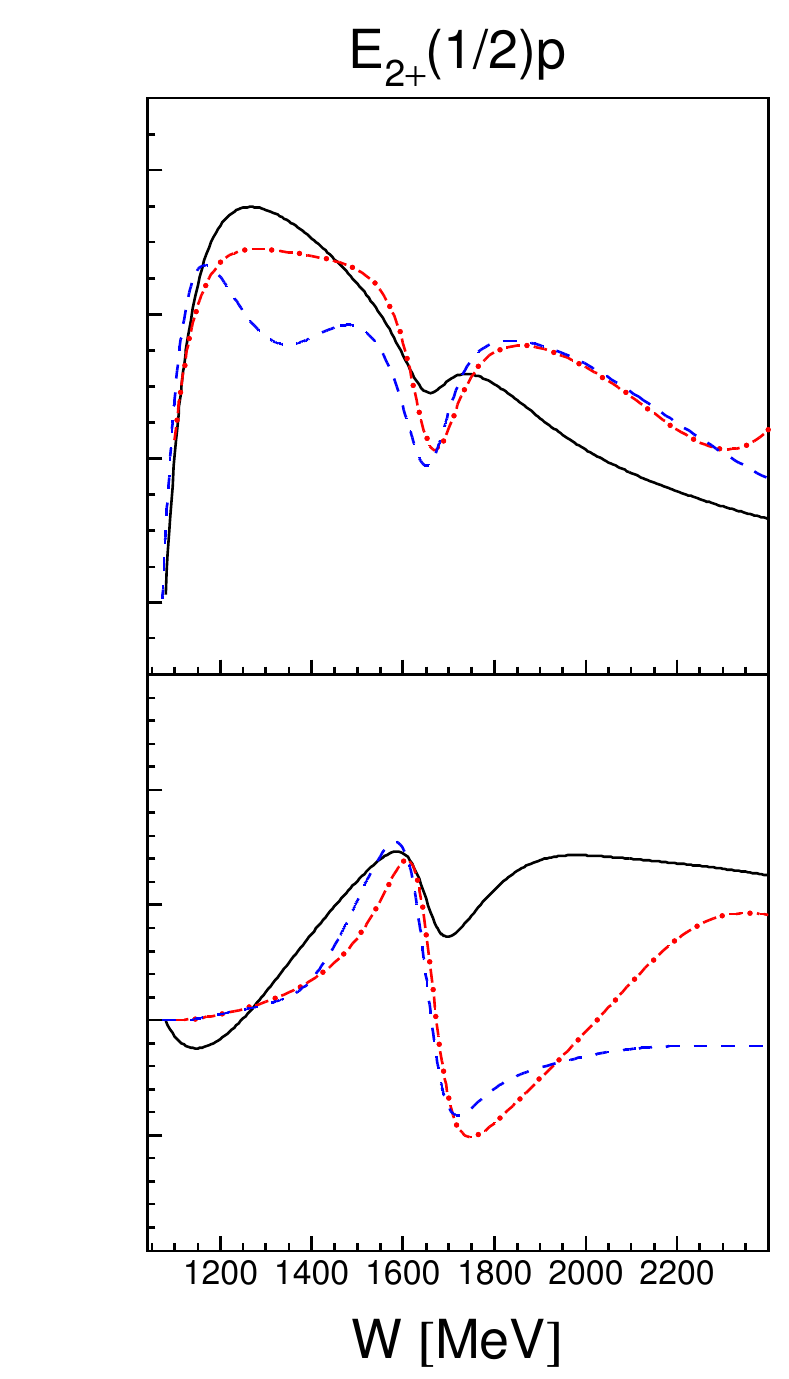,width=0.28\textwidth,height=0.42\textwidth}
}
\centerline{\epsfig{file=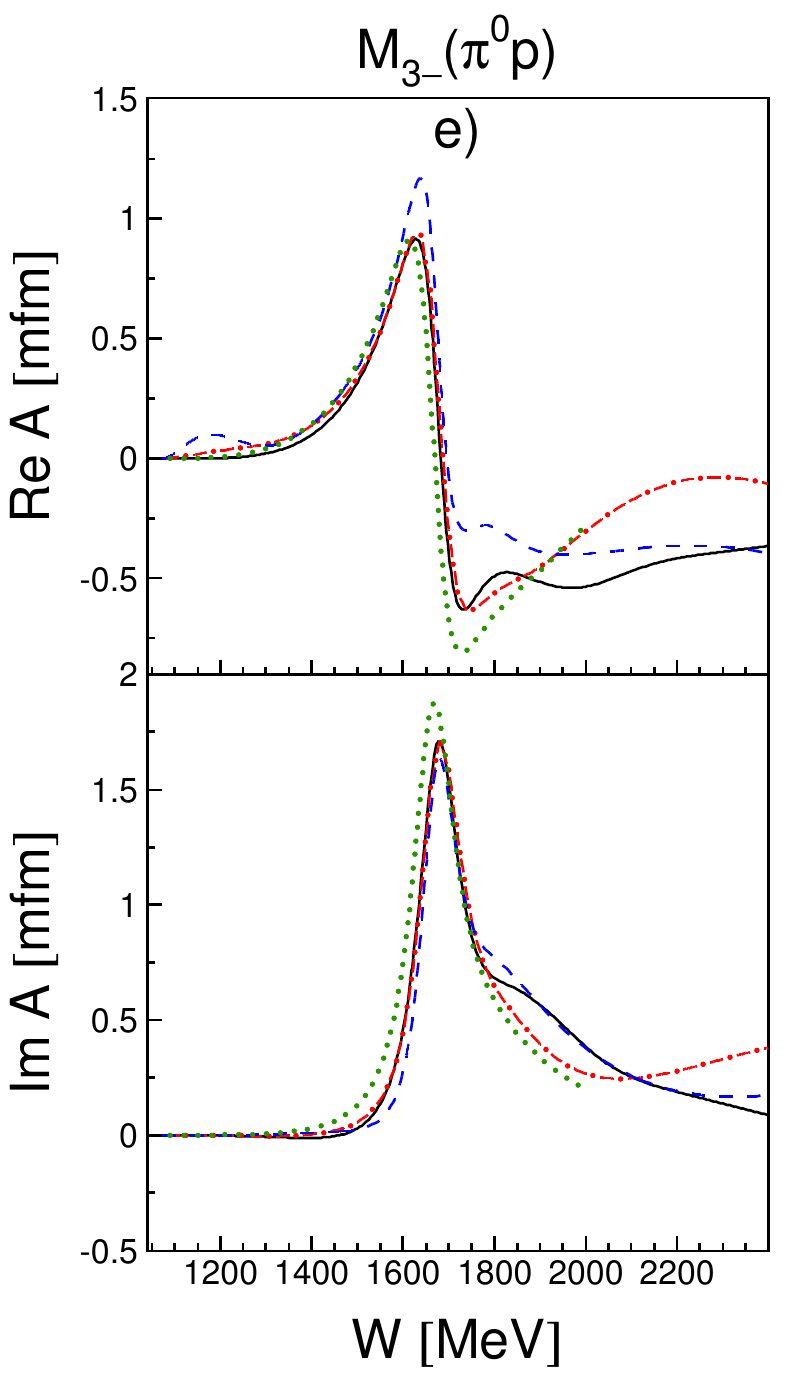,width=0.28\textwidth,height=0.42\textwidth}
\hspace{-12mm}\epsfig{file=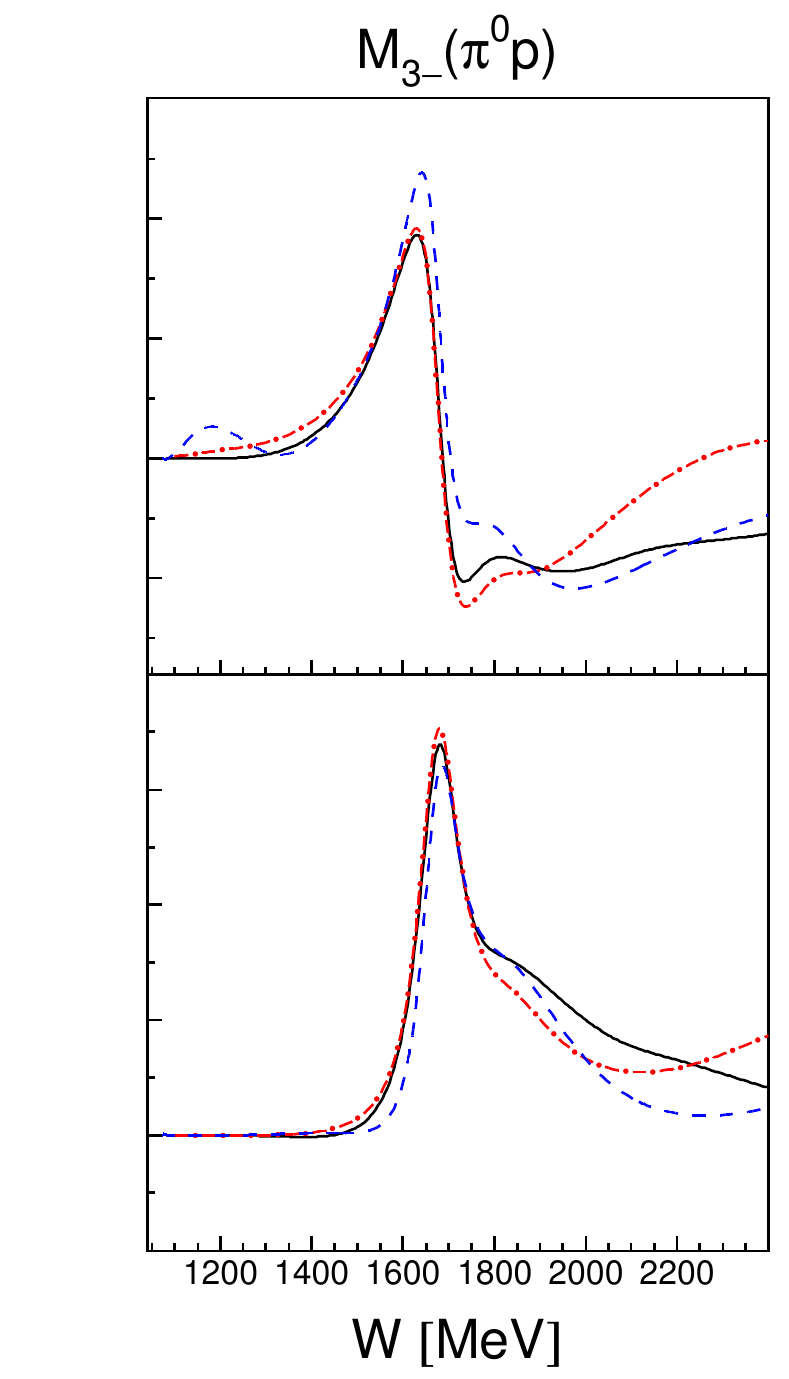,width=0.28\textwidth,height=0.42\textwidth}
              \epsfig{file=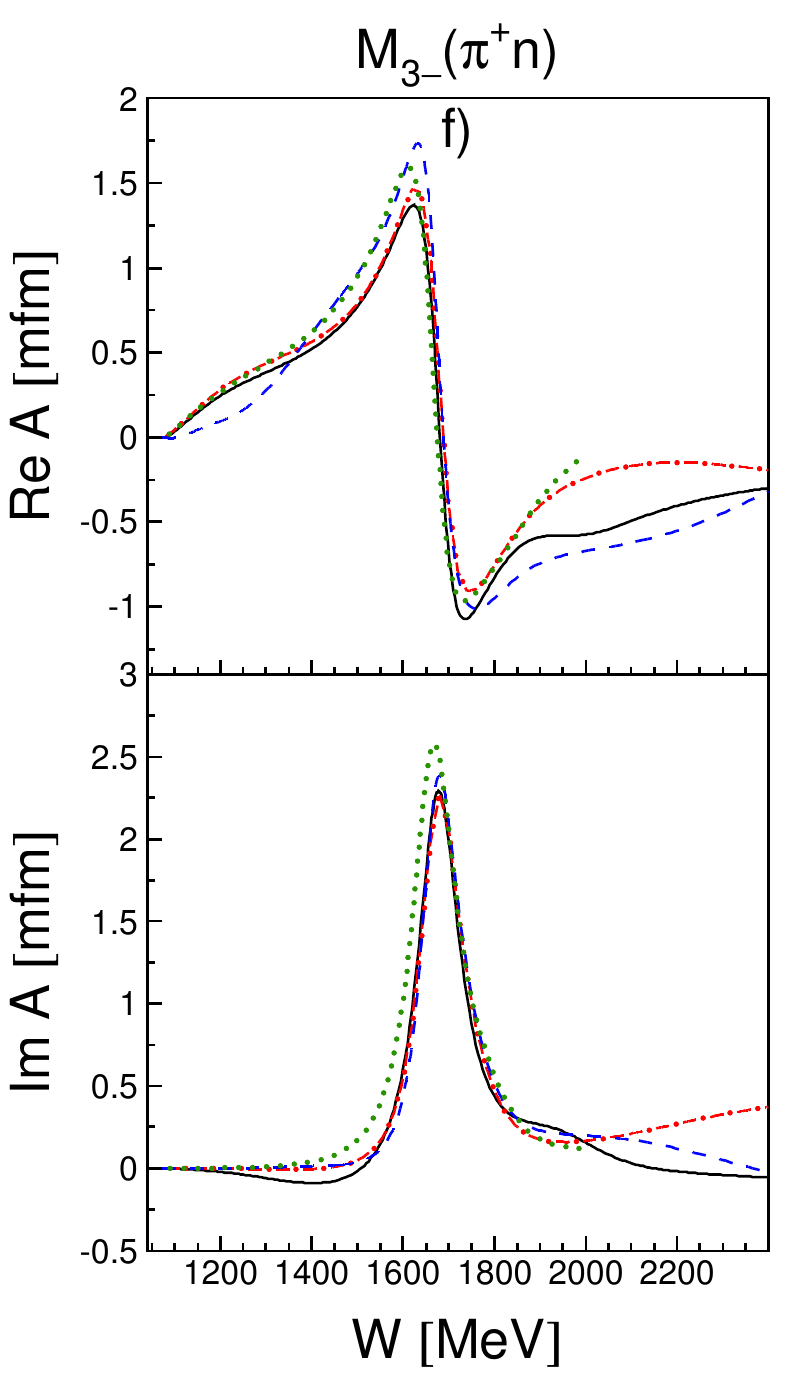,width=0.28\textwidth,height=0.42\textwidth}
\hspace{-12mm}\epsfig{file=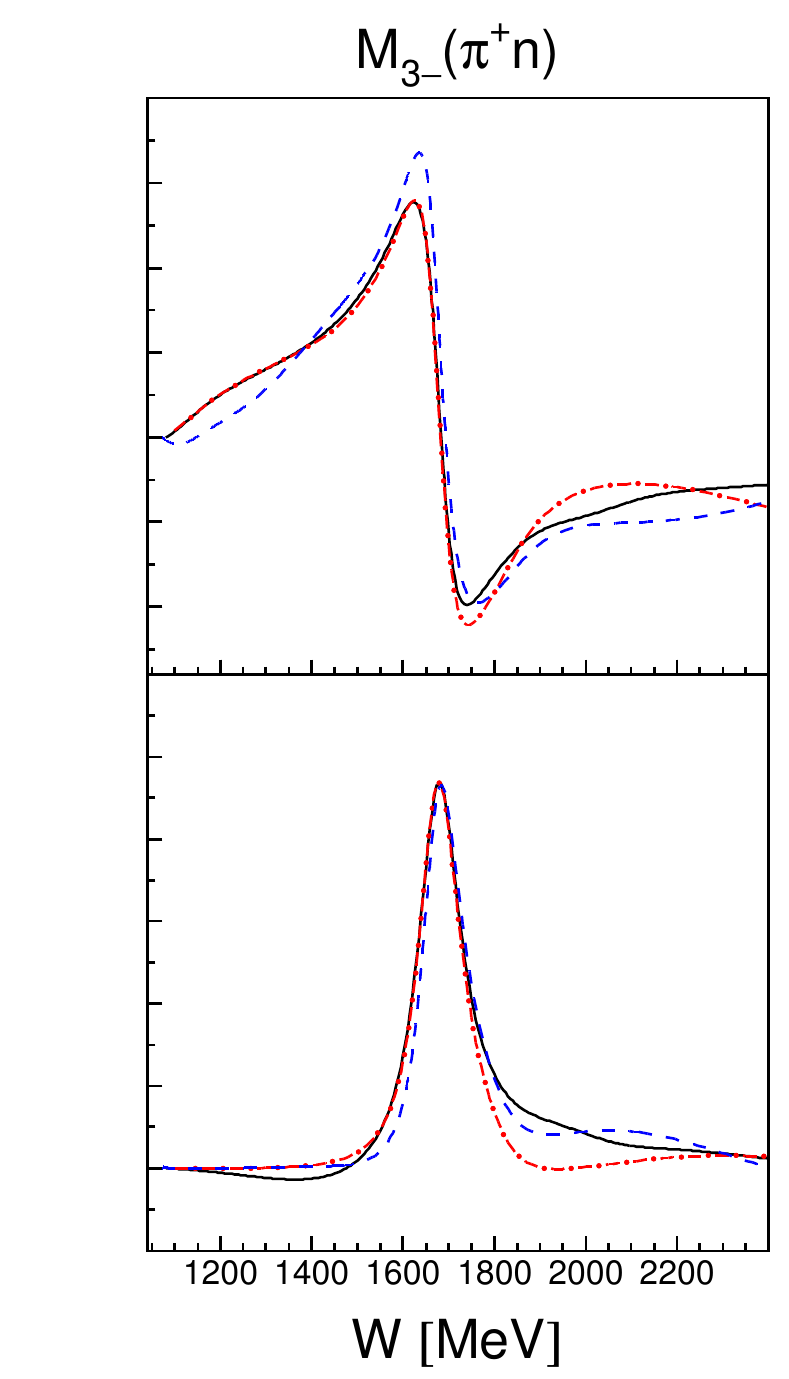,width=0.28\textwidth,height=0.42\textwidth}
}
\caption{\label{part3}Each block presents the real (top) and imaginary (bottom) part of multipoles for
$\gamma p\to \pi N$, before (left) and after (right) including new data. Black solid line: BnGa, blue dashed: J\"uBo, red dashed dotted: SAID, green dotted: MAID. Blocks a and c (b and d)  present the $I=3/2$ ($I=1/2$) multipoles; block e
shows the $\gamma p\to \pi^0 p$ multipole, f that for $\gamma p\to \pi^+ n$.
}
\end{figure*}

\begin{figure*}
\centerline{\epsfig{file=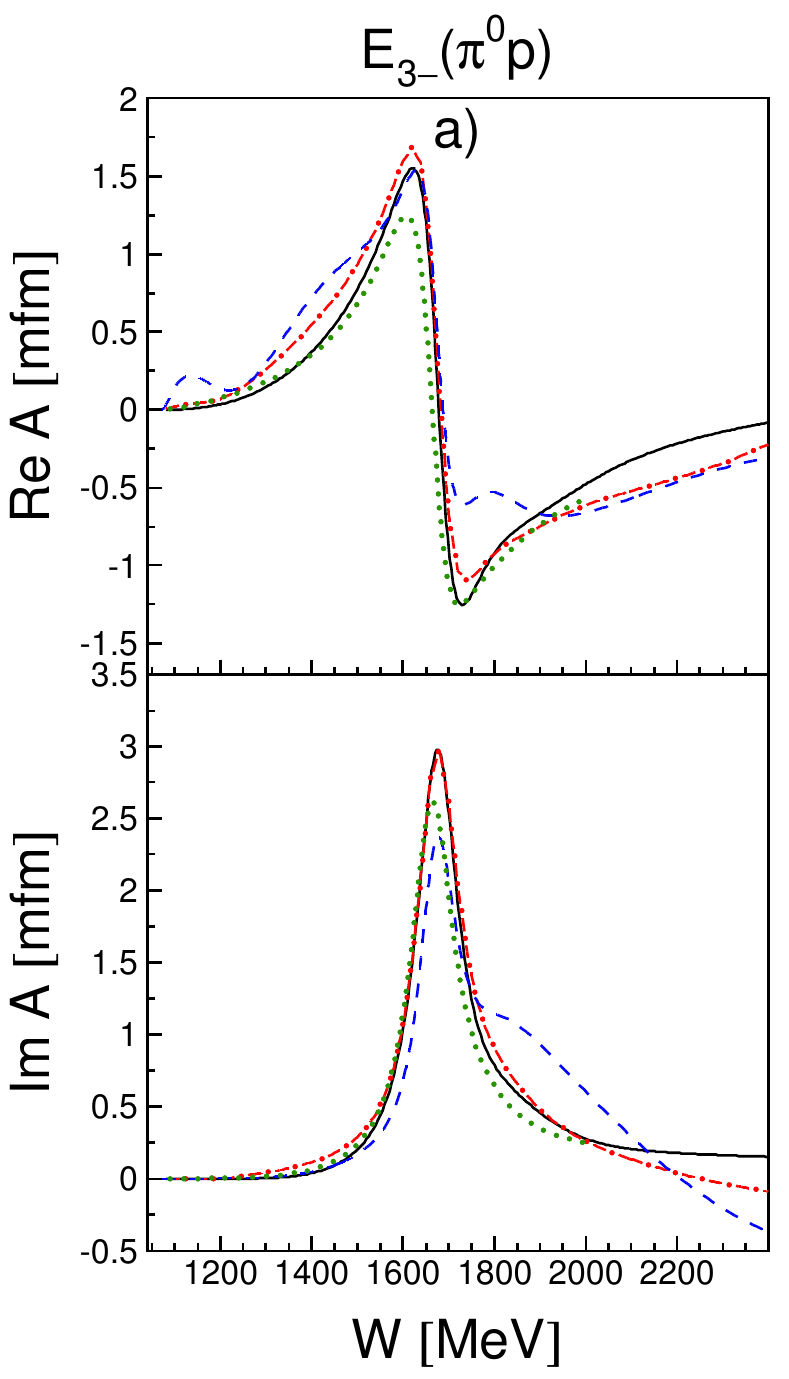,width=0.28\textwidth,height=0.42\textwidth}
\hspace{-12mm}\epsfig{file=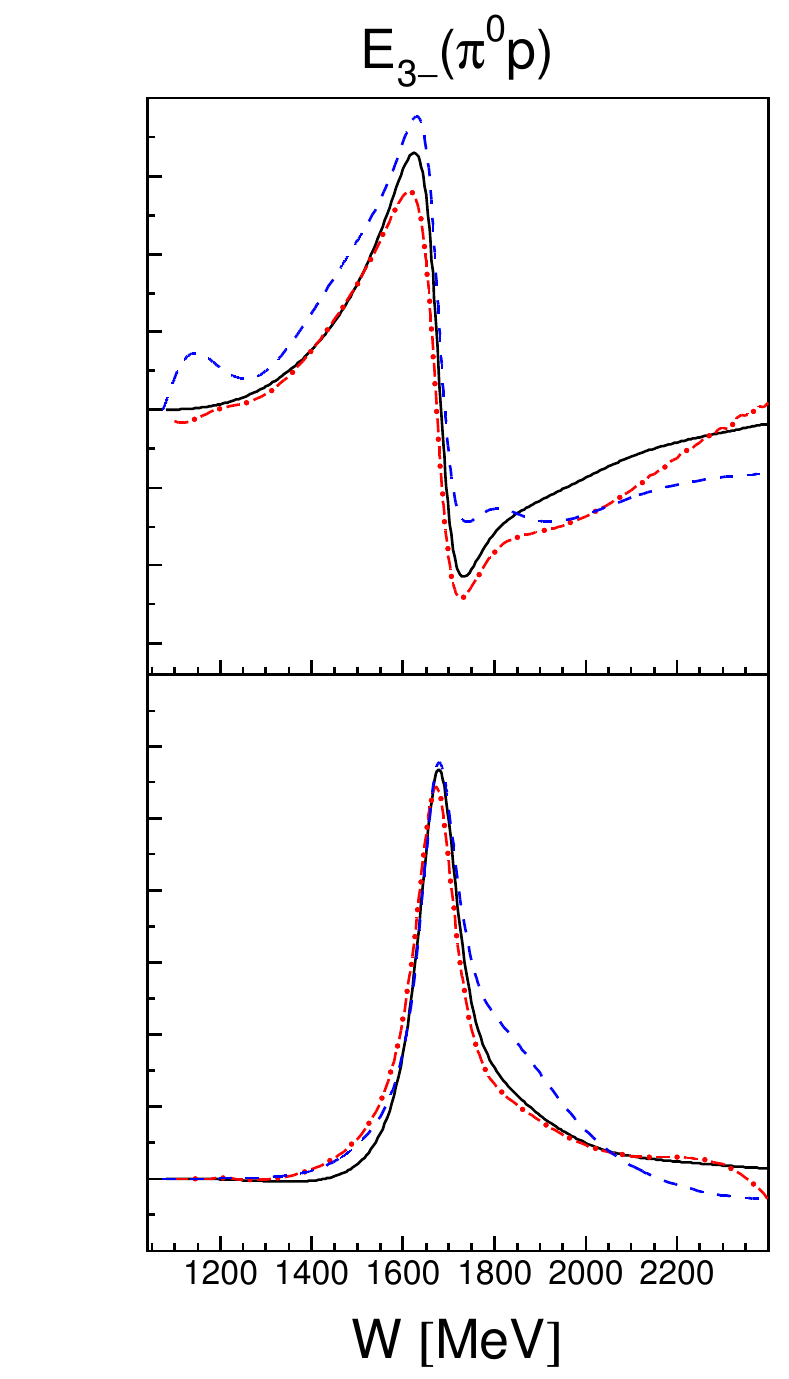,width=0.28\textwidth,height=0.42\textwidth}
              \epsfig{file=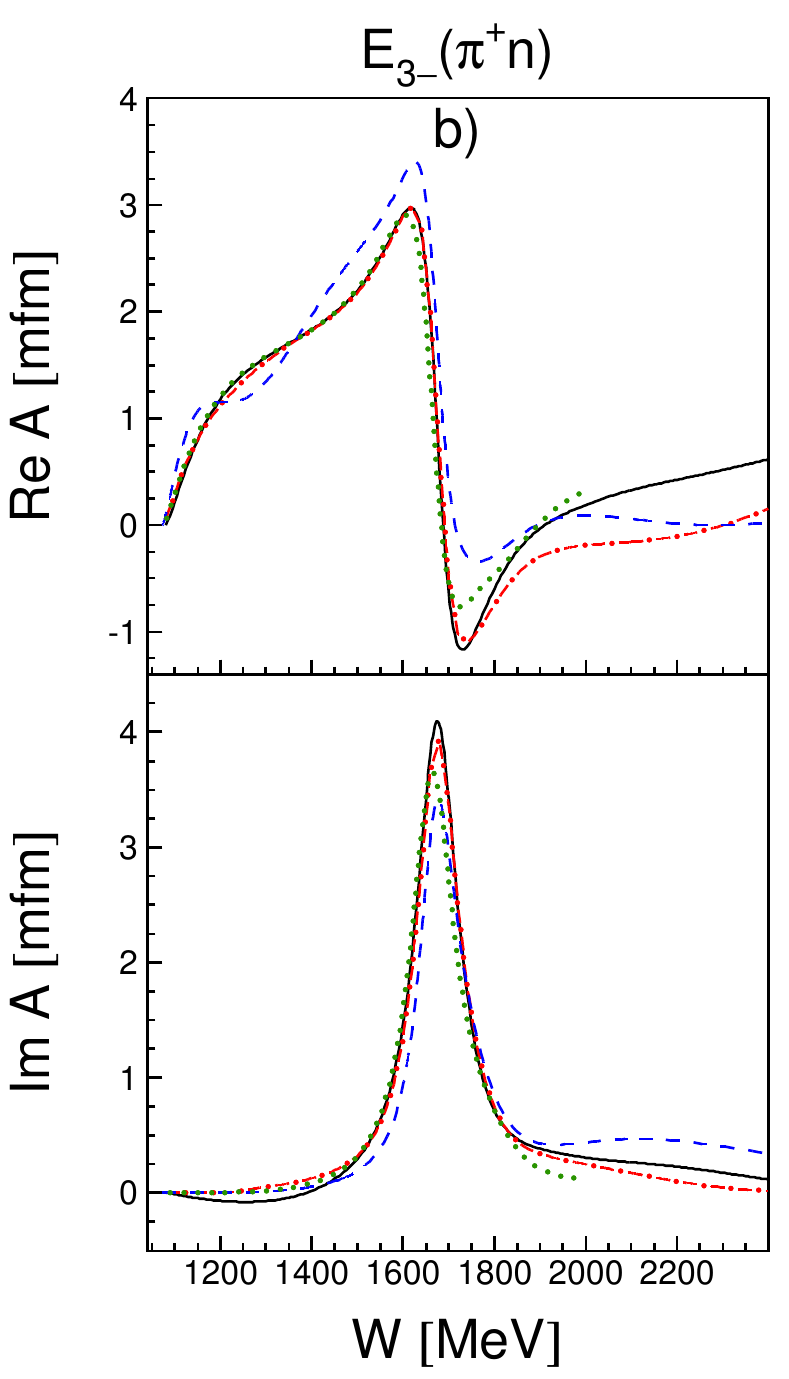,width=0.28\textwidth,height=0.42\textwidth}
\hspace{-12mm}\epsfig{file=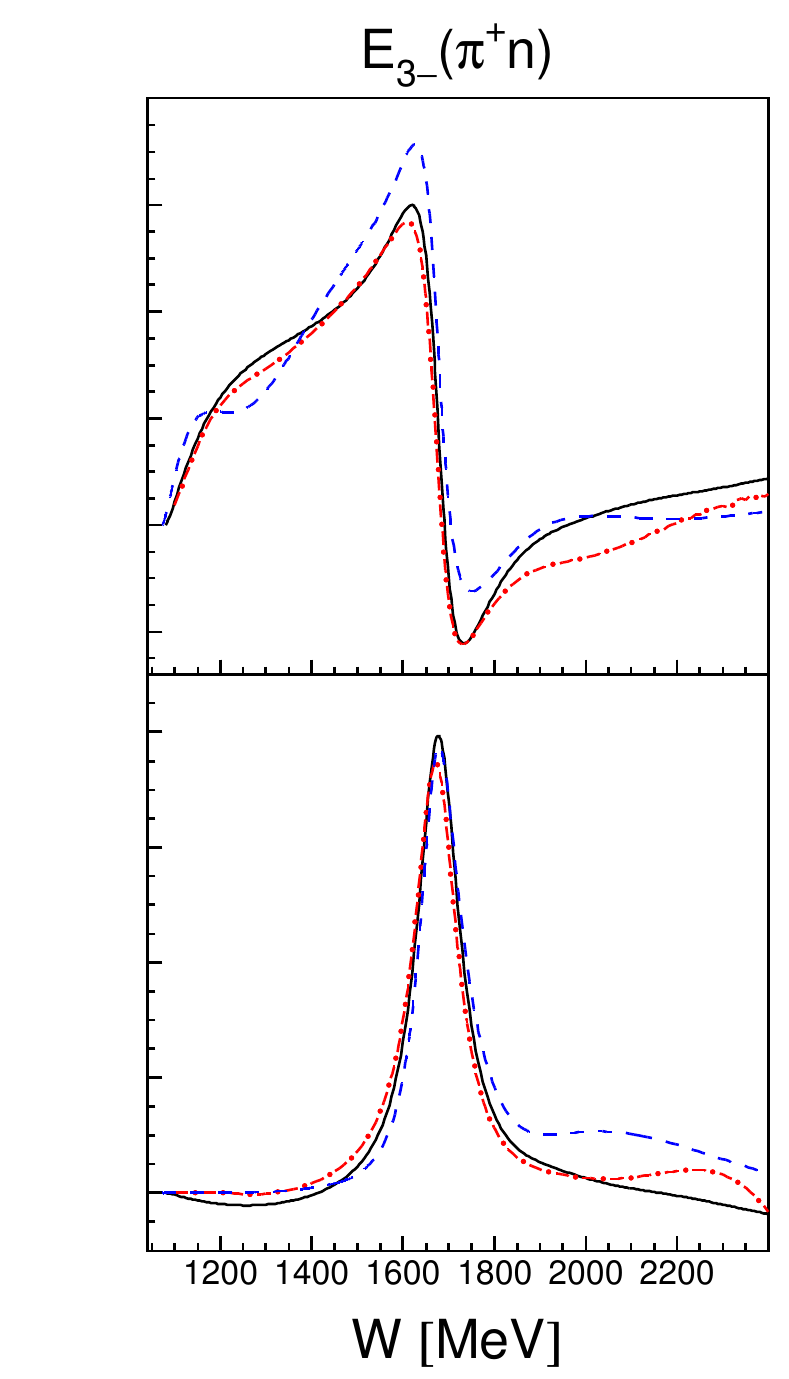,width=0.28\textwidth,height=0.42\textwidth}
}
\centerline{\epsfig{file=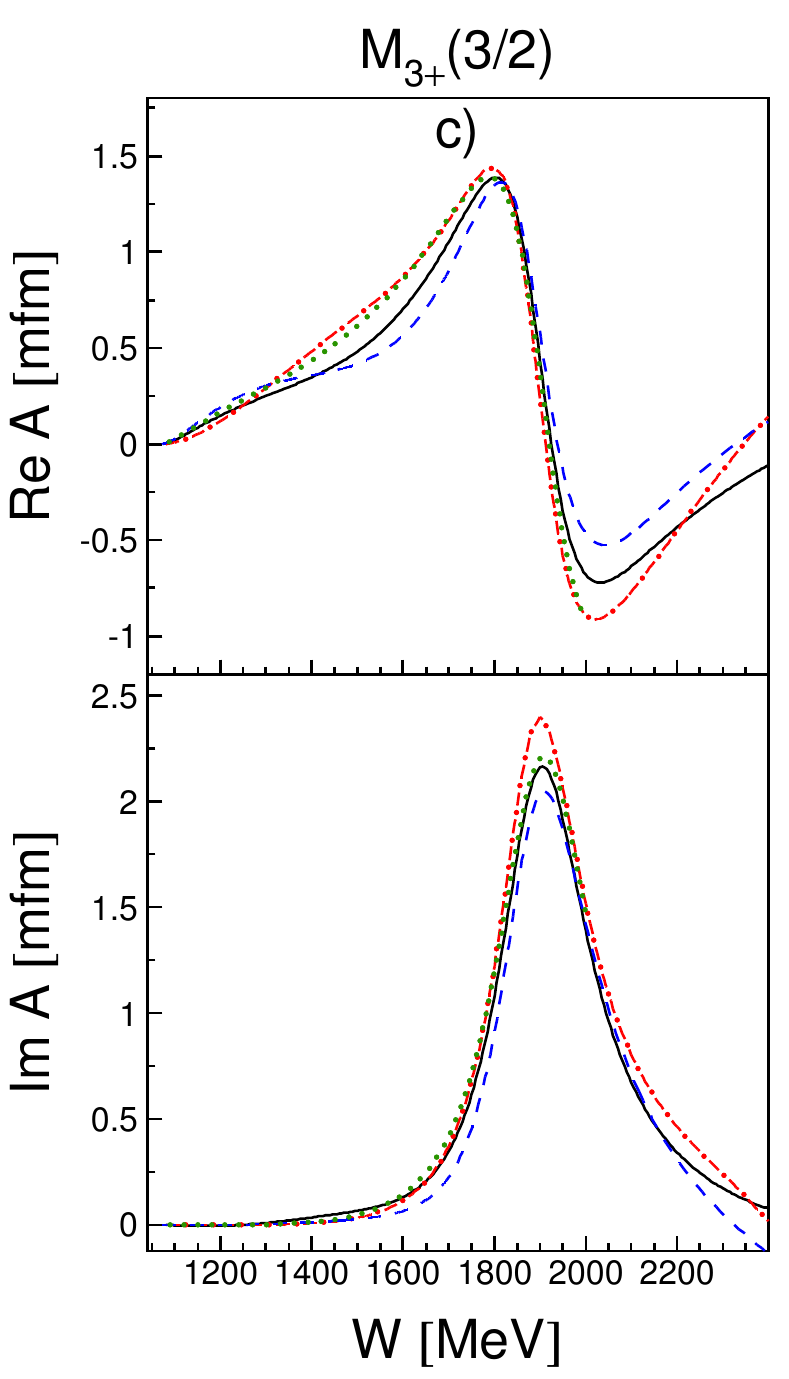,width=0.28\textwidth,height=0.42\textwidth}
\hspace{-12mm}\epsfig{file=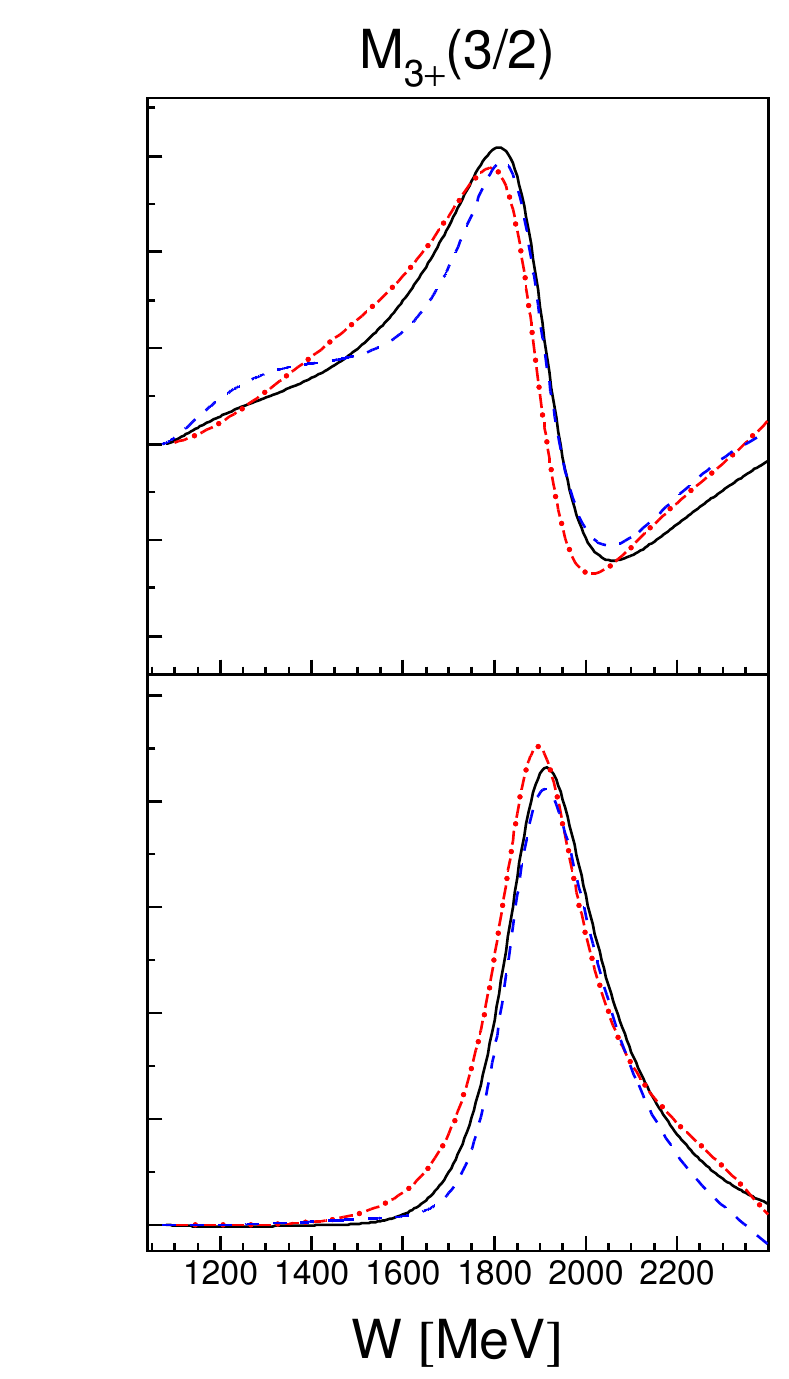,width=0.28\textwidth,height=0.42\textwidth}
              \epsfig{file=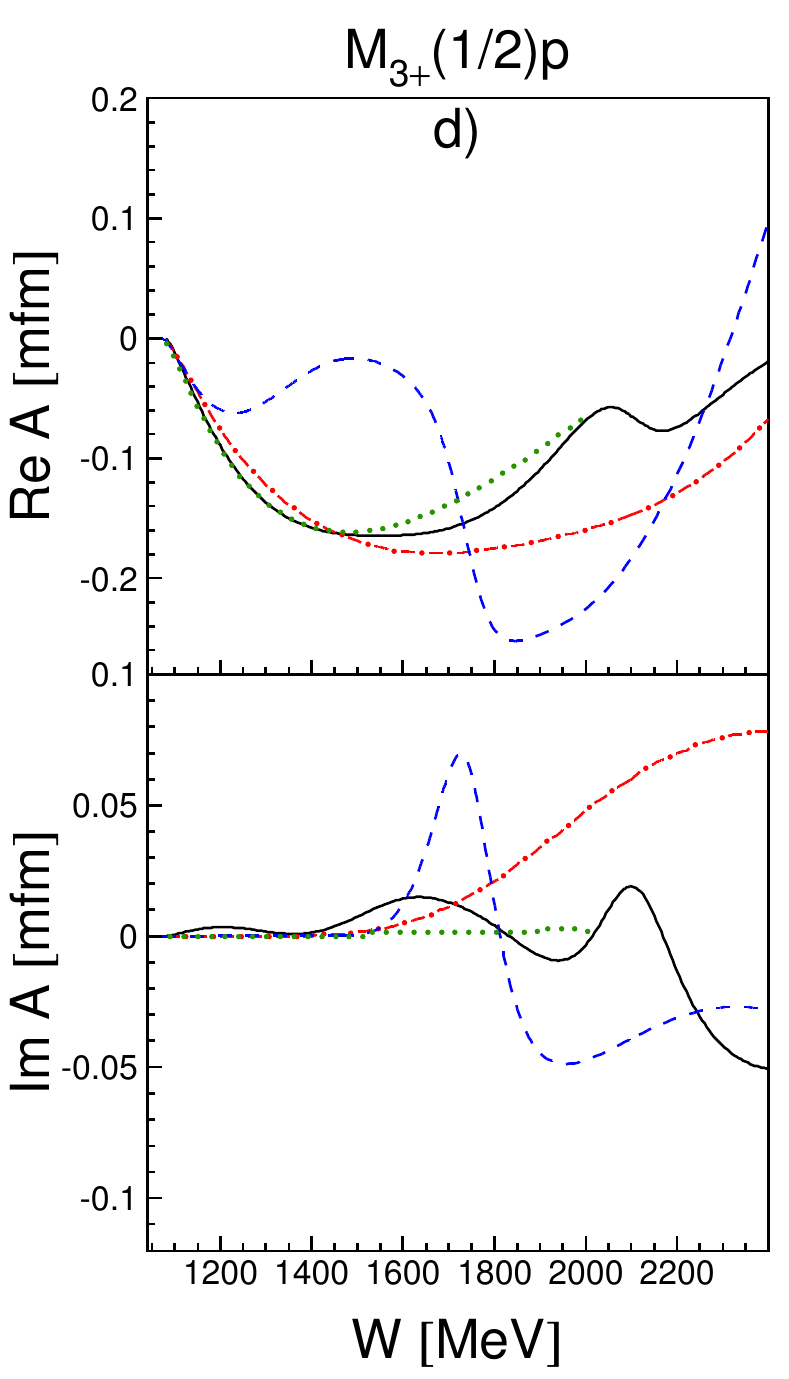,width=0.28\textwidth,height=0.42\textwidth}
\hspace{-12mm}\epsfig{file=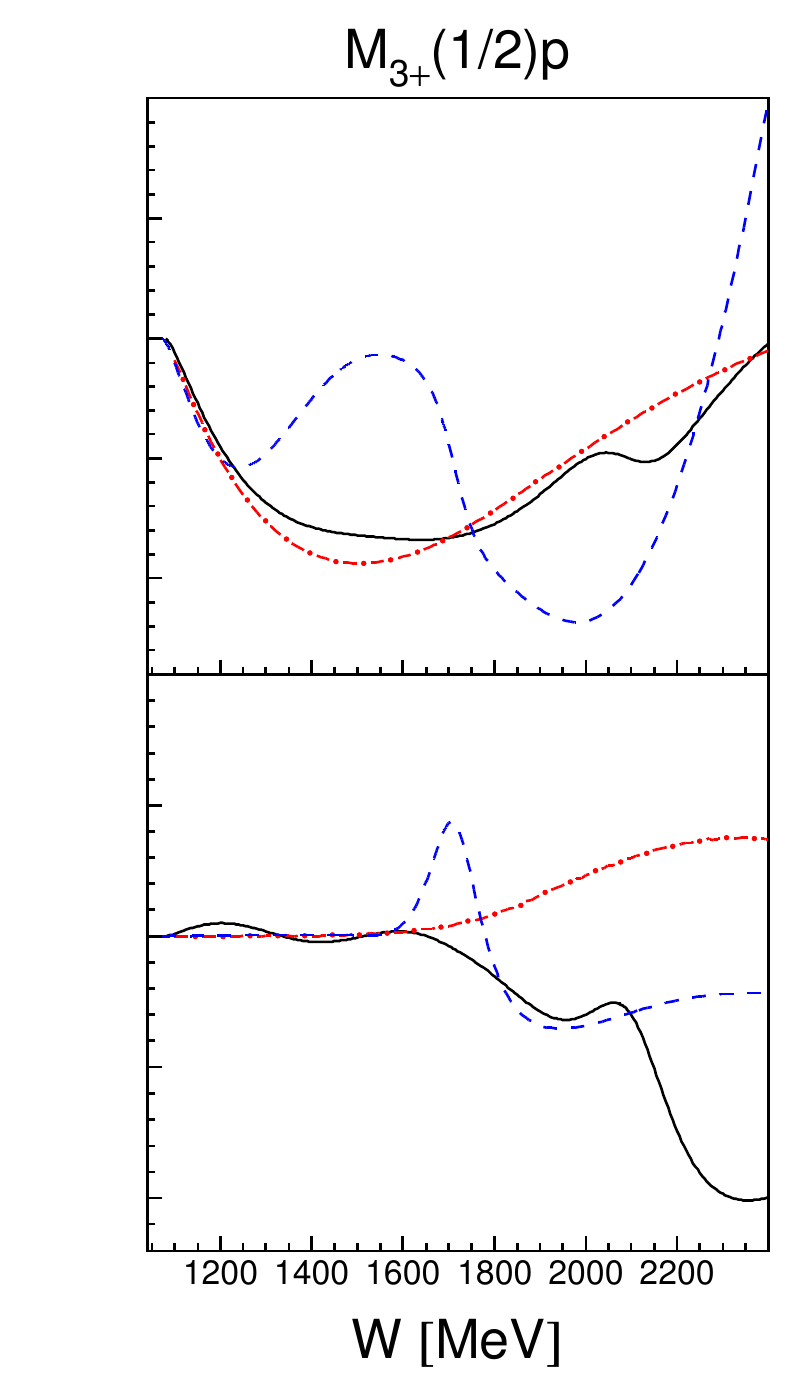,width=0.28\textwidth,height=0.42\textwidth}
}
\centerline{\epsfig{file=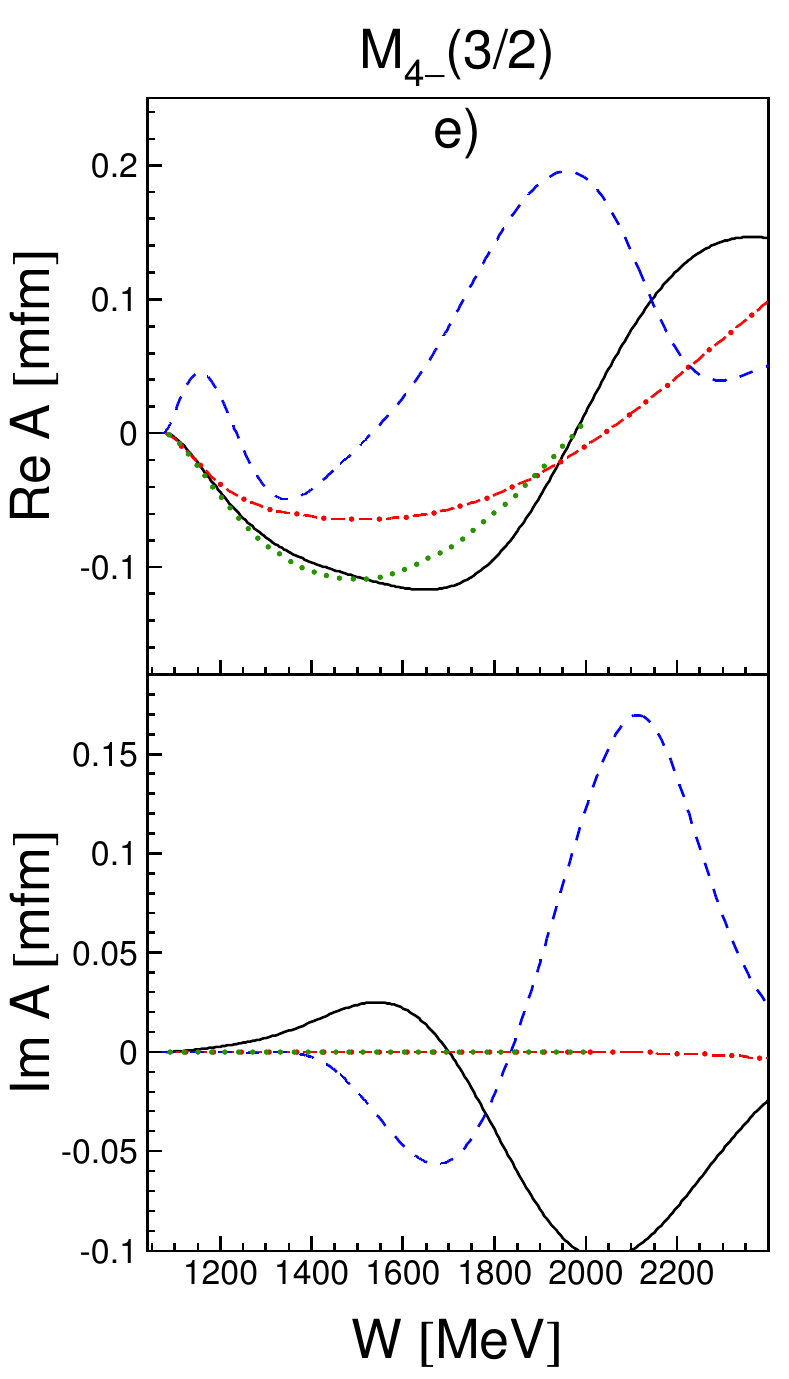,width=0.28\textwidth,height=0.42\textwidth}
\hspace{-12mm}\epsfig{file=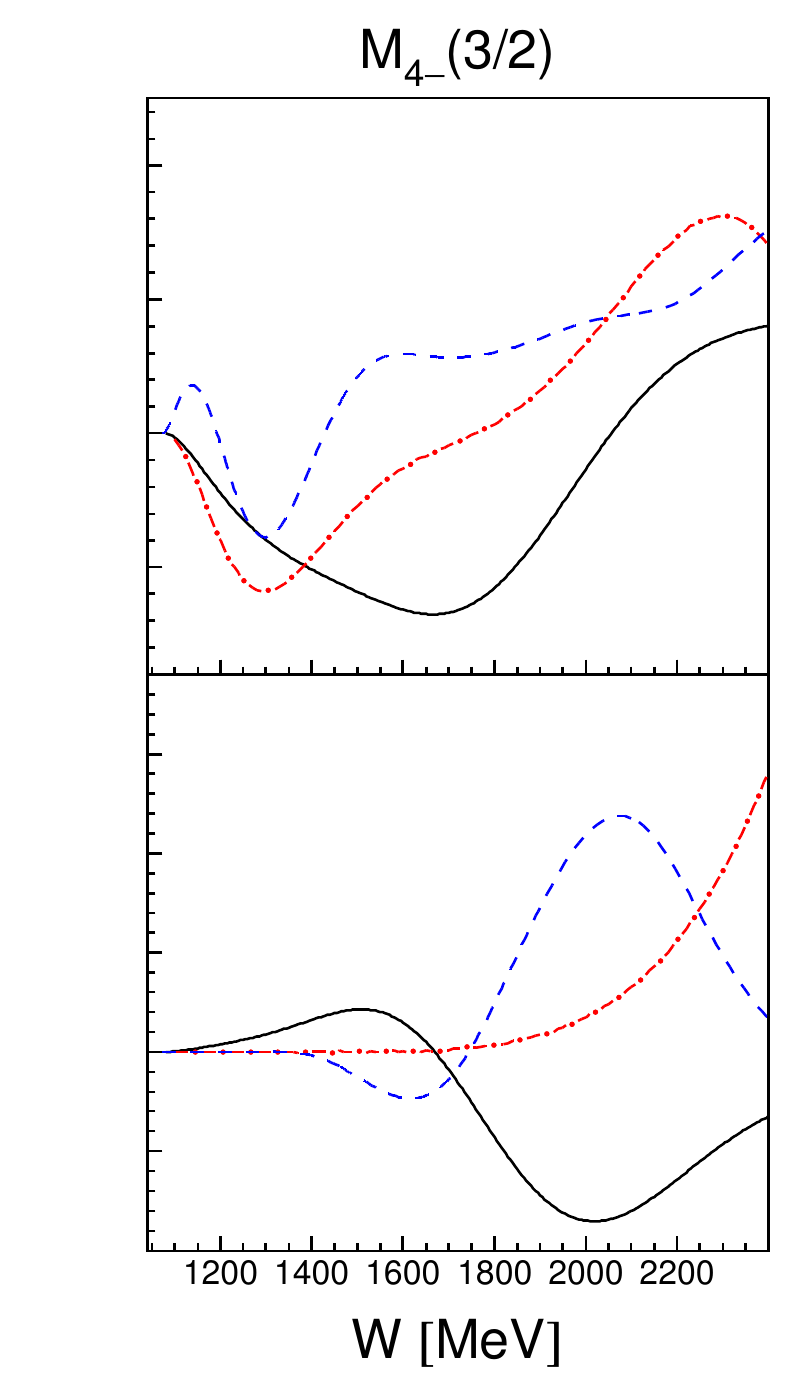,width=0.28\textwidth,height=0.42\textwidth}
              \epsfig{file=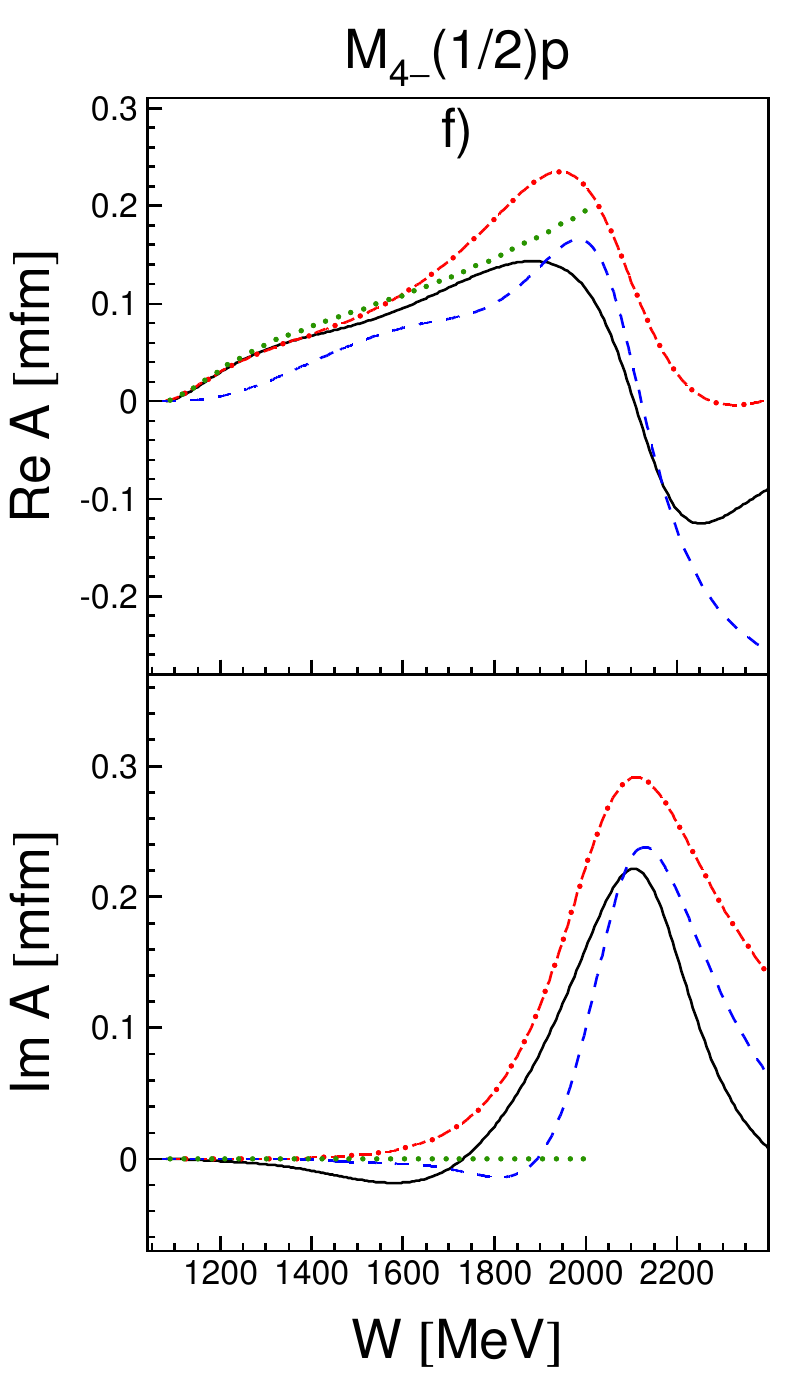,width=0.28\textwidth,height=0.42\textwidth}
\hspace{-12mm}\epsfig{file=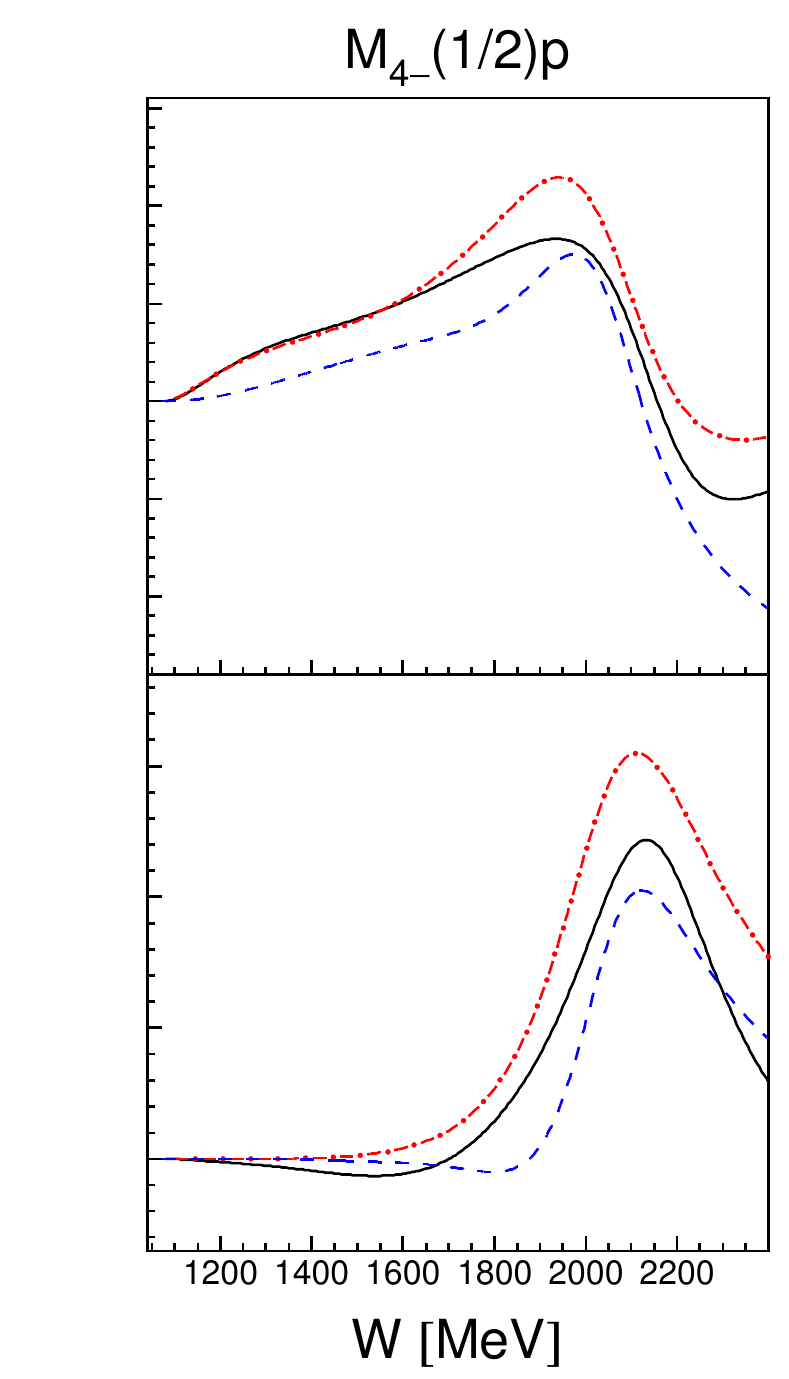,width=0.28\textwidth,height=0.42\textwidth}
}
\caption{\label{part4}Each block presents the real (top) and imaginary (bottom) part of multipoles for
$\gamma p\to \pi N$, before (left) and after (right) including new data. Black solid line: BnGa, blue dashed: J\"uBo, red dashed dotted: SAID, green dotted: MAID. Block a shows the $\gamma p\to \pi^0 p$ multipole, b that for $\gamma p\to \pi^+ n$.
Blocks c and e (d and f) present the $I=3/2$ ($I=1/2$) multipole.
}
\end{figure*}

The two reactions $\gamma p\to\pi^0p$ and $\gamma p\to\pi^+n$ can be described by 
multipoles in two different bases, in the physical channels, for example $E_{0+}(p\pi^0)$ and 
$E_{0+}(n\pi^+)$, or by the two isopin contributions, $I=1/2$ leading to the nucleon resonances 
$N^*$ and $I=3/2$ leading to the $\Delta$ resonances $\Delta^*$. In Figs.~\ref{part1} - \ref{part4}, 
we do not use an identical basis for all multipoles. For example, for the $\Delta(1232)3/2^+$ 
resonance we show the contributing multipoles $M_{1+}$ and $E_{1+}$ in the isospin basis 
$I=3/2$ and $I=1/2$. The $\Delta(1232)3/2^+$ resonance has an intrinsic quark spin $S=3/2$. 
Hence one quark in the proton has to undergo a spin-flip, the driving multipole is a magnetic 
multipole $M_{1+}$, and the electric multipole $E_{1+}$ is very small. The same argument
holds, e.g., for $N(1675)5/{2^-}$ where the contributing multipoles are the magnetic 
multipole $M_{2+}$ and the electric multipole $E_{2+}$. The $N(1675)5/{2^-}$ has an intrinsic orbital 
angular momentum $L_q=1$ (and at most very 
small components with higher angular momenta), hence its intrinsic quark spin must also be $S=3/2$, and
the relevant multipole driving the $N\to N(1675)5/2^-$ transition is a magnetic one, $M_{2+}$. Hence we 
present the multipoles in the isospin basis, $M_{2+}(I=1/2)$ and $M_{2+}(I=3/2)$.  
On the contrary, $N(1680)5/2^+$, excited by $E_{3-}$ and $M_{3-}$, 
is a member of a spin doublet, with the $N(1720)3/2^+$ resonance as its spin partner. Here both, 
electric and magnetic multipoles contribute with similar magnitude, and we show the physical 
multipoles $M_{3-}(p\pi^0)$, $M_{3-}(n\pi^+)$ and $E_{3-}(p\pi^0)$, $E_{3-}(n\pi^+)$. 
The selection of the basis, isospin  or physical channels, allows the reader to see the 
reaction-dependence of the different PWAs.

Figure~\ref{part1} shows the $E_{0+}$ multipole for the reactions $\gamma p\to \pi^0p$ (a) and
$\gamma p\to \pi^+n$ (b). The multipole leads to $\pi N$ in a relative
S-wave. The real part of the $\gamma p\to \pi^+n$ multipole reaches 40~mfm at threshold;
the multipole is much smaller for $\gamma p\to \pi^0p$. 
The sharp peak at low masses in $\pi^+n$ is due to $t$-channel pion exchange which is forbidden for
$\pi^0p$ production. All three PWAs yield the same pattern: a strong threshold enhancement
in $\gamma p\to \pi^+n$ and a peak in the imaginary part of the multipole due to $N(1535)1/2^-$. 
The real part of the multipoles for $\pi^0p$ and for $\pi^+n$ production exhibits a sharp spike 
at the $\eta$ production threshold.  
The multipoles show a very significant interference pattern between $N(1535)1/2^-$ and $N(1650)1/2^-$. The
latter resonance is much more pronounced in the $E_{0+}$ multipole for $\pi^+n$ production. 
When the different PWA solutions are
compared, it can be seen that the spread in the imaginary part of the multipole is reduced 
substantially when the new polarization data are taken into account. However, significant 
discrepancies remain, in particular in the low-mass region. The background contributions show 
a much wider spread than the resonant contributions.
However, also the (small) $N(1650)1/2^-$ contributions differ in the different PWAs.

The $M_{1-}$ multipole  (Fig.~\ref{part1}c,d) drives the excitation of the
$J^P=1/2^+$ partial wave containing the Roper $N(1440)1/2^+$ resonance, the three-star $N(1710)1/2^+$ 
resonance, the one-star $\Delta(1750)1/2^+$,
and the four-star $\Delta(1910)1/2^+$. The imaginary part of the $M_{1-}$ multipole evidences clearly
$N(1440)1/2^+$, the
contributions from the higher-mass resonances are small. The new data lead to a small improvement
of the consistency of the results for the imaginary part of the multipole.  In the real part a 
significant improvement can be observed.

The largest multipole is the $I=3/2$ $M_{1+}$ multipole (Fig.~\ref{part1}e), with 
the $\Delta(1232)3/2^+$ as the most prominent feature.
The imaginary part peaks above 1200\,MeV, and the real part goes through zero.
In this mass region, the different methods gave already very similar
predictions: very precise data and the Watson theorem were sufficient to determine the multipole
reliably (if the latter is fulfilled). However, small discrepancies are seen above the 
$\Delta(1232)3/2^+$. The real part of the amplitude is slightly smaller
in the BnGa analysis, this might be due to a small difference in the overall phase. It should be noted that 
BnGa includes the $\Delta(1600)3/2^+$ while in J\"uBo this state
emerges as dynamically generated. Also the SAID approach finds a resonance pole in the 
corresponding energy region. The consistency of the four
different analyses is hardly improved by the inclusion of the new data. The electric contribution, the
$E_{1+}$ multipole, is much smaller, see Fig.~\ref{part2}a. Also the $I=1/2$ contributions to the
$M_{1+}$ and $E_{1+}$ multipoles are small. The agreement between the different
PWAs increased when the new polarization data were taken into account.

The $E_{2-}$ and $M_{2-}$ multipoles drive resonances with $J^P=3/2^-$. The prominent peaks 
in the imaginary part of the amplitudes in Fig.~\ref{part2} c-f are due to the $N(1520)3/2^-$.
Its description by the different PWAs has converged to a nearly unique solution. Above, contributions are
expected from $N(1700)3/2^-$, $\Delta(1700)3/2^-$, $N(1875)3/2^-$ (a new entry
in the RPP and graded as three-star resonance), and $\Delta(1940)3/2^-$.

\begin{figure}[pb]
\begin{center}
\epsfig{file=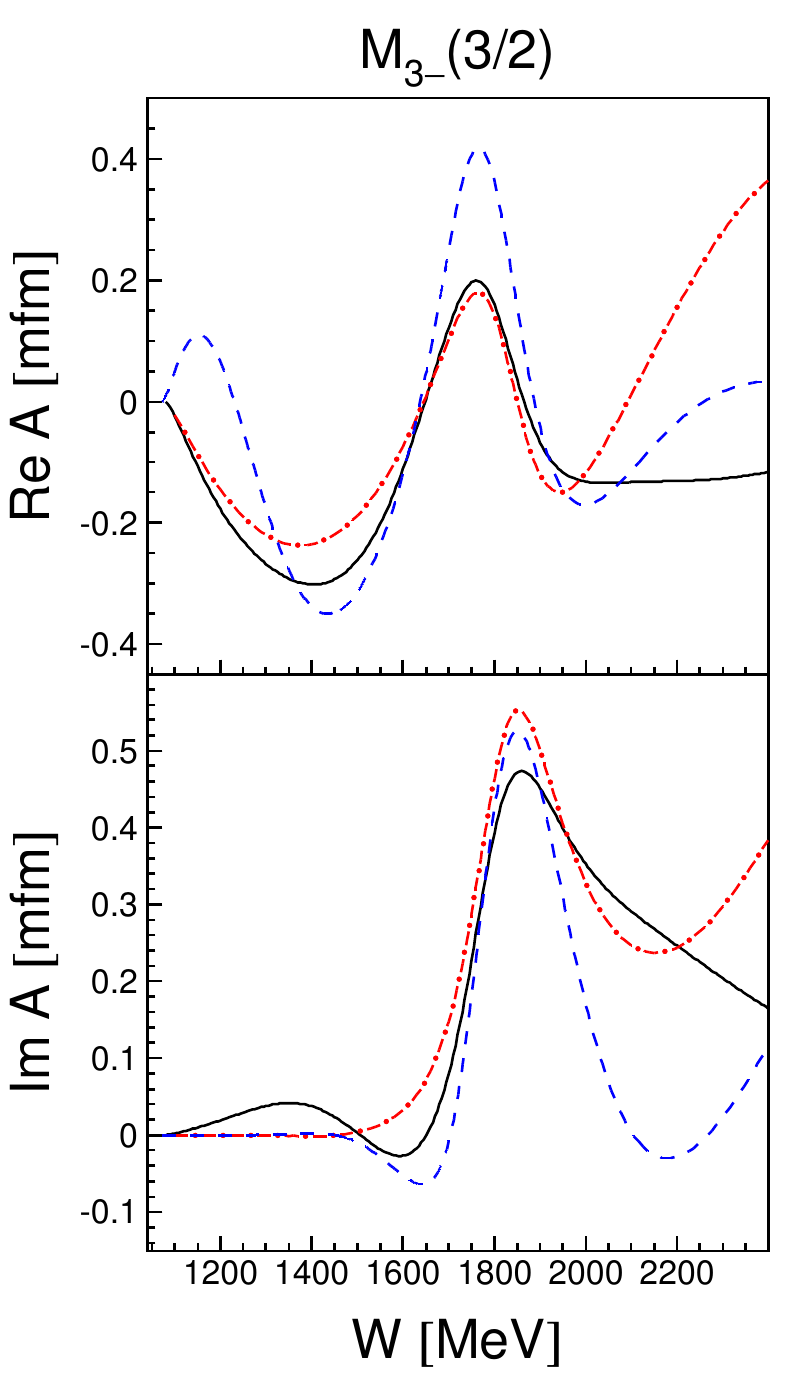,width=0.3\textwidth,height=0.5\textwidth}
\end{center}
\caption{\label{fivehalfplus}
The real and imaginary part of the $M_{3-}$ multipole for $\gamma p\to \pi N$ after including the new data 
(black solid line: BnGa, blue dashed: J\"uBo, red dashed dotted: SAID) evidences contributions from the $\Delta(1905)5/2^+$. 
}
\end{figure}

These four resonances contribute little to photoproduction
of single pions, their main decay modes lead to $\pi\pi N$ final states. Their treatment is different 
in the different approaches, so it is understandable that the multipoles do not show a consistent 
behavior above 1600\,MeV.

The $M_{2+}$ and $E_{2+}$ multipoles in Fig.~\ref{part3} are again separated according to their 
isospin content. The imaginary part of the $M_{2+}$ $I=1/2$ multipole peaks at about 1700\,MeV 
(Fig.~\ref{part3}b). The peak is due to the $N(1675)5/2^-$. In the quark model, this resonance has 
an intrinsic spin of $S=3/2$, and hence the spin of one quark of the hit proton must flip. Thus 
we expect a dominant contribution from the magnetic multipole. This expectation
is clearly met by the multipoles derived from the fits to the data. A second observation is that 
the spread of results from the different PWAs in the imaginary and real part of the $M_{2+}$ multipole
resonance is reduced visibly in the region of the $N(1675)5/2^-$ when the new polarization data are included
in the fits, while the $I=3/2$ background and the $E_{2+}$ multipole have a large variance.

The multipoles $M_{3-}$ and $E_{3-}$ both show a clear resonant behavior due to photoexcitation of
the $N(1680)5/2^+$ (Figs.~\ref{part3}e,f and~\ref{part4}a,b). We thus expect that the resonance must belong
to a quark spin $S=1/2$ doublet (with the $N(1720)$ $3/2^+$ as its spin partner). The consistency of 
the different PWA fits is rather good, and still improved with the new polarization data. In particular 
on the resonance position, the agreement between
the different PWAs is excellent. In this case, the multipoles are shown in the particle base, 
the $\Delta(1905)5/2^+$ is hardly seen. Figure~\ref{fivehalfplus} shows the multipoles in the isospin
basis. There is clear evidence for the $\Delta(1905)5/2^+$.

Figs.~\ref{part4}c,d show the $I=3/2$ and $I=1/2$ $M_{3+}$ multipoles. The $I=3/2$ $M_{3+}$ multipole
shows a very significant resonance, the $\Delta(1950)7/2^+$, which requires a quark spin flip. The electric
$E_{3+}$ multipole is correspondingly small and does not contain any evidence for a resonant structure.
Likewise, the $E_{4+}$, $M_{4+}$ -- leading to $J^P=9/2^+$ -- and higher multipoles are small and their
behavior is smooth.

The $M_{4-}$ multipoles are small (Fig.~\ref{part4}e,f), $E_{4-}$ (not shown)
is even smaller. BnGa, J\"uBo, and GWU/SAID agree that the $N(2100)7/2^-$
resonance plays some role in photoproduction.

\subsection{Consistency of the results}

\begin{figure}[pt]
\epsfig{file=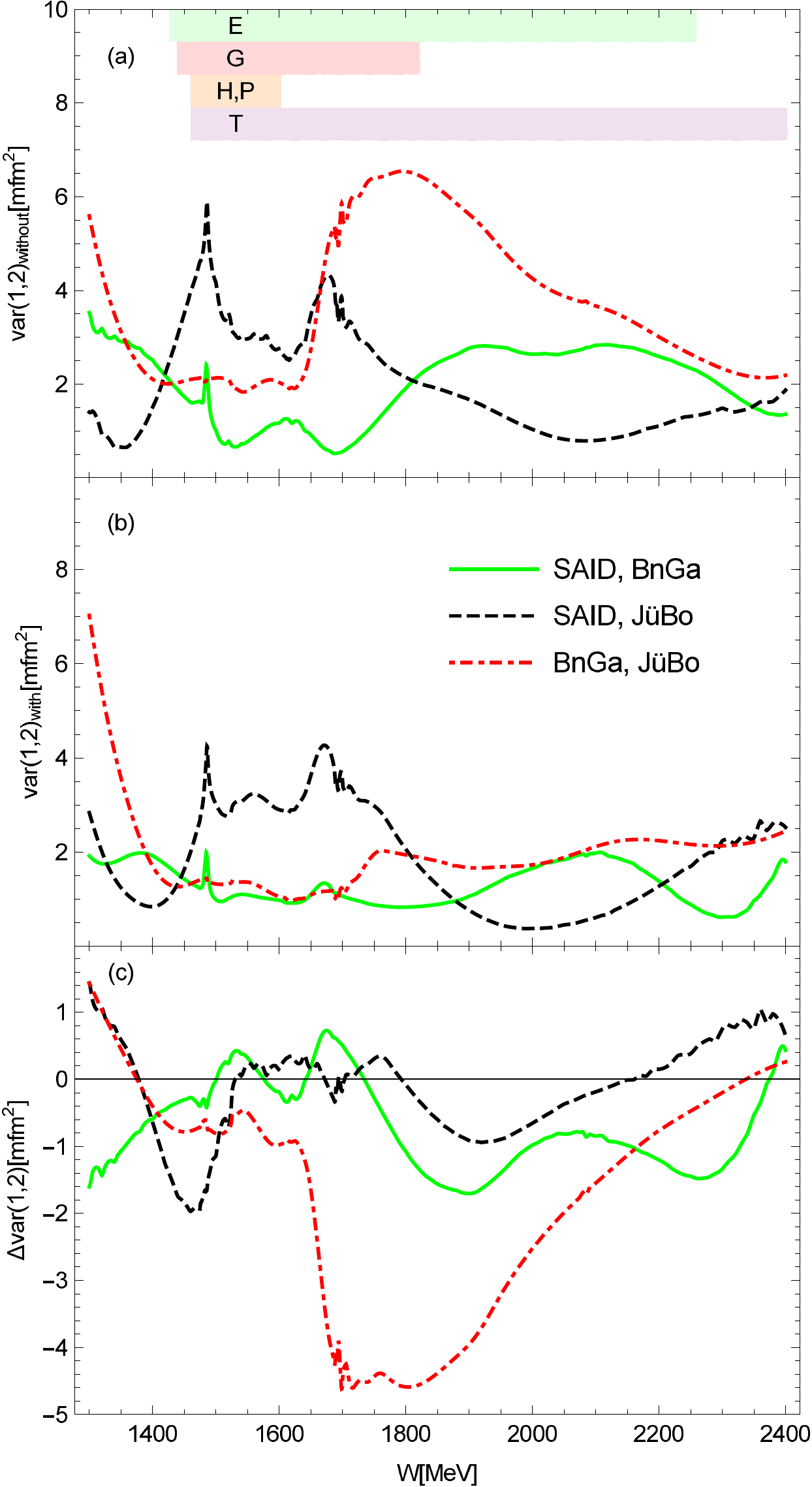,width=0.48\textwidth,height=0.64\textheight}
\caption{\label{compare}
The variances taken pairwise between two PWAs summed over all $\gamma p\to \pi^0 p$ multipoles up
to $L=4$. Solid (green): BnGa - SAID; dashed (black): SAID - J\"uBo; dashed-dotted (red): J\"uBo - BnGa.
a: before including the new data, b: after including the new data, c:
differences between a and b. The range covered by the new double polarization observables
\cite{Hartmann:2015kpa,Thiel:2015tbd,Gottschall:2015tbd} is indicated by shaded areas.
}
\end{figure}

As one might expect, the four partial wave analyses yield different amplitudes.
Ideally, the amplitudes should, in the limit of a complete data base with 
accurate data and full angular coverage, converge to the physical solution. In Fig.~\ref{compare} the different amplitudes are
compared. For this purpose, we calculate the variance 
between model 1 and 2 as the sum over the squared differences of the
 16 (complex) $\gamma p\to \pi^0 p$ multipoles ${\cal M}$ up to $L=4$:
\be
\hspace{-3mm}\mathrm{var}(1,2) = \frac12 \sum_{i=1}^{16}({\cal M}_1(i) 
- {\cal M}_2(i))({\cal M}_1^*(i) - {\cal M}_2^*(i))\, .
\ee
This quantity is plotted in Fig.~\ref{compare}a for the amplitudes before
and in Fig.~\ref{compare}b for the amplitudes after the new data were included.
The spike in Fig.~\ref{compare}a slightly below $W=1.5$\,GeV reflects the discrepancies 
in the description of the  $\eta p$ cusp between the approaches. Indeed, this is also directly 
visible for $E_{0+}$ shown in Fig.~3.  Once the new data are included, this discrepancy becomes 
smaller (Fig.~\ref{compare}b). A wider peak below $W=1.7$\,GeV  might
stem from slightly different $N(1680)5/2^+$ properties used in the three PWAs. Also
the wider peak becomes less pronounced when the new data are included in the fits. Large discrepancies
are observed in the BnGa-J\"uBo comparison which are reduced very significantly in the
new fits. Quite in general, all pairwise differences have become significantly smaller with the new data.
With the new data included, the  BnGa, J\"uBo, and SAID multipoles are now in closer
agreement at energies beyond 1.7~GeV. In the region from 1500 to 1700\,MeV, the BnGa prediction 
falls in between the J\"uBo and SAID predictions, thus BnGa agrees well with J\"uBo and SAID while a larger
discrepancy remains between the latter two models. The improvement can be made visible in a
figure (Fig.~\ref{compare}c) which displays the difference between 
Figs.~\ref{compare}a and b: it shows
negative values indicating that the situation has been improved.

\begin{figure}[pt]
\epsfig{file=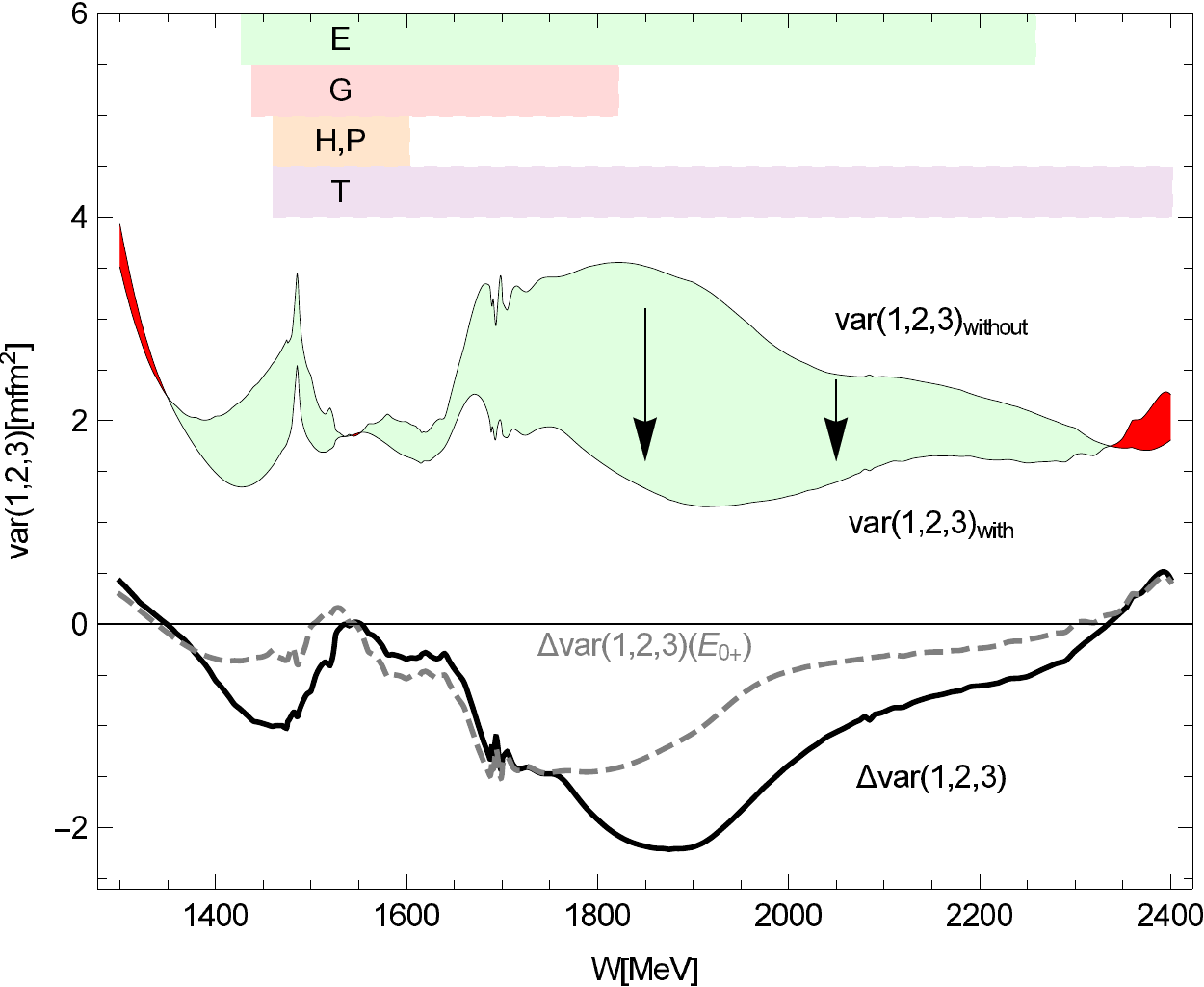,width=0.48\textwidth}
\caption{\label{allthree}
The variance of all three PWAs summed over all $\gamma p\to \pi^0 p$ multipoles   up
to $L=4$. The range covered by the new double polarization observables
are indicated by shaded areas. Over the largest part of the energy range
the new data have enforced an improvement of the overall consistency. The
improvement is displayed as light green area and, separately as difference
of the variance. The contribution to the improvement from the $E_{0+}$ wave
is shown as the dashed curve. Ranges with an overall deterioration
are marked in red.
}
\end{figure}

Figure~\ref{allthree} shows the reduction of the overall spread of the three partial wave analyses. 
Overall this spread is reduced considerably due to the impact of the new polarization variables
\cite{Hartmann:2015kpa,Thiel:2015tbd,Gottschall:2015tbd}. A significant fraction of the improvement 
stems from the $E_{0+}$ multipole exciting the $J^P=1/2^-$ wave (and thus the resonances $N(1535)$, 
$\Delta(1620)$, $N(1650)$, $N(1895)$, and $\Delta(1900)$).

\section{\label{Sum}Summary and conclusions}
We have presented a comparison of the multipoles derived from data on pion photoproduction
in four different partial wave analyses performed by the groups Bonn-Gatchina (BnGa), J\"ulich-Bonn
(J\"uBo), MAID from Mainz, and SAID from GWU. We have compared the multipoles from fits
made before new polarization data became available and from fits which included the
new data in the fitted data base. We find that the new data force the multipoles to get
closer to each other, the variance is reduced by about a factor of two.

Even more important seems to be that the multipoles converge to similar values
in the region of leading resonances while the ``background'' and the contribution
of higher-mass resonances remain less constrained by the new data. Clearly,
the aim is to get very similar answers also in the mass range which contains higher-mass
resonances. This task will require more precise data, in particular more precise
data on polarization observables. \\

{\small{\bf Acknowledgements:} This work was supported by the DFG (SFB/TR~16, SFB~1044, 
and the Sino-German CRC~110), the
Department of Energy (Offices of Science and Nuclear Physics)
award no. DE-FG02-99-ER41110, the National Science Foundation through the NSF/PIF grant No. PHY 1415459,  NSF/Career grant No. 1452055 and  the Chinese Academy of Sciences (CAS) President's 
International Fellowship Initiative (PIFI) (Grant No. 2015VMA076).
M.~D., U.-G.~M. and D.~R. gratefully acknowledge the
computing time granted on the supercomputer JURECA at the J\"ulich Supercomputing Centre (JSC). A.V.A., V.N. and A.S. acknowledge the support from RSF grant 16-12-10267.

}
\end{document}